\documentclass[12pt]{style/spieman}  
\usepackage{amssymb,amsbsy,amsmath,amsfonts,amscd,mathrsfs,upgreek}
\usepackage{graphicx}
\usepackage{setspace}
\usepackage{tocloft}
\usepackage{lineno}
\usepackage{placeins}
\usepackage{color}
\usepackage{xcolor}
\usepackage{soul}
\usepackage{tabularx}
\usepackage{makecell}

\title{I-FENN with DeepONets: accelerating simulations in coupled multiphysics problems}

\author[a,b,c]{Fouad M. Amin} 
\author[b,d]{Diab W. Abueidda}
\author[b]{Panos Pantidis}
\author[a,b]{Mostafa E. Mobasher}

\affil[a]{Civil and Urban Engineering Department, Tandon School of Engineering, New York University, Brooklyn, NY11201, US}
\affil[b]{Civil and Urban Engineering Department, New York University Abu Dhabi, Abu Dhabi, P.O. Box 129188, UAE}
\affil[c]{Structural Engineering Department, Faculty of Engineering, Mansoura University, Mansoura 35516, Egypt}
\affil[d]{National Center for Supercomputing Applications, University of Illinois at Urbana-Champaign, Urbana, IL, USA}

\begin{document}

\maketitle

\begin{abstract}
Coupled multiphysics simulations for high-dimensional, large-scale problems can be prohibitively expensive due to their computational demands. This article presents a novel framework integrating a deep operator network (DeepONet) with the Finite Element Method (FEM) to address coupled thermoelasticity and poroelasticity problems. This integration occurs within the context of I-FENN, a framework where neural networks are directly employed as PDE solvers within FEM, resulting in a hybrid staggered solver. In this setup, the mechanical field is computed using FEM, while the other coupled field is predicted using a neural network (NN). By decoupling multiphysics interactions, the hybrid framework reduces computational cost by simplifying calculations and reducing the FEM unknowns, while maintaining flexibility across unseen scenarios. The proposed work introduces a new I-FENN architecture with extended generalizability due to the DeepONets ability to efficiently address several combinations of natural boundary conditions and body loads. A modified DeepONet architecture is introduced to accommodate multiple inputs, along with a streamlined strategy for enforcing boundary conditions on distinct boundaries. We showcase the applicability and merits of the proposed work through numerical examples covering thermoelasticity and poroelasticity problems, demonstrating computational efficiency, accuracy, and generalization capabilities. In all examples, the test cases involve unseen loading conditions. The computational savings scale with the model complexity while preserving an accuracy of more than 95\% in the non-trivial regions of the domain.

\end{abstract}


{\noindent \footnotesize\textbf{*} Fouad Amin, \linkable{fma9357@nyu.edu} }
{\noindent \footnotesize\textbf{$\dag$} Diab W. Abueidda, \linkable{da3205@nyu.edu} }
\newline
{\noindent \footnotesize\textbf{$\ddag$} Panos Pantidis, \linkable{pp2624@nyu.edu} }
{\noindent \footnotesize\textbf{\S} Mostafa E. Mobasher, \linkable{mostafa.mobasher@nyu.edu} }

\section{Introduction}
\label{section:introduction}

\subsection{Literature review}
\label{section:introduction:literature_review}

A wide range of natural phenomena incorporates multiphysics interactions, often characterized by the coupling between the mechanical behavior and other physical fields. Thermomechanical coupling, for example, plays a critical role in diverse applications such as additive manufacturing \cite{santi_multiphysics_2023}, thermal protection systems \cite{xu_thermomechanical_2019}, and metal solidification \cite{koric_multiphysics_2010}. Hydromechanical coupling is essential in seepage and consolidation problems \cite{guo_uncertainty_2021}, fluid-structure interaction \cite{kloppel_fluidstructure_2011}, and slope stability analysis \cite{moradi_comparing_2024}. More complex multiphysics processes exist in several domains, including infrastructure, energy, and manufacturing \cite{han_thermalhydraulicmechanicalchemical_2023, abdullah_two-phase_2024, churakov_position_2024}. The large-scale and intricate character of these phenomena naturally limits the scope of experimental approaches for their investigation, rendering computational methods the primary pathway for studying the evolution of the phenomena. Modeling of these multiphysics systems is often based on conventional numerical approaches such as Finite Element Method (FEM), Finite Difference/Volume Method (FDM/FVM) \cite{keyes_multiphysics_2013}, among others \cite{wang_review_2022, liu_virtual_2023, chen_review_2022, diehl_comparative_2022}. These techniques are often associated with high computational costs, which can be prohibitively expensive for complex, high-dimensional, large-scale problems \cite{moradinia_navigating_2024, meinecke_data-driven_2024}. To address these challenges, researchers have developed a range of numerical remedies, including domain decomposition, parallel computing, staggered solvers, and adaptive remeshing techniques \cite{tang_review_2021,permann_moose_2020, castellazzi_staggered_2021, stershic_adaptively_2025}. Advancing existing computational capabilities remains a vibrant and rapidly evolving area of research, with a profound impact on our understanding of multiphysics real-world problems. 

Numerous machine learning (ML) approaches have been proposed to address the computational challenges associated with coupled mechanics problems. These approaches can be categorized into two main groups based on training strategies: (1) data-driven methods, and (2) physics-informed methods. Data-driven methods rely solely on supervised learning using labeled datasets to train the ML models. On the other hand, physics-informed methods incorporate physics into loss functions or manipulate the ML model output to obey physical laws \cite{karniadakis_physics-informed_2021}. Another categorization approach for ML adoption in computational mechanics can be based on the methodology: (1) simulation substitution, and (2) simulation enhancement \cite{herrmann_deep_2024}. Simulation substitution methods aim to replace the entire numerical simulation with an ML (surrogate) model to predict all response fields, whereas simulation enhancement focuses on improving a specific component of the simulation, including preprocessing, modeling, and postprocessing. 

Neural networks (NNs) play a central role in the successful application of ML-based methods to computational mechanics problems. NNs are nonlinear approximators of functions, mapping input to output variables through a set of learnable parameters\cite{pinkus_approximation_1999}. Several types of networks have been developed over the years, including multi-layer perceptrons (MLPs), sequence models (RNNs, GRUs, and transformers), and others \cite{chung_empirical_2014, vaswani_attention_2017}. Recently, to further enhance the expressivity and generalization capabilities of ML techniques, operator-based architectures have emerged and are being actively developed. Lu et al. \cite{lu_learning_2021} proposed the deep operator network (DeepONet), showcasing its ability to learn both explicit operators (such as integrals) and implicit operators (such as differential equations). Alongside DeepONet, alternative deep operator networks have also been introduced \cite{raonic_convolutional_2023, sarkar_spatio-spectral_2025, li_fourier_2020}. A typical DeepONet architecture consists of two main components: (1) a branch network, and (2) a trunk network. The branch encodes the discretized input function at fixed scattered locations called "sensors", while the trunk encodes the domain of the output function described by a set of target locations that can be selected arbitrarily. The branch and trunk outputs are then aggregated through a dot product operation to produce the final discretized output at the target locations. Given its advanced generalization capabilities, DeepONet has been used in a wide range of applications, including large-scale carbon storage operations \cite{kadeethum_improved_2024} and domain decomposition \cite{wang_time-marching_2025}, using an innovative hybrid NN-FEM framework. In addition, several variations of DeepONet have been introduced, aiming to enhance its generalization capabilities and overcome training challenges. For instance, the multiple-input deep neural operators (MIONet) \cite{jin_mionet_2022} extends DeepONet to handle multiple input functions. Another example is Fusion DeepONet\cite{peyvan_fusion-deeponet_2025}, which implements data transfer between the branch and trunk hidden layers.  

Despite the efficient computational performance of ML models, they suffer from significant limitations like limited interpretability \cite{sun_data-driven_2022, manfren_data-driven_2022}, generalizability \cite{fuhg_physics-informed_2022}, reliability \cite{pantidis_error_2023, daw_rethinking_2022}, sensitivity to hyperparameters \cite{pantidis_error_2023, markidis_old_2021}, network convergence \cite{shin_convergence_2020, mishra_estimates_2022}, and training stability \cite{krishnapriyan_characterizing_2021, wang_understanding_2021}. Therefore, the simulation substitution approach struggles in large-scale applications because it demands massive datasets and expensive training. In addition, without enforcing governing equations, time-history predictions can suffer from uncontrolled error propagation. In this context, it is evident that there is a need for more robust and reliable hybrid approaches that can combine the strengths of both numerical methods and ML models, while addressing their limitations. To this end, several hybrid formulations have been developed, such as the Integrated Finite Element Neural Network (I-FENN) proposed by the authors \cite{pantidis_integrated_2023}, as well as other approaches which integrate neural networks and operator networks with FEM for multiple applications such as FEMIN \cite{thel_introducing_2024}, hybrid FEM-NN \cite{mitusch_hybrid_2021, meethal_finite_2023}, NNFE \cite{zhang_simulation_2022}, and hybrid PI-DeepONet with FEM \cite{wang_time-marching_2025}. I-FENN is a framework designed to accelerate and improve the accuracy of multiphysics simulations, and it is conceptually distinct from the frameworks mentioned above. The core novelty of I-FENN is to employ NNs as PDE approximators in conjunction with a conventional FEM solver, establishing a hybrid scheme that is conceptually similar to staggered approaches for coupled problems. This approach is different from the other hybrid approaches \cite{thel_introducing_2024, mitusch_hybrid_2021, meethal_finite_2023,zhang_simulation_2022,wang_time-marching_2025}, where a NN is used to approximate solutions for a specific part of the mesh or a specific term in a balance law, or as a surrogate model trained using a differentiable FEM representation. Over the past years, the authors have demonstrated the applicability of I-FENN across several linear and non-linear problems, including phase-field fracture\cite{pantidis_integrated_2025}, non-local gradient damage\cite{pantidis_i-fenn_2024}, and thermoelasticity\cite{abueidda_i-fenn_2024, abueidda_variational_2024}, while investigating the error convergence analysis and model performance\cite{pantidis_error_2023}. However, each of these efforts had its unique limitations and challenges, including the need for expensive training, data-driven dependencies, and others. One of the main issues was the poor generalization capabilities, with methods often restricted to a single snapshot in time or a specific load history. 

\subsection{Scope and Outline}
\label{section:introduction:scope_and_outline}

In this study, we expand the scope and modeling capabilities of I-FENN, targeting two- and three-dimensional thermoelasticity and poroelasticity problems under varying (body and surface), unseen loading conditions. For the first time, we demonstrate the feasibility and efficiency of integrating a neural operator (DeepONet) as the network of choice within I-FENN, a development that significantly expands the framework's generalization capability. We propose a modified MIONet architecture to accommodate multiple input functions. In addition, we introduce a simplified approach for enforcing boundary conditions across different boundary segments. The proposed architecture comprises a fully connected network (MLP) in the trunk and a gated recurrent unit (GRU) in the branch, with the latter capturing temporal dependencies without requiring a predetermined number of prior input time steps. \cite{chung_empirical_2014}. The framework is integrating a finite element solver built upon the deal.II library \cite{africa_dealii_2024}. Deal.II is a widely used open-source C++ library that supports creating finite element solvers with the capability of parallel processing. The framework implementation enables high-performance computing (HPC) deployment while maintaining efficient data transfer and memory management.

The proposed framework is applied to three key case studies: (1) a thermoelasticity problem with a spatially and temporally varying thermal body load, (2) a thermoelasticity problem with a spatially and temporally varying thermal surface load, and (3) a poroelasticity problem featuring a time-dependent fluid flux. The examples are provided for various geometric shapes spanning both two-dimensional (2D) and three-dimensional (3D) domains, demonstrating the proposed framework's ability to predict over different geometries and loading conditions. Importantly, all testing examples are provided for load conditions that were never used during training, demonstrating the proposed framework's ability to generalize to unseen scenarios. Additionally, examples of mesh refinement are provided to illustrate the framework's scalability and independence from specific mesh resolutions.

The rest of the document is structured as follows: Section~\ref{section:problem_statement} presents a concise overview of the mathematical formulation for thermoelasticity and poroelasticity problems. Section~\ref{section:methodology} details the theoretical foundation of the I-FENN framework, including its mathematical formulation for thermoelasticity and poroelasticity problems. In addition, this section outlines the implementation aspects, including data management and transfer between the DeepONet model and the FEM solver. Section~\ref{section:deeponet_for_ifenn} delves into the DeepONet architecture developed in this work. Section~\ref{section:numerical_examples} presents the numerical examples used to test the proposed framework. Finally, Section~\ref{section:summary_conclusion} summarizes this work and offers a future outlook.
\section{Problem Statement} 
\label{section:problem_statement}

In this section, we introduce the mathematical formulation for thermoelasticity and poroelasticity problems. For the sake of brevity, here we introduce only the time-discretized weak forms, and a more detailed description of the strong and weak formulations is provided in Appendix~\ref{section:appendix:math}. 

\subsection{Thermoelasticity}
\label{section:problem_statement:thermoelasticity}

For the thermoelasticity problem under consideration, the unknown variables are the displacement field $\boldsymbol{u}$ and the temperature profile $T$ across the domain, which are coupled through the governing system of equations shown in Eq.~\eqref{eq:th_s_blm}-\eqref{eq:th_s_energy}. The finite element analysis for the thermoelasticity problem is implemented using the discretized weak form, where the simulation is incremented from time step $n$ to time step $n+1$. Using an implicit Euler scheme, the incremental weak form of the balance of linear momentum equation reads as:
\begin{align} \label{eq:th_w_blm_time}
    \int_{\Omega} \nabla^{s} \boldsymbol{\widehat{w}} : (\boldsymbol{\hat{C}} : \boldsymbol{\varepsilon}_{n+1}) \, d\Omega  \;- \int_{\Omega} \nabla^{s} \boldsymbol{\widehat{w}} : (\boldsymbol{\hat{C}} : \alpha (T_{n+1}-T_0)\boldsymbol{\mathit{I}}) \, d\Omega \nonumber 
    \\
    = \int_{\Gamma_t} \boldsymbol{\widehat{w}} \cdot \boldsymbol{\overline{t}}_{n+1} \, d\Gamma 
    \;+ \int_{\Omega} \boldsymbol{\widehat{w}} \cdot \boldsymbol{b}_{n+1} \, d\Omega 
    \quad \forall \, \boldsymbol{\widehat{w}} \in \mathcal{W}_u
\end{align}
\noindent
where $\boldsymbol{\nabla}$ is the gradient operator and $\boldsymbol{\nabla}\cdot$ is the divergence operator, $T_{o}$ is the reference temperature\, 
and $\boldsymbol{\varepsilon}$ is the strain defined as a function of displacement ($\boldsymbol{u}$) as $\boldsymbol{\varepsilon} = \boldsymbol{\nabla}^s \boldsymbol{u} =\frac{1}{2}(\boldsymbol{\nabla} \boldsymbol{u}+\boldsymbol{\nabla} \boldsymbol{u}^T)$ for small deformations, 
$\boldsymbol{\overline{t}}$ is the boundary traction, 
$\boldsymbol{b}$ is the mechanical body force vector, 
$\boldsymbol{\hat{C}}$ is the elasticity tensor, 
$\alpha$ is the coefficient of thermal expansion, 
$\Omega$ is the physical domain, 
$\Gamma$ is the domain boundary, 
$\widehat{\boldsymbol{w}}$ is the displacement test function, 
and $\mathcal{W}_u$ denotes the displacement function space.

The time-discretized weak form of the conservation of energy equation can be written as:
\begin{align}\label{eq:th_w_heat_time}
    &\int_{\Omega} \widehat{T} \rho C_{\varepsilon} \dfrac{T_{n+1}-T_n}{\Delta t} \, d\Omega 
    \;+ \int_{\Omega} \nabla \widehat{T} \cdot \left( k \nabla T_{n+1} \right) \, d\Omega 
    \;+ \int_{\Omega} \widehat{T} \alpha (n_{dim}\lambda + 2\mu) \text{tr} \left(\dfrac{\boldsymbol{\varepsilon}_{n+1}-\boldsymbol{\varepsilon}_n}{\Delta t} \right) T_0  \, d\Omega \nonumber 
    \\
    &= - \int_{\Gamma_q} \widehat{T} \overline{Q}_{n+1} \, d\Gamma 
    \;+ \int_{\Omega} \widehat{T} r_{n+1} \, d\Omega 
    \quad \forall \, \widehat{T} \in \mathcal{W}_T
\end{align}

\noindent
where $\overline{Q}$ is the boundary heat flux, 
$r$ is the heat body source or sink, 
$\rho$ is the material mass density, 
$\mu$ and $\lambda$ are Lamé constants, 
$k$ is the thermal conductivity of the material, 
$C_{\varepsilon}$ is the specific heat per unit mass at constant strain, 
$n_{dim}$ is the number of spatial dimensions, 
$\widehat{T}$ is the temperature test function, 
and $\mathcal{W}_T$ is the temperature function space.

For the given thermoelasticity problem (described in Eq.~\eqref{eq:th_w_blm_time} and Eq.~\eqref{eq:th_w_heat_time}), an FEM solver is developed using the deal.II\cite{africa_dealii_2024}. We implement the monolithic approach to solve the coupled system of equations, where the solver concurrently computes the displacement ($\boldsymbol{u}_{n+1}$) and temperature ($T_{n+1}$) fields given the response from the previous time step.

\subsection{Poroelasticity}
\label{section:problem_statement:poroelasticity}
Poroelasticity describes the interaction between the mechanical and fluid flow behaviors in a porous medium, where deformations and fluid pressure mutually influence one another. The poromechanical properties of the porous medium can depend on both space and time. For the considered poroelasticity formulation, the unknown variables are the displacements ($\boldsymbol{u}$) and fluid pressure ($p$), coupled through the governing system in Eq.~\eqref{eq:poro_st_blm_2}-\eqref{eq:poro_st_cont_2}. Adopting the implicit Euler scheme for time discretization and considering incremental analysis, the weak form of the linear momentum equilibrium equation can be written as:
\begin{align} \label{eq:poro_w_blm_time}
    &\int_{\Omega} \nabla^{s} \boldsymbol{\widehat{w}} : \boldsymbol{\hat{C}} : \boldsymbol{\varepsilon}_{n+1} \, d\Omega  \;- \int_{\Omega} \nabla^{s} \boldsymbol{\widehat{w}} : (\alpha^\prime p_{n+1} \boldsymbol{\mathit{I}}) \, d\Omega \nonumber 
    \\
    &= \int_{\Gamma_t} \boldsymbol{\widehat{w}} \cdot \boldsymbol{\overline{t}}_{n+1} \, d\Gamma 
    \;+ \int_{\Omega} \boldsymbol{\widehat{w}} \cdot \boldsymbol{b}_{n+1} \, d\Omega 
    \quad \forall \, \boldsymbol{\widehat{w}} \in \mathcal{W}_u
\end{align}
\noindent
where $\alpha^\prime$ is the Biot's coefficient. All other quantities are the same as in the thermoelasticity formulation.

Assuming isotropic properties, the weak form of the fluid flow continuity equation can be written as:
\begin{align}\label{eq:poro_w_cont_time}
    &\int_{\Omega} \widehat{p}\; \dfrac{p_{n+1}-p_{n}}{ \Delta t M} \, d\Omega 
    \;+ \int_{\Omega} \widehat{p} \alpha^\prime \text{tr}\left(\dfrac{\boldsymbol{\varepsilon}_{n+1}-\boldsymbol{\varepsilon}_{n}}{\Delta t} \right)  \, d\Omega
    \;+ \int_{\Omega} \nabla \widehat{p} \cdot \left(\dfrac{ K_I \boldsymbol{\mathit{I}}}{\mu_f} \cdot \nabla p_{n+1} \right) \, d\Omega 
     \nonumber 
    \\
    &= -\int_{\Omega} \nabla \widehat{p} \cdot \left(\dfrac{ K_I \boldsymbol{\mathit{I}}}{\mu_f} \cdot \gamma_f \boldsymbol{i}_g \right) \, d\Omega - \int_{\Gamma_q} \widehat{p} \;{\overline{q}_f}_{n+1} \, d\Gamma \;+ \int_{\Omega} \widehat{p} \; {Q_f}_{n+1} \, d\Omega 
    \quad \forall \, \widehat{p} \in \mathcal{W}_p
\end{align}
\noindent
where $\overline{q}_f $ is the boundary fluid flux,
$Q_f$ represents any fluid source or sink, 
$M$ is the Biot's modulus,
$K_I$ is the isotropic permeability,
$\gamma_f$ is the weight density,
$\mu_f$ is the fluid viscosity,
$\boldsymbol{i}_g$ is the unit vector parallel to gravity, but in the opposite direction,
$\widehat{p}$ is the fluid pressure test function, and $\mathcal{W}_p$ is the fluid pressure function space. An FEM solver is built using the deal.II library to solve the poroelasticity problem in a monolithic approach. For each time step, the solver, monolithically, satisfies Eq.\eqref{eq:poro_w_blm_time} and Eq.\eqref{eq:poro_w_cont_time} to compute the displacement ($\boldsymbol{u}_{n+1}$) and pressure ($p_{n+1}$) fields, given the response at the previous time step.  

\section{I-FENN Formulation and Implementation}
\label{section:methodology}

\subsection{I-FENN Formulation}
\label{section:methodology:ifenn_formulation}

\begin{figure}[hbt!]
\centering
\includegraphics[width=0.85\textwidth]{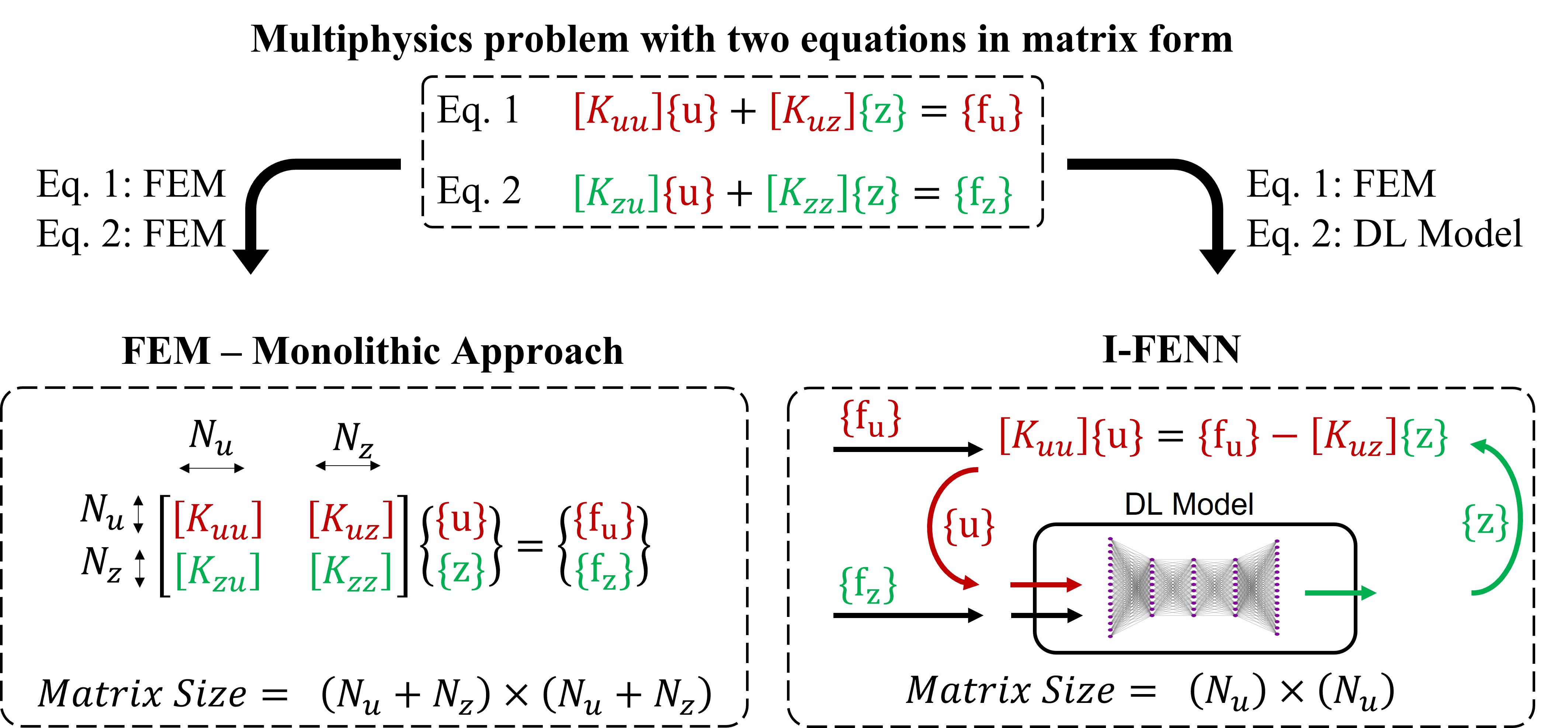}
\caption{Explanation of FEM vs. I-FENN approach for multiphysics problems.}
\label{fig:ifenn_approach}
\end{figure}

The Integrated Finite Element Neural Network (I-FENN) framework is a computational methodology introduced to accelerate the numerical simulation of multiphysics problems \cite{pantidis_integrated_2023}. An illustration of the I-FENN setup is shown in Fig.~\ref{fig:ifenn_approach}. For a problem with two governing equations and two coupled variables (displacement field $u$ and another variable $z$), the matrix form of the governing equations can be expressed as shown in Fig.~\ref{fig:ifenn_approach}. One standard option to solve the system is the FEM monolithic approach (refer to Fig.~\ref{fig:ifenn_approach}), where the two equations are solved simultaneously. This approach requires assembling the system matrix and the right-hand side vector, including all degrees of freedom, which can be computationally expensive and/or memory-intensive, especially for large-scale problems. I-FENN, on the other hand, is conceptually similar to FEM-staggered approaches, where the two equations are decoupled and solved iteratively (see Fig.~\ref{fig:ifenn_approach}). I-FENN utilizes a conventional FEM solver for the balance of linear momentum to compute the displacement variable $u$, while the other variable is predicted using a Neural Network (NN). The second coupled field ($z$) is explicitly introduced to the first equation at each time step, therefore constantly informing the displacement field of its updated profile. This hybrid formulation yields several advantages. First, for the FEM-based part, it results in a symmetric system matrix and right-hand side vector that only include the degrees of freedom related to the displacement variable $u$. The symmetry of the system matrix $K_{uu}$ facilitates the convergence of the linear solver. In addition, given the reduced matrix size and degrees of freedom, I-FENN provides faster assembly and lower memory requirements for the FEM solver, while still ensuring the minimization of the displacement residuals. Second, by focusing on predicting only one unknown variable ($z$), which constitutes a single learning task, the NN achieves greater training efficiency than approaches that attempt to predict all variables concurrently~\cite{manav2024phase} (i.e., simulation substitution). Third, since the NN-based prediction of the second variable is much faster than an FEM-based computation, I-FENN yields considerable savings overall compared to either an FEM-monolithic or an FEM-staggered method, which is demonstrated through previous studies \cite{pantidis_i-fenn_2024}.

\subsection{I-FENN for Thermoelasticity and Poroelasticity}
\label{section:methodology:ifenn_thermo_poro}
In this section, the I-FENN formulation is applied to two multiphysics problems, where the second coupled variable, denoted $z$, represents the temperature field ($T$) in thermoelasticity and the fluid pressure field ($p$) in poroelasticity. The governing equations of the thermoelasticity problems are listed in Section~\ref{section:problem_statement:thermoelasticity}. Utilizing FEM, I-FENN solves the equilibrium equation (Eq.~\eqref{eq:th_w_blm_time}) and computes the displacement field $\boldsymbol{u}$, while the temperature field $T$ in Eq.~\eqref{eq:th_w_heat_time} is predicted using the NN. Equation \eqref{eq:th_w_heat_time} shows that the temperature $T$ is coupled with the displacement $u$ through the trace of the strain tensor $\text{tr}(\boldsymbol{\varepsilon})$. Utilizing this remark enables the use of simplified inputs to the NN. Rather than relying on the full displacement field $\boldsymbol{u}$ (a vector quantity), the model can be trained to predict $T$ based on the trace of the strain tensor $\text{tr}(\boldsymbol{\varepsilon})$, which is a scalar quantity. This simplification reduces the complexity of the input parameters for the NN, making it more efficient and easier to train. In addition to strain trace $\text{tr}(\boldsymbol{\varepsilon})$, thermal loads depicted in Eq.~\eqref{eq:th_w_heat_time} are also included as input training parameters. The thermal loads can be domain loads (body loads $r$) or surface loads (Neumann boundary conditions $\overline{Q}$) . Depending on the specific problem setup, either or both types of thermal loads can serve as input features for the NN. 

I-FENN for poroelasticity employs a methodology similar to that used in thermoelasticity, illustrating the framework's adaptability for various multiphysics problems. First, the balance of linear momentum equation (Eq.~\eqref{eq:poro_w_blm_time}) is solved using an FEM solver, and the continuity equation (Eq.~\eqref{eq:poro_w_cont_time}) is solved by the NN to predict the fluid pressure field ($p$). By inspecting the coupled equations (Eq.~\eqref{eq:poro_w_blm_time} and Eq.~\eqref{eq:poro_w_cont_time}), it can be seen that the pressure field is coupled with the displacement field through strain trace. This implies that the NN needs only this scalar quantity to sufficiently represent the mechanical effects (deformations). In addition to the strain trace, the NN inputs should also represent the fluid flux $(\overline{q}_f)$ over the boundaries and the fluid sinks or sources $(Q_f)$ throughout the domain. The design of the NN model (DeepONet) will be presented in detail in Section~\ref{section:deeponet_for_ifenn}. In this paper, it is important to note the difference in numerical strategies employed in I-FENN versus the fully coupled FEM solver used for data generation and performance comparisons. The I-FENN setup inherently follows a staggered scheme, whereas the fully coupled FEM solver employs a monolithic approach. The choice of a monolithic solver for FEM is driven by the need to achieve optimal performance for the strongly coupled transient multiphysics problems. 

\subsection{I-FENN Implementation}
\label{section:methodology:ifenn_implementation}

\begin{figure}[hbt!]
\centering
\includegraphics[width=.8\textwidth]{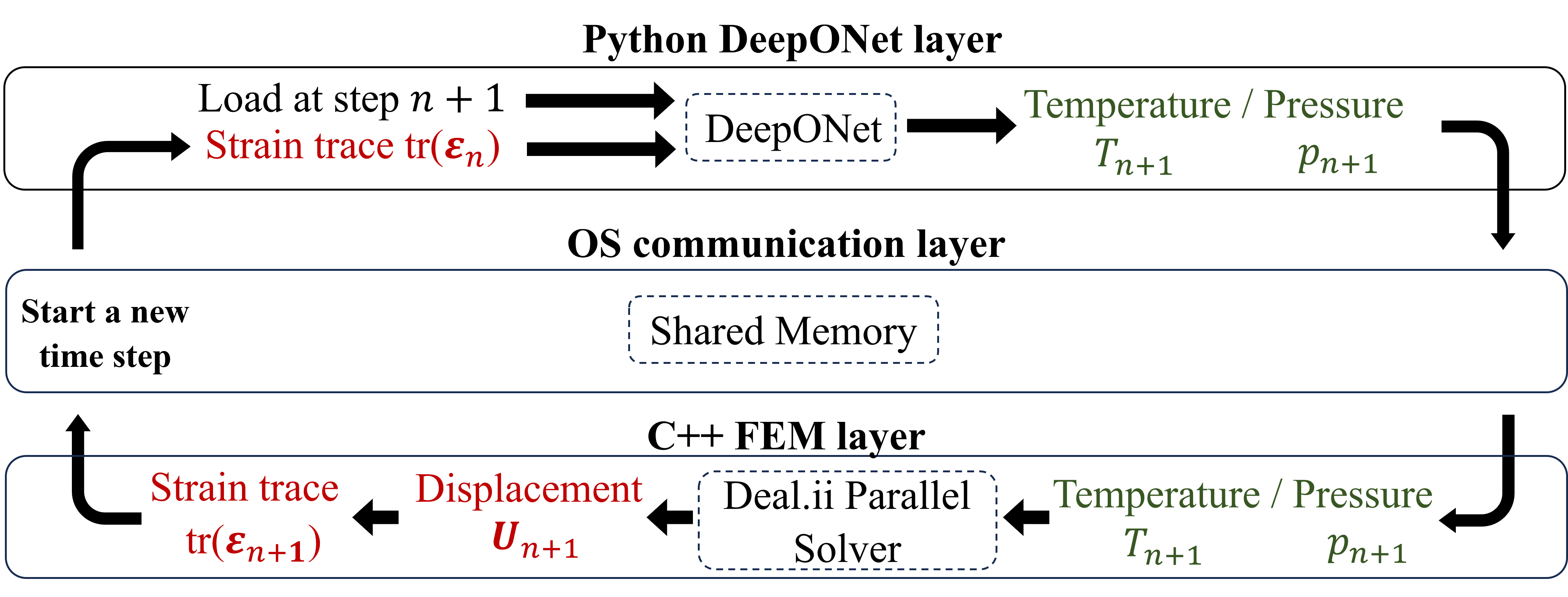}
\caption{I-FENN implementation for thermoelasticity/poroelasticity problems.}
\label{fig:ifenn_implementation}
\end{figure}

In this section, we illustrate the integration of the NN of choice (in this case, a DeepONet) within the I-FENN framework. Fig.~\ref{fig:ifenn_implementation} depicts the framework architecture, which consists of three main layers: (1) Python DeepONet layer, (2) Operating system communication layer, and (3) C++ FEM layer. The DeepONet layer, written in Python, works as a server running on the machine's graphics processing unit (GPU). It contains the trained DeepONet model, which is loaded once and receives inputs from the communication layer at every time step. The second layer serves as an intermediate communication layer that leverages the operating system's shared memory. Both the first and third layers utilize libraries for interprocess communication via shared memory and semaphore-based synchronization. In this context, a semaphore is a coordination mechanism that controls access to shared resources by blocking some processes until another process signals that it has completed its operation.

The third layer primarily consists of an FEM solver implementation based on the deal.II library~\cite{africa_dealii_2024}, with PETSc~\cite{petsc-web-page} integration for large-scale systems parallel processing. The third layer's FEM solver is used to solve mechanical systems governed by the coupled balance of linear momentum equations, as discussed in Section~\ref{section:methodology}. Specifically, an FEM solver is created to solve Eq.~\eqref{eq:th_w_blm_time} and Eq.~\eqref{eq:poro_w_blm_time} for thermoelasticity and poroelasticity problems, respectively. Depending on the problem, temperature or pressure fields are provided to a specific FEM solver as inputs. These fields are then internally incorporated when solving for the displacement field.

The execution of the framework initiates with the activation of the first computational layer, which monitors a shared memory space and waits for input from the FEM solver. At each discrete time step $n+1$, the FEM solver outputs strain trace data from the previous time step $\text{tr}(\boldsymbol{\varepsilon}_n)$ and writes it into the shared memory. In a parallel computing environment, the computational mesh is distributed among multiple processors, each contributing its local strain response to the shared memory. Once all processors have successfully completed this data exchange, a synchronization mechanism, implemented via a semaphore, is triggered to signal the DeepONet module that it may proceed.

Upon activation, the DeepONet module performs inference using the strain data $\text{tr}(\boldsymbol{\varepsilon}_n)$ retrieved from the shared memory in conjunction with the prescribed input loads for step $n+1$. The resulting outputs ($T_{n+1}$ or $p_{n+1}$) are then written back into the shared memory. Another semaphore is subsequently activated to notify the FEM solver that it can resume computation. The FEM solver then reads the updated input fields ($T_{n+1}$ or $p_{n+1}$), computes the corresponding displacement field ($\boldsymbol{u}_{n+1}$), and performs post-processing operations to obtain the strain trace field $\text{tr}(\boldsymbol{\varepsilon}_{n+1})$, thus completing one iteration of the coupled workflow.

This framework is designed with high versatility and flexibility, effectively leveraging both GPU acceleration for the DeepONet model (using PyTorch library) and parallel processing capabilities for the FEM solver (using deal.II library). Notably, the implementation achieves efficient resource utilization by loading the DeepONet inference model into memory only once, even when FEM is executed across multiple processors. This design choice optimizes memory utilization, avoiding memory overhead, and enabling scalability and computational efficiency in distributed systems typically required for computationally demanding multiphysics problems.

\section{DeepONet for I-FENN}
\label{section:deeponet_for_ifenn}

In this section, we present a detailed overview of the proposed DeepONet, including the operator architecture, GRU design and integration, boundary conditions handling, training setup, loss functions, and testing metrics.

\subsection{DeepONets General Architecture}
\begin{figure}[hbt!]
\centering
\includegraphics[width=.7\textwidth]{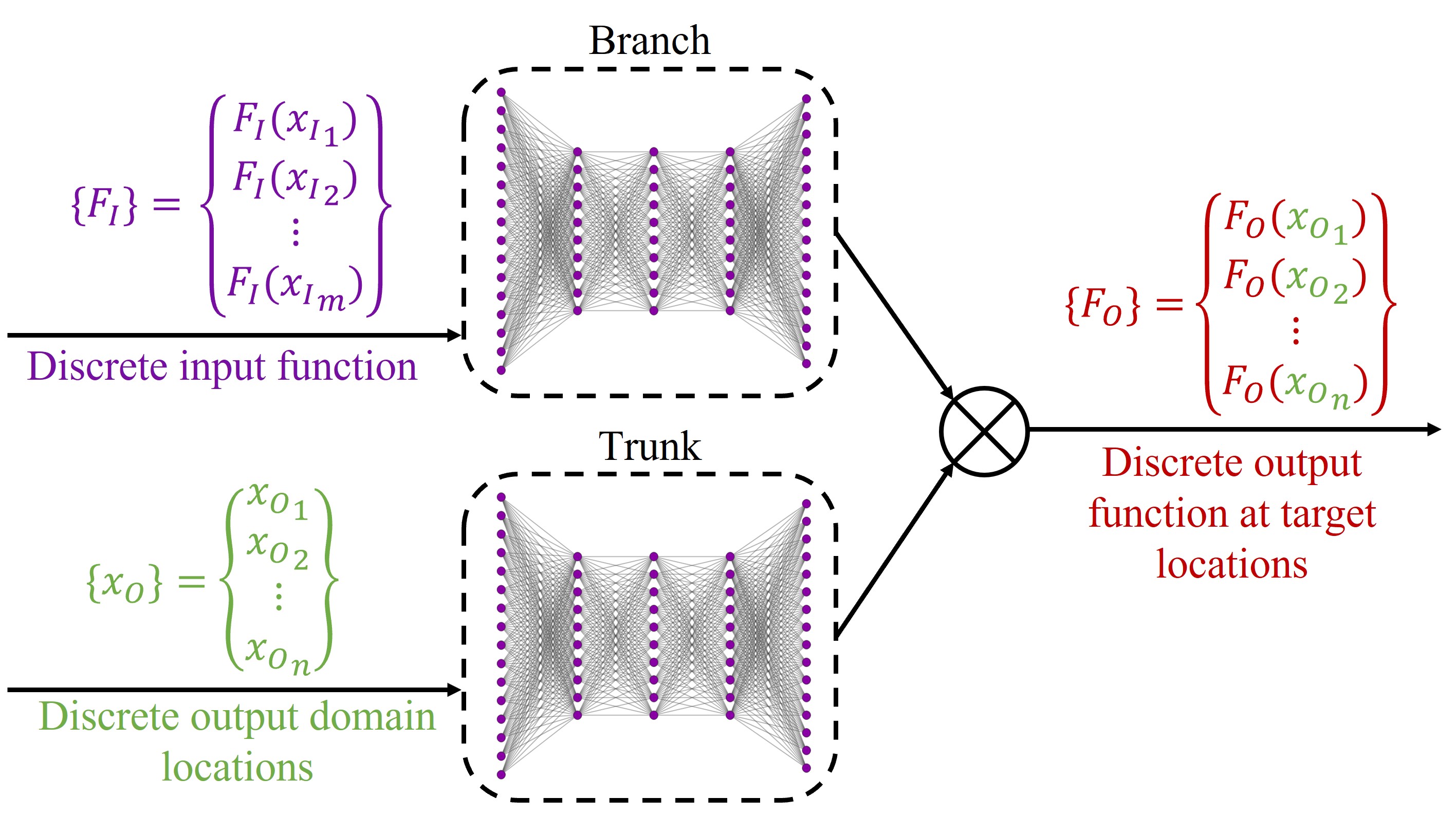}
\caption{DeepONet general architecture.}
\label{fig:deeponet_general}
\end{figure}

The DeepONet architecture was introduced as a network approximating an operator $(\mathcal{G})$ that can map an input function $(F_I)$ to an output function $(F_O)$ \cite{lu_comprehensive_2022, lu_learning_2021}. A DeepONet model consists mainly of two sub-networks, denoted as "branch" and "trunk", with each network assigned a specific functionality. As shown in Fig.~\ref{fig:deeponet_general}, the branch network is used to encode the discretized input function $F_I(x_I)$, while the trunk encodes the target locations in the output domain $x_O$. The final output of the DeepONet network is computed through an element-wise multiplication process defined as:
\begin{align}
    \label{eq:deeponet_output}
    \mathcal{G}(F_I)(x_O) = \sum_{k=1}^{D_{out}} b_k(F_I) t_k(x_O) + b_0
\end{align}
\noindent
where $b_0  \in \mathbb{R}$ is a bias,  $D_{out}$ is the number of outputs from both branch and trunk networks, $b(F_I)$ and $t(x_O)$ are the output vectors, of size $D_{out}$, from the branch and trunk, respectively.  

For an input function $F_I$ and an output function $F_O$ that are defined over domains $D_I \subset \mathbb{R}^{d_I}$ and  $D_O \subset \mathbb{R}^{d_O}$, respectively, the operator mapping can be defined as:
\begin{align}
    F_I : D_I \ni x_I &\mapsto F_I(x_I) \in \mathbb{R}
    \\
    F_O : D_O \ni x_O &\mapsto F_O(x_O) \in \mathbb{R}
    \\
    \mathcal{G}:\mathcal{F_I} \ni F_I &\mapsto F_O \in \mathcal{F_O}
\end{align}
\noindent
where $\mathcal{F_I}$ and $\mathcal{F_O}$ are the spaces for $F_I$ and $F_O$ defined over domains $D_I$ and $D_O$, respectively. It should be noted that in this approach, the input function $F_I$ is discretized by substitution at a set of locations $\{{x_I}_1, {x_I}_2, ..., {x_I}_m\}$ to get a set of evaluations $\{F_I({x_I}_1), F_I({x_I}_2), ..., F_I({x_I}_m)\}$. MIONet\cite{jin_mionet_2022} extends the DeepONet architecture by introducing multiple branches, each processing a separate input function, with their outputs subsequently combined with the trunk output through a tensor product. A further development, Fourier-MIONet~\cite{jiang_fourier-mionet_2024}, introduces element-wise summation, instead of multiplication, to merge branch outputs. More recently, the Image Generator Enhanced Deep Operator Network (IGE-DeepONet)~\cite{liu_image_2025} explored hybrid strategies, combining data concatenation with data fusion (transfer) between the hidden layers of the branch and trunk. Notably, in IGE-DeepONet, concatenation-based fusion was shown to achieve higher accuracy than multiplication-based approaches. 

\begin{figure}[hbt!]
\centering
\includegraphics[width=1.0\textwidth]{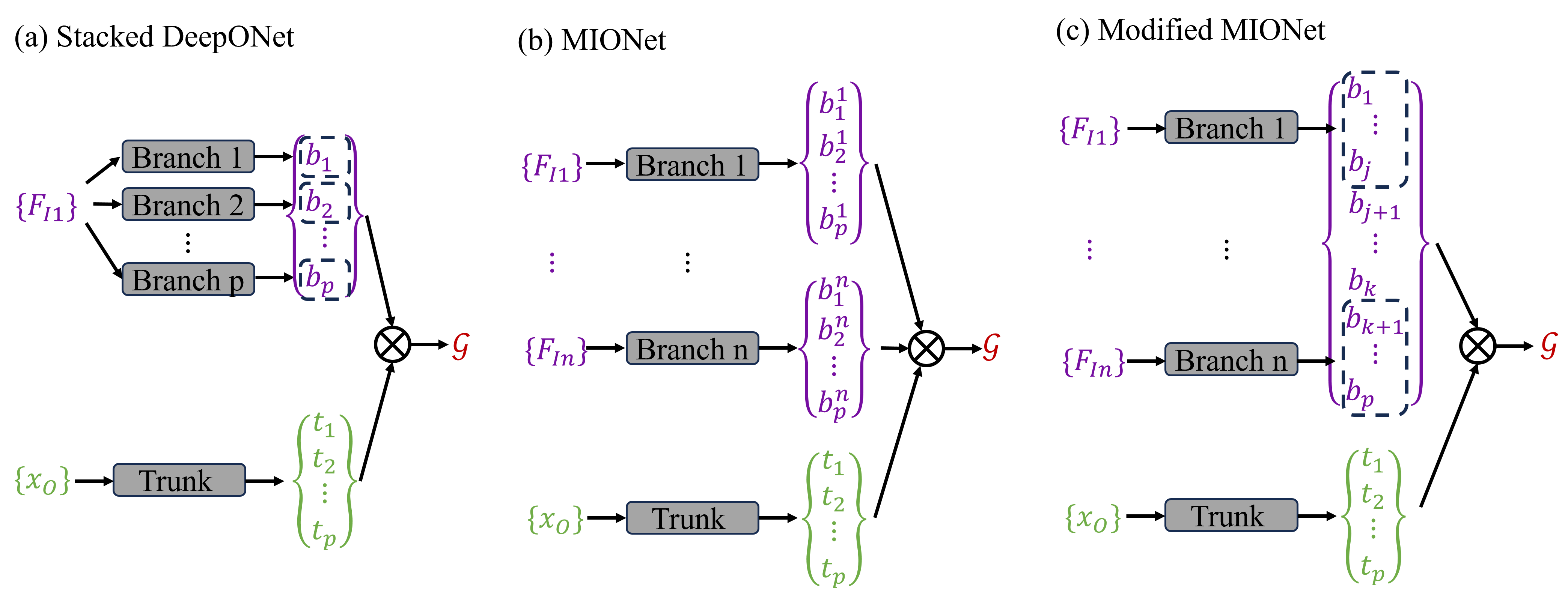}
\caption{General architecture of a) Stacked DeepONet, b) MIONet, and c) our proposed modified MIONet (with $1<j<k<p$).}
\label{fig:deeponet_arch_comparison}
\end{figure}

Building on these ideas, our method employs multiple branches to accommodate multiple inputs and concatenates their outputs into a unified representation. For clarification, three different architectures are depicted in Fig.~\ref{fig:deeponet_arch_comparison}. First, the stacked DeepONet approach (see Fig.~\ref{fig:deeponet_arch_comparison}a), introduced by Lu et al. \cite{lu_learning_2021}, employs several branches to encode the same input function, with each branch producing a scalar quantity. Second, MIONet (Fig.~\ref{fig:deeponet_arch_comparison}b) aggregates branch outputs through a tensor product with the trunk output. Finally, our proposed architecture (Fig.~\ref{fig:deeponet_arch_comparison}c), a modified version of MIONet, concatenates the outputs of the individual branches. This design provides flexibility in tailoring the branch network size for each input. Moreover, it preserves the full output of each branch prior to reduction, thereby offering a richer representation for integration with the trunk output. For clarity, we will hereafter refer to this modified version of MIONet as the "DeepONet model".

\subsection{Gated Recurrent Unit (GRU) Networks}
The gated recurrent unit (GRU) architecture was introduced as an extension to the recurrent neural network (RNN). GRU uses a relatively sophisticated approach, named the gating mechanism, to process the temporal dependencies along sequences of data \cite{chung_empirical_2014, cho_properties_2014}.  For an input $\mathbf{x}_t \in \mathbb{R}^d$, at time step $t$, the GRU output hidden state $\mathbf{h}_t \in \mathbb{R}^h$, is computed through the following computational steps \cite{tsantekidis_recurrent_2022, zhang_dive_2023}:
\begin{align}
\mathbf{r}_t &= \sigma(\mathbf{W}_{xr} \mathbf{x}_t + \mathbf{W}_{hr} \mathbf{h}_{t-1} + \mathbf{b}_r)
\\
\mathbf{z}_t &= \sigma(\mathbf{W}_{xz} \mathbf{x}_t + \mathbf{W}_{hz} \mathbf{h}_{t-1} + \mathbf{b}_z)
\\
\mathbf{h}^\prime_t &= \tanh(\mathbf{W}_{hh} (\mathbf{r}_t \odot \mathbf{h}_{t-1}) + \mathbf{W}_{xh} \mathbf{x}_t + \mathbf{b}_h)
\\
\mathbf{h}_t &= \mathbf{z}_t \odot \mathbf{h}_{t-1} +  (1 - \mathbf{z}_t)\odot \mathbf{h}^\prime_t
\end{align}

\noindent
where symbol $\odot$ is the Hadamard (element-wise) product operator, $\mathbf{r}_t$ is the reset gate, $\mathbf{z}_t$ is the update gate, $\mathbf{h}^\prime_t$ is the candidate hidden state, $\sigma$ is the sigmoid activation function, $\mathbf{W}_{xr},\mathbf{W}_{xz},\mathbf{W}_{xh} \in \mathbb{R}^{d \times  h}$ and $\mathbf{W}_{hr},\mathbf{W}_{hz},\mathbf{W}_{hh} \in \mathbb{R}^{h \times  h}$ are weight parameters, $\mathbf{b}_r, \mathbf{b}_z, \mathbf{b}_h \in \mathbb{R}^{1 \times h}$ are the bias.

\subsection{DeepONet with GRUs: detailed implementation}
\label{section:deeponet_for_ifenn:detailed}
\begin{figure}[hbt!]
\centering
\includegraphics[width=0.85\textwidth]{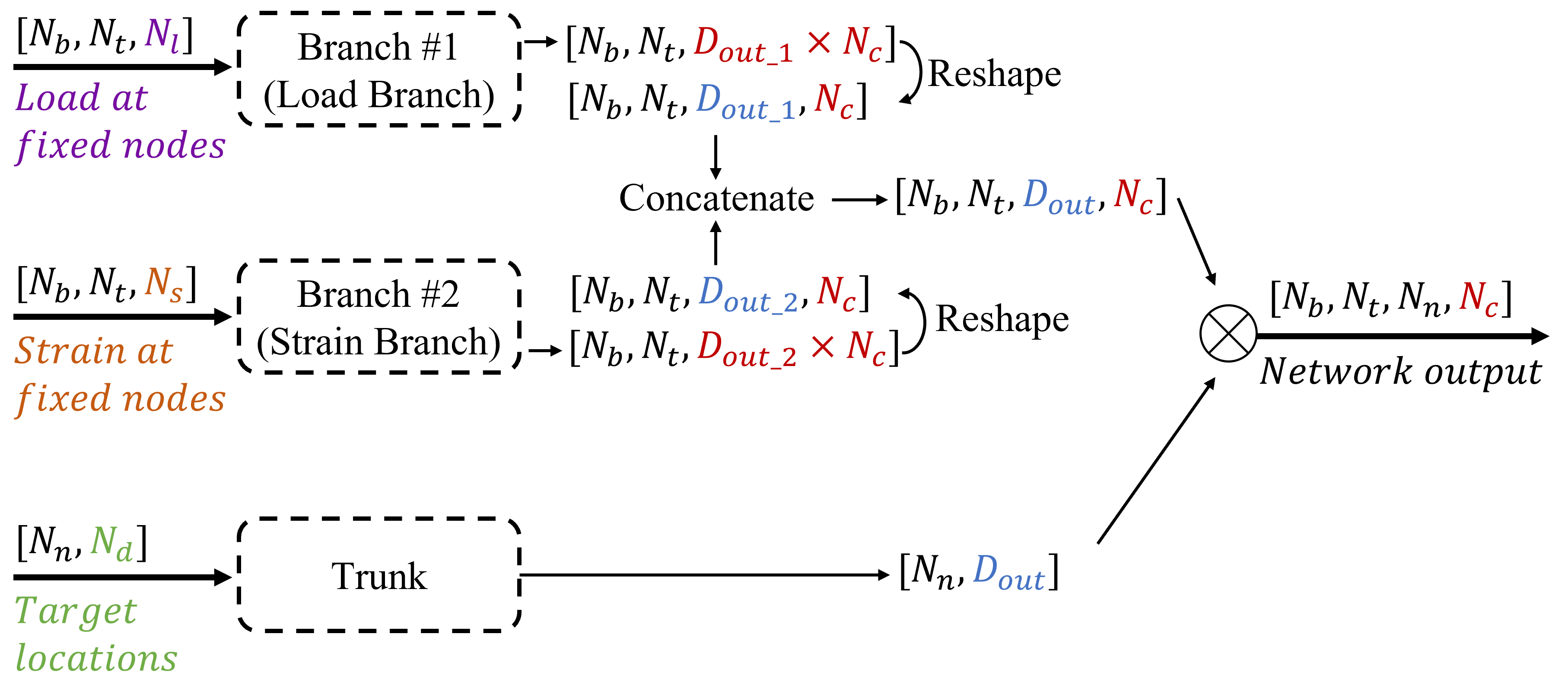}
\caption{Detailed explanation of data flow through the DeepONet model designed for multiphysics problems.}
\label{fig:deeponet_detailed_1}
\end{figure}

Based on the analysis of the coupled problems and the I-FENN approach introduced in Section~\ref{section:methodology}, a generic architecture of the DeepONet is proposed as depicted in Fig.~\ref{fig:deeponet_detailed_1}. Load and strain inputs are processed through separate branches, allowing for greater flexibility in network size and enabling independent control over the network's learning dynamics for each input. A comparable strategy was adopted in the literature \cite{kadeethum_improved_2024}, involving the use of two distinct branches to differentiate between homogeneous and heterogeneous inputs. 

\begin{figure}[hbt!]
\centering
\includegraphics[width=0.45\textwidth]{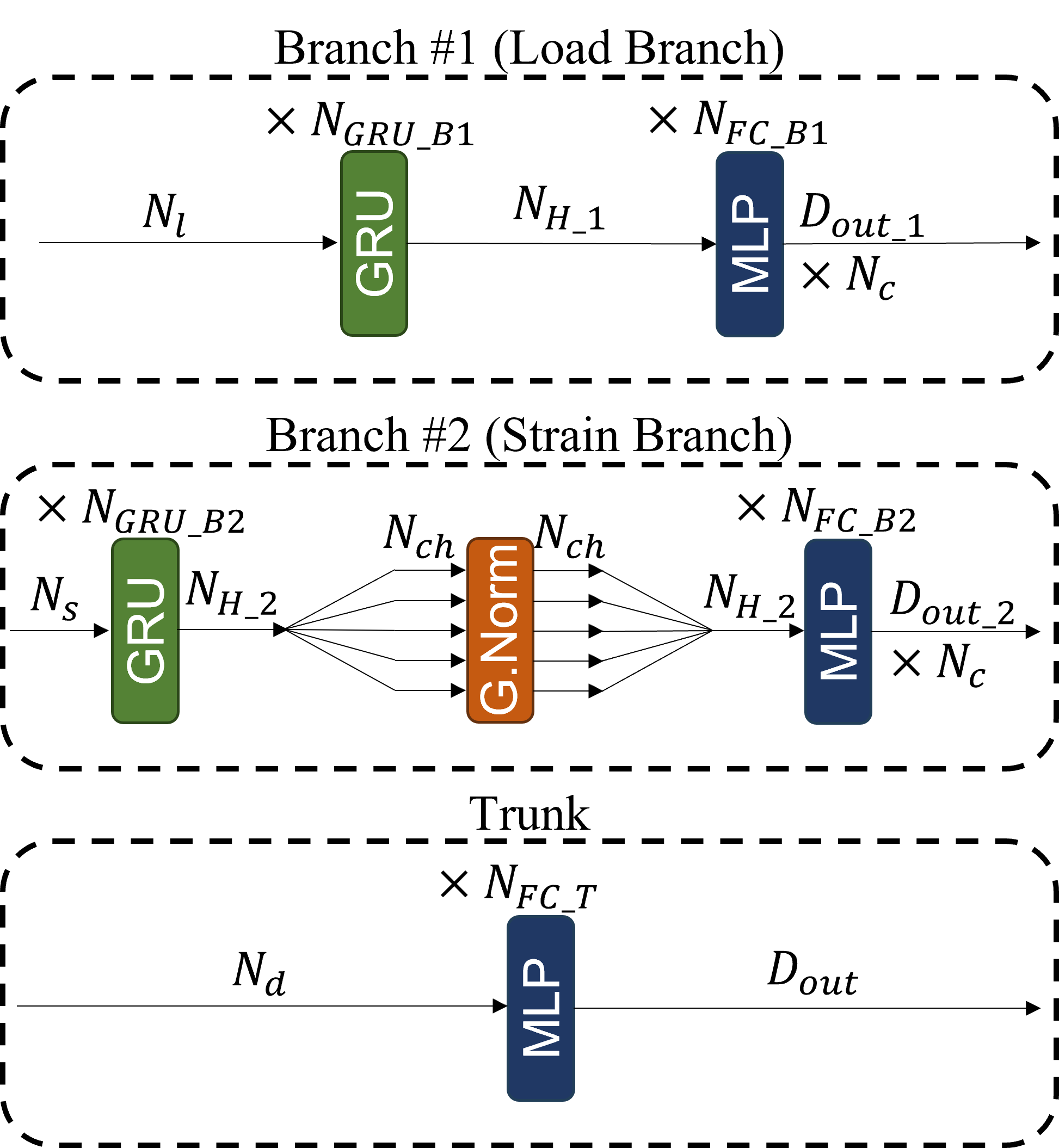}
\caption{Detailed architecture of different components of the DeepONet model.}
\label{fig:deeponet_detailed_2}
\end{figure}

In the proposed model (refer to Fig.~\ref{fig:deeponet_detailed_2}), each branch incorporates a stacked GRU (multiple consecutive GRUs) to capture temporal dependencies across time steps. The GRU outputs are subsequently passed through a fully connected feedforward multilayer perceptron (MLP) to effectively encode spatial relationships among the discretized inputs. The trunk network only consists of a fully connected feedforward MLP \cite{lu_comprehensive_2022}. 

For the first branch, a load input of size $N_l$ is encoded through the GRU, which outputs a vector of size $N_{H\_1}$,  which is then processed through the MLP with a final layer size $D_{out\_1}$. The second branch encodes the strain input of size $N_s$ through the GRU to get the hidden output of size $N_{H\_2}$. The hidden output in the second branch is split into $N_{ch}$ channels to be normalized through a group normalization layer, then reshaped before proceeding to the MLP, as depicted in Fig.~\ref{fig:deeponet_detailed_2}. The additional group normalization layer is crucial for the I-FENN framework stability, as discussed in Section~\ref{section:numerical_examples:cube:stability}. 

If the network is trained to predict multiple $N_c$ components, not a single component, the output of the branches is set to be $D_{out\_1} \times N_c$ and $D_{out\_2} \times N_c$, respectively. For example, if the network is trained to predict the temperature field only, $N_c$ is set to unity, while if it is trained to predict temperature and all displacement components, $N_c$ will be four. Noting that, for I-FENN implementation in this paper, $N_c$ will always be unity; however, for a case study on simulation substitution, $N_c$ will be set to four. Finally, for each branch, $N_{GRU\_Bi}$ and $N_{FC\_Bi}$ represent the number of stacked GRUs and fully connected layers, respectively, for the $ith$ branch.

Regarding the trunk, an input of size $N_d$ (number of coordinates per node) is processed through an MLP with $N_{FC\_T}$ layers, ending with an output of size $D_{out}$. The output of the branches is reshaped and concatenated to have the same dimension as the trunk output $D_{out\_1} + D_{out\_2} = D_{out}$, as shown in Fig.~\ref{fig:deeponet_detailed_1}. The final output is computed by applying the element-wise multiplication (Eq.~\eqref{eq:deeponet_output}) along the output vectors of length $D_{out}$ for $N_c$ times to compute the output for all components. Further details on the input and output dimensions, as well as the training procedure, are provided in Section~\ref{section:deeponet_for_ifenn:training_testing}.

\subsection{Enforcing boundary conditions}
\label{section:deeponet_for_ifenn:bcs}
Lu et al. \cite{lu_learning_2021} introduced the following approach to enforce Dirichlet boundary conditions (BCs). A Dirichlet BC $g(x_O)$ defined over the boundary part $\Gamma_D$ can be automatically satisfied by the DeepONet as follows:
\begin{align}
    \label{eq:deeponet_output_bc_lulu}
    \mathcal{G}(F_I)(x_O) &= \ell(x_O) \mathcal{N}(F_I)(x_O) + g(x_O) 
    \nonumber
    \\
    \ell(x_O) &= \begin{cases}
    0, & x_O \in \Gamma_D, 
    \\
    > 0, & \text{otherwise}.
    \end{cases} 
\end{align}
\noindent
where $\mathcal{N}(F_I)(x_O)$ is the resultant of the tensor product of the trunk and branch outputs.

In case of complex BCs where different functions are defined over different boundary parts, additional extensions are required. Assume we have $n$ boundary parts $ \{\Gamma_{D_1}, \Gamma_{D_n}, ..., \Gamma_{D_n} \}$ with well defined Dirichlet BC functions $\{g_1(x_O), g_2(x_O), ..., g_n(x_O)\}$. Sukumar and Srivastava~\cite{sukumar_exact_2022} proposed enforcing these functions as follows:

\begin{align}
    \label{eq:deeponet_output_bc_sukumar}
    \mathcal{G}(F_I)(x_O) &=   \mathcal{N}(F_I)(x_O) \prod_{i=1}^n \phi_i(x_O) + \sum_{i=1}^n w_ig_i(x_O)   
\end{align}

\noindent where $\phi_i$ is the approximate distance function for the boundary segment $\Gamma_{D_i}$, and $w_i$ is the corresponding transfinite interpolation weight. Further details on the construction of these distance functions and the interpolation scheme can be found in literature~\cite{sukumar_exact_2022, dyken_transfinite_2009}.

Inspired by approaches in Eq.~\eqref{eq:deeponet_output_bc_lulu} and Eq.~\eqref{eq:deeponet_output_bc_sukumar} we propose another extension as follows:
 
\begin{align}
    \label{eq:deeponet_output_bc}
    \mathcal{G}(F_I)(x_O) &=   \mathcal{N}(F_I)(x_O) \prod_{i=1}^n \ell_i(x_O) + \sum_{i=1}^n (1-\ell_i(x_O))g_i(x_O)  
    \nonumber
    \\
    \ell_i(x_O) &=
        \begin{cases}
        0, & x_O \in \Gamma_{D_i}, 
        \\
        1, & x_O \in \Gamma \setminus \Gamma_{D_i}, 
        \\
        \in (0,1], & \text{otherwise}.
        \end{cases} 
\end{align}

\begin{figure}[hbt!]
\centering
\includegraphics[width=1.0\textwidth]{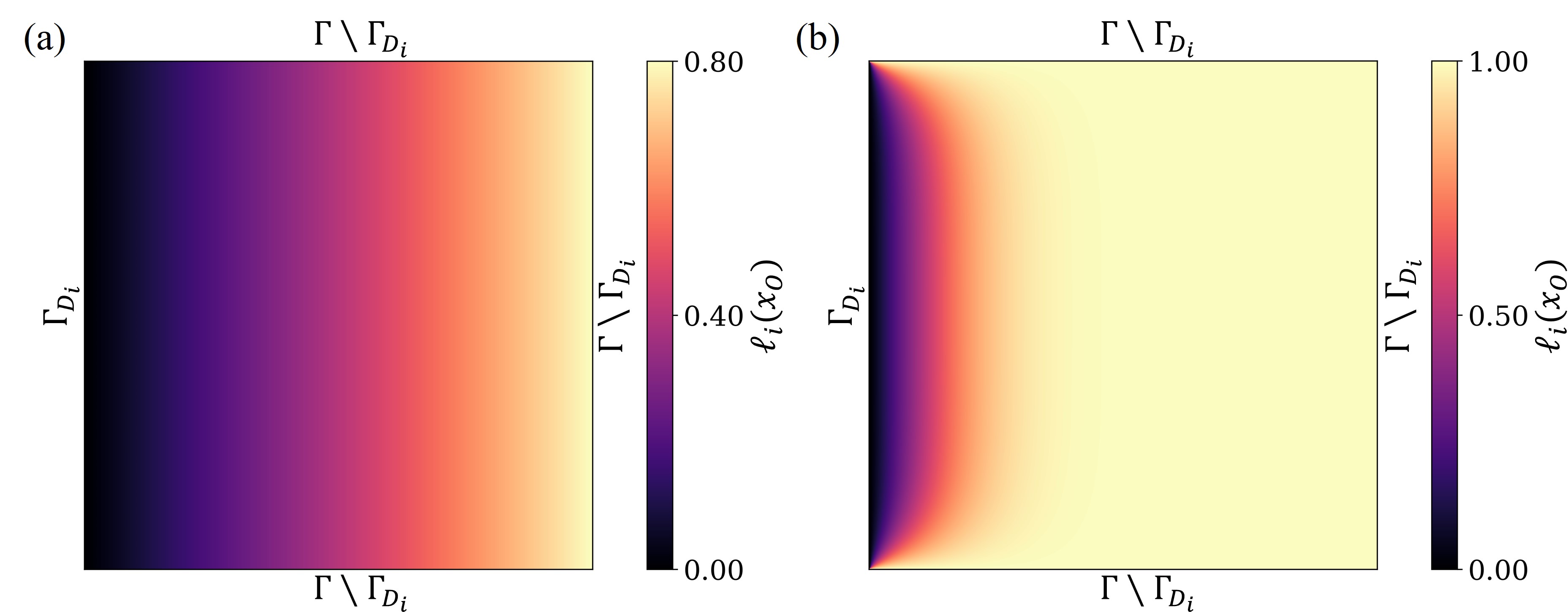}
\caption{Comparison between the $\ell_i(x_O)$ definitions for the boundary part $\Gamma_{D_i}$ in: a) original approach introduced by Lu et al. \cite{lu_learning_2021}, b) our approach  }
\label{fig:DeepONet_BCs}
\end{figure}

Fig.~\ref{fig:DeepONet_BCs} depicts the main difference between the original approach introduced by Lu et al. \cite{lu_learning_2021} and our approach. In the original approach (Eq.~\eqref{eq:deeponet_output_bc_lulu}), $\ell_i(x_O)$ is zero on the boundary part $\Gamma_{D_i}$ and can be any positive value on other boundaries ($\Gamma \setminus \Gamma_{D_i}$). In our approach, the function $\ell_i(x_O)$ is constructed to vanish on the boundary part $\Gamma_{D_i}$, thereby canceling the contribution of $\mathcal{N}(F_I)(x_O)$ on $\Gamma_{D_i}$. Simultaneously, it is set to be unity on the remaining boundary $\Gamma \setminus \Gamma_{D_i}$, effectively nullifying the influence of $g_i(x_O)$ outside of $\Gamma_{D_i}$. However, it should be mentioned that enforcing boundary conditions through the proper selection of $\ell_i(x_O)$ and $g_i(x_O)$ functions can be challenging in the presence of complex geometries, highlighting the importance of the  alternative methods\cite{liu_unified_2022, sukumar_exact_2022, horie_physics-embedded_2022}, and the need for more versatile approaches in future research.

\subsection{DeepONet training and testing}
\label{section:deeponet_for_ifenn:training_testing}

The DeepONet model is trained in a data-driven approach where labeled pairs of inputs and outputs are generated before model training. Using the fully coupled FEM solvers, explained in Section~\ref{section:problem_statement:thermoelasticity} and Section~\ref{section:problem_statement:poroelasticity}, datasets are generated for several load histories (load cases). Afterwards, the DeepONet model training is conducted using the PyTorch library written in Python. For a feed-forward path in the DeepONet, three arrays are used as inputs for the load branch, the strain branch, and the trunk, as depicted in Fig.~\ref{fig:deeponet_detailed_1}. The shapes of the three input tensors for the first branch, the second branch, and the trunk are $[N_b, N_t, N_l]$, $[N_b, N_t, N_s]$, and $[N_n, N_d]$, respectively, where $N_b$ is the batch size (number of load cases per batch), $N_t$ is the number of time steps, $N_l$ is the number of load points (load sensors), $N_s$ is the number of strain points (strain sensors), $N_n$ is the number of target nodes (locations), $N_d$ is the number of spatial dimensions. The trunk encodes the same set of node coordinates for every load case and time step. To eliminate redundancy and enhance the model performance\cite{lu_comprehensive_2022}, the batch size ($N_b$) and number of time steps ($N_t$) are excluded from the input dimensions of the trunk. The overall network output for a given batch is a tensor with dimensions $[N_b, N_t, N_n, N_c]$. Each feed-forward operation is followed by backpropagation and weight updates to minimize the errors.

With backpropagation, the training process can be described as an optimization task, defined as follows:
\begin{align}
    \phi^* = \arg \min_{\phi} \mathcal{L}(\phi)
\end{align}
\noindent
where $\phi$ represents the model trainable parameters, and $\phi^*$ is the optimized set of weights to reduce the loss function $\mathcal{L}$.  Several definitions for the loss function were tested. Here, we list the main ones that were used for the following examples in Section~\ref{section:numerical_examples}:

\begin{align}
    e &= y_{true} -y_{pred}
    \\
    \mathcal{L}_{L_2} &= \left\| e \right\|_2
    \\
    \mathcal{L}_{L_2 Norm} &= \left\| e \right\|_2 / \left\| y_{true} \right\|_2
    \\
    \mathcal{L}_{\text{SSE}} &= \sum_{i=1}^{n} e_i^2
    \\
    \mathcal{L}_{\text{MSE}} &= \frac{1}{n} \sum_{i=1}^{n} e_i^2
\end{align}

\noindent
where $\left\| . \right\|_2$ is the second norm of a vector, $y_{pred}$ is the predicted response, $y_{true}$ is the fully coupled FEM response considered as the true value.

A clear distinction should be made between the training loss function $\mathcal{L}$ and the testing loss metric. To make consistent and fair comparisons, a unified system for testing error reporting is adopted as follows:
\begin{align}
        L_2^t &= {\left\| e^{(t)} \right\|_2} / {\left\| y^{(t)}_{true} \right\|_2} 
    \\
        L_2^{LC} &= {\left\| \{L_2^{t=1},L_2^{t=2},...,L_2^{t=N_t}\} \right\|_2}/{\sqrt{N_t}}
    \\
        L_2^{All\ LCs} &= {\left\| \{L_2^{LC=1},L_2^{LC=2},...,L_2^{LC=N_L}\} \right\|_2}/{\sqrt{N_L}}
    \\
        \epsilon_{rel} &= {(y_{true} -y_{pred})}/{(y_{true} + \epsilon_{tol})} 
\end{align}
\noindent
where $L_2^t$ is the normalized response error for a specific component at a specific time step $t$, $L_2^{LC}$ is the error for a specific load case (LC) over its whole time history, $L_2^{All\ LCs}$ is the overall error for all testing load cases. Finally, $\epsilon_{rel}$ is the relative error, with the tolerance $\epsilon_{tol}$ added for numerical stability. Due to having different response components that vary widely in magnitude, ranging from $10^{-5}$ to $10^5$, a consistent $\epsilon_{tol}$ is set as $10^{-5}$ of the maximum absolute $y_{true}$ value per component.
\section{Numerical Examples}
\label{section:numerical_examples}

In this section, we assess the proposed framework through different examples demonstrating its flexibility, generalizability, and scalability. Three examples are provided: (1) thermoelasticity model of a 3D cube with domain thermal loads, (2) thermoelasticity model of a 3D thick-walled tube with thermal surface loads, and (3) poroelasticity problem of a 2D excavation setup under fluctuating dewatering flux. 

For all examples, a systematic approach is adopted to reach the I-FENN implementation example. First, a sensitivity analysis is conducted to define the initial problem setup, including total time, time step increment, and mesh size. Afterwards, labeled datasets are generated for different load cases. Datasets are split into training, validation, and testing subsets. Extensive optimization trials are then undergone to reach an optimized DeepONet network with reasonable accuracy. However, it should be noted that despite persistent efforts to optimize the DeepONet models, there remains room for improvement and further optimization, often constrained by limitations in resources and time. 

After training, certain testing load cases exhibiting unique physics attributes are selected for I-FENN implementation and further model assessment. In the following discussions, these load cases will be referred to as "featured" load cases. In addition, the DeepONet testing error is evaluated for all the load cases within the testing dataset. The 10th, 50th (median), and 90th error-percentile cases are also chosen for I-FENN implementation. For performance comparisons, I-FENN and fully coupled FEM simulations are conducted on a workstation running Ubuntu 24.04, equipped with an Intel Xeon W3-2435 processor (16 cores), 128 GB of RAM, and an NVIDIA RTX A2000 GPU with 12 GB of dedicated memory. 

\subsection{Thermoelasticity example: 3D Cube with thermal body load}
\label{section:numerical_examples:cube}
\subsubsection{Problem setup}
\label{section:numerical_examples:cube:setup}
\begin{figure}[hbt!]
\centering
\includegraphics[width=.5\textwidth]{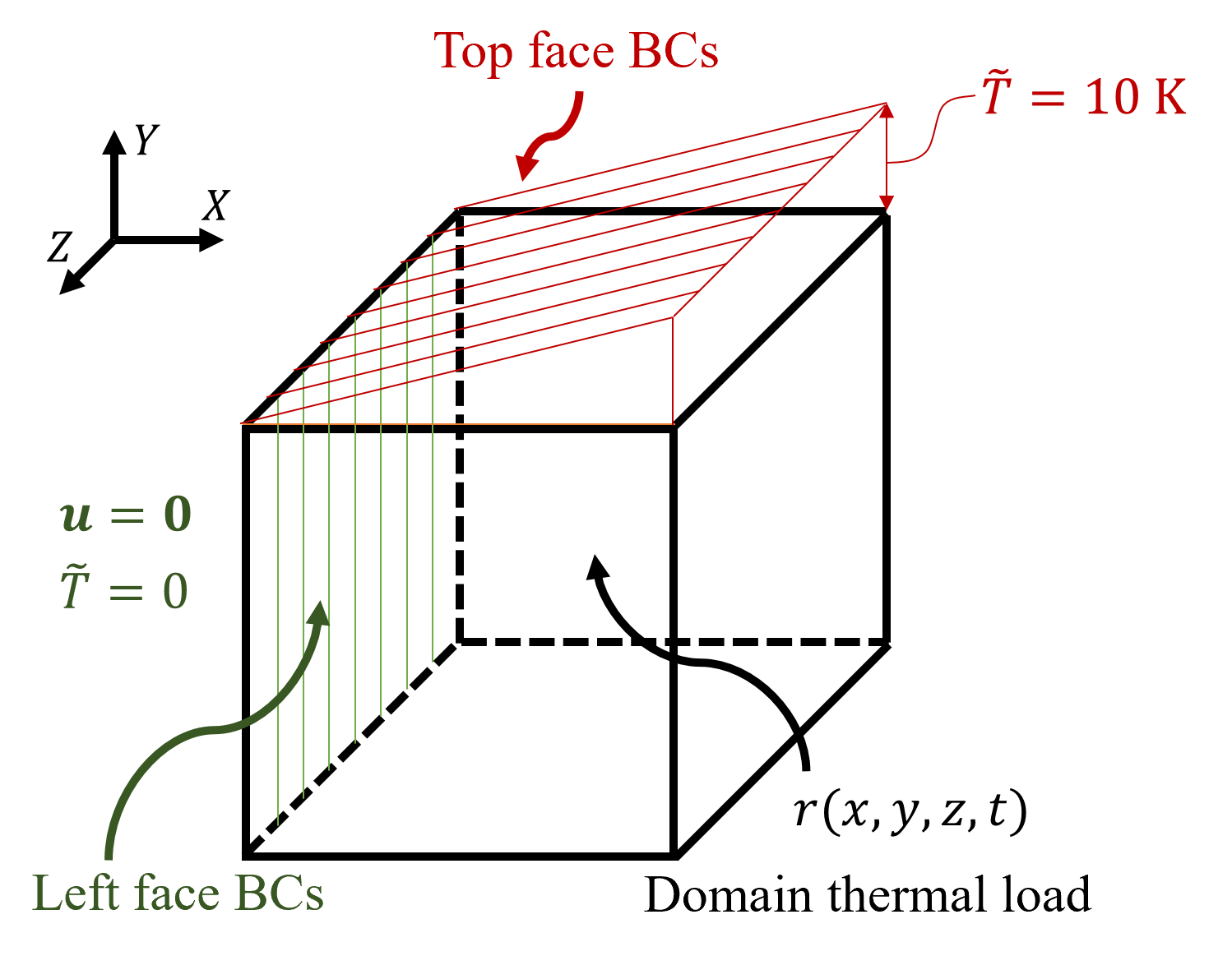}
\caption{Schematic representation of the geometry and boundary conditions
of the 3D cube problem.}
\label{fig:cube_setup}
\end{figure}

The first example is a 3D cube with a unit length along each direction. The material properties are set as $\lambda=40$GPa, $\mu=27$ GPa, $\alpha=2.31\times10^{-5}$ m/(m.K), $C_{\varepsilon}=910$ J/(kg.K), $\rho=2700$ kg/m$^3$, and $k=237$ W/(m.K).  As shown in Fig.~\ref{fig:cube_setup}, the cube has zero essential boundary conditions ($\boldsymbol{u}_0{}=\boldsymbol{0}$ and $\widetilde{T} = 0$) on the left face, where $\widetilde{T} = T - T_0$, and the reference temperature $T_0 = 293$ K. On the top face, a temperature boundary condition, function of space and time, is imposed as $\widetilde{T}=10x\gamma$, where $\gamma = \text {time in seconds}/1800 \le 1$. The total duration is $18,000$ seconds, which means that the temperature on the top face stabilizes after 10\% of the total duration. A Gaussian random thermal body load is applied throughout the whole domain. The simulation is conducted over uniformly spaced 100 time increments. 

\subsubsection{Training and testing results}
\label{section:numerical_examples:cube:training}
A total of 1000 load cases were generated: 900 of these cases used a coarse mesh of $11 \times 11 \times 11$ nodes ($N_n = 1331$) and were split into training and validation sets with a ratio of 4:1. The remaining 100 load cases were reserved for testing using a fine mesh $(31 \times 31 \times 31)$ nodes. Table~\ref{tab:cube_deeponet_arch} lists the hyperparameters describing the DeepONet model architecture. An input of size 512 is used for both load and strain branches. The input points (sensors) are spaced equally in a space grid, with 8 points per direction. A group normalization layer is added to the second branch to stabilize the framework's performance, as discussed in Subsection~\ref{section:numerical_examples:cube:stability}.

\begin{table}[hbt!]
\renewcommand{\arraystretch}{1.2}
\centering
\caption{Hyperparameters of the DeepONet model trained for the 3D cube example}
\begin{tabular}{ccccccc}
\hline
& Input Size & $N_{GRU}$ & $N_H$ & $N_{ch}$& $N_{FC}$ & $D_{out}$ \\
\hline
Branch \# 1 &$N_l=$512 & 2 & 200 & -  & 1 & 200 \\
Branch \# 2 &$N_b=$512 & 2 & 50  & 25 & 1 & 50  \\
Trunk       &$N_d=$3   & - & 200 & -  & 4 & 250 \\
\hline
\end{tabular}
\label{tab:cube_deeponet_arch}
\end{table}

\begin{figure}[hbt!]
\centering
\includegraphics[width=1.0\textwidth]{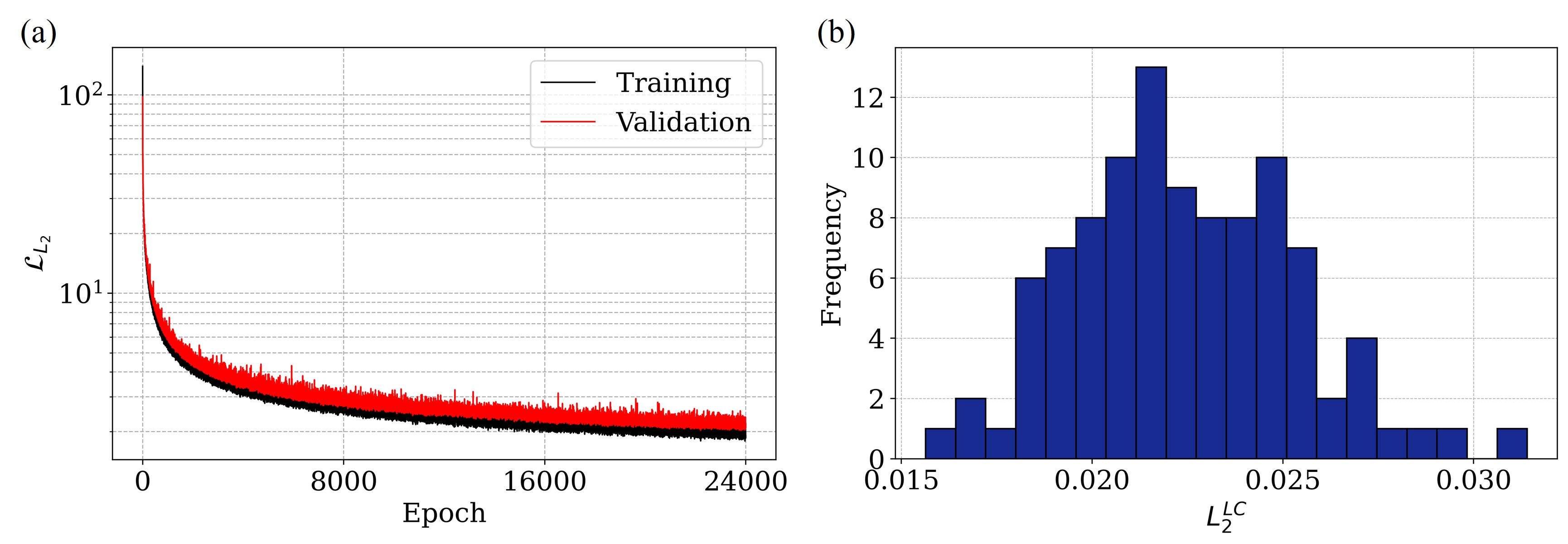}
\caption{a) DeepONet model training loss $\mathcal{L}_{L_2}$ for the 3D cube example, b) Histogram of the DeepONet model testing loss $L_2^{LC}$ for different load cases of the 3D cube example.}
\label{fig:cube_training_testing_loss}
\end{figure}

Training is done over $24,000$ epochs with the $L_2$ norm function $\mathcal{L}_{L_2}$ selected for loss evaluation. Training vs. validation loss values are depicted in Fig.~\ref{fig:cube_training_testing_loss}a, where both curves are decreasing fairly smoothly, showing that the model is learning effectively. The training loss is relatively lower than the validation curve, which is typically expected. Training required a total time of 9 hours and 4 minutes. Although increasing the batch size could further reduce training time, we kept the current configuration (batch size = 16) because earlier experiments showed that larger batches negatively affected accuracy. That said, there are still several avenues for improving training efficiency. Potential directions include recalibrating the training hyperparameters, applying mixed precision, experimenting with alternative optimizer variants, or adjusting data loading configurations \cite{migacz_performance_2025, janson_pylo_2025}.

After training on a coarse mesh, the model is tested for 100 load cases (LCs) using a fine mesh of equally spaced $N_n=29791$ nodes (31 per direction). A histogram of the normalized $L_2$ norm of error per load case ($L_2^{LC}$) is plotted in Fig.~\ref{fig:cube_training_testing_loss}b. The median, 10th, and 90th error-percentile load cases are selected for I-FENN integration. The results for the featured and the median load cases are presented in the following subsection, while the 10th and the 90th error-percentile load cases are presented in the appendix (\ref{section:numerical_examples:cube:appendix:10-90}).

\subsubsection{I-FENN results}
\label{section:numerical_examples:cube:results}

\begin{figure}[hbt!]
\centering
\includegraphics[width=1.0\textwidth]{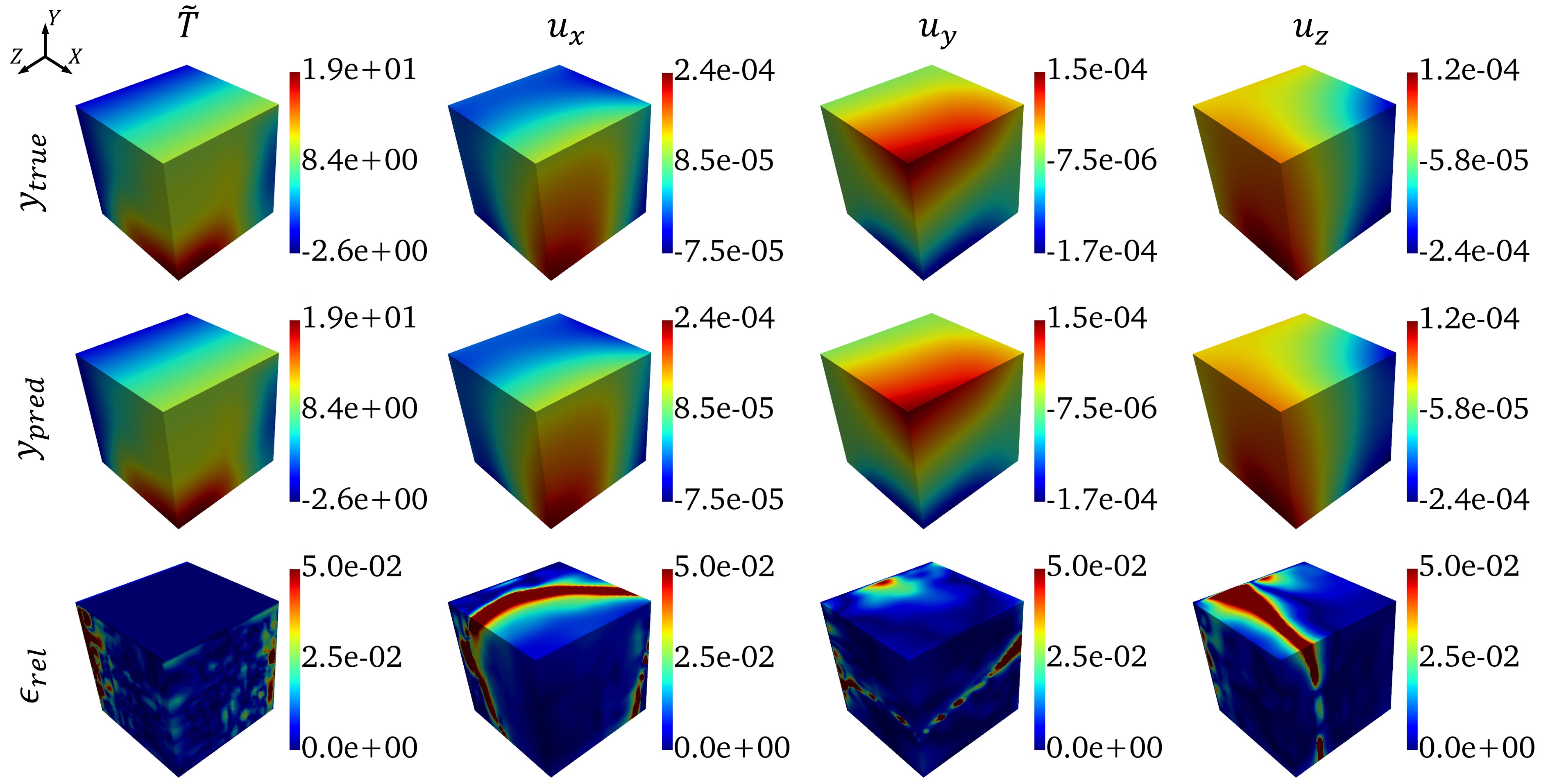}
\caption{Solution obtained using the fully coupled FEM solver ($y_{true}$) and I-FENN ($y_{pred}$) for the median load case at the 100th time step for the 3D cube example.}
\label{fig:cube_median}
\end{figure}

\begin{figure}[hbt!]
\centering
\includegraphics[width=1.0\textwidth]{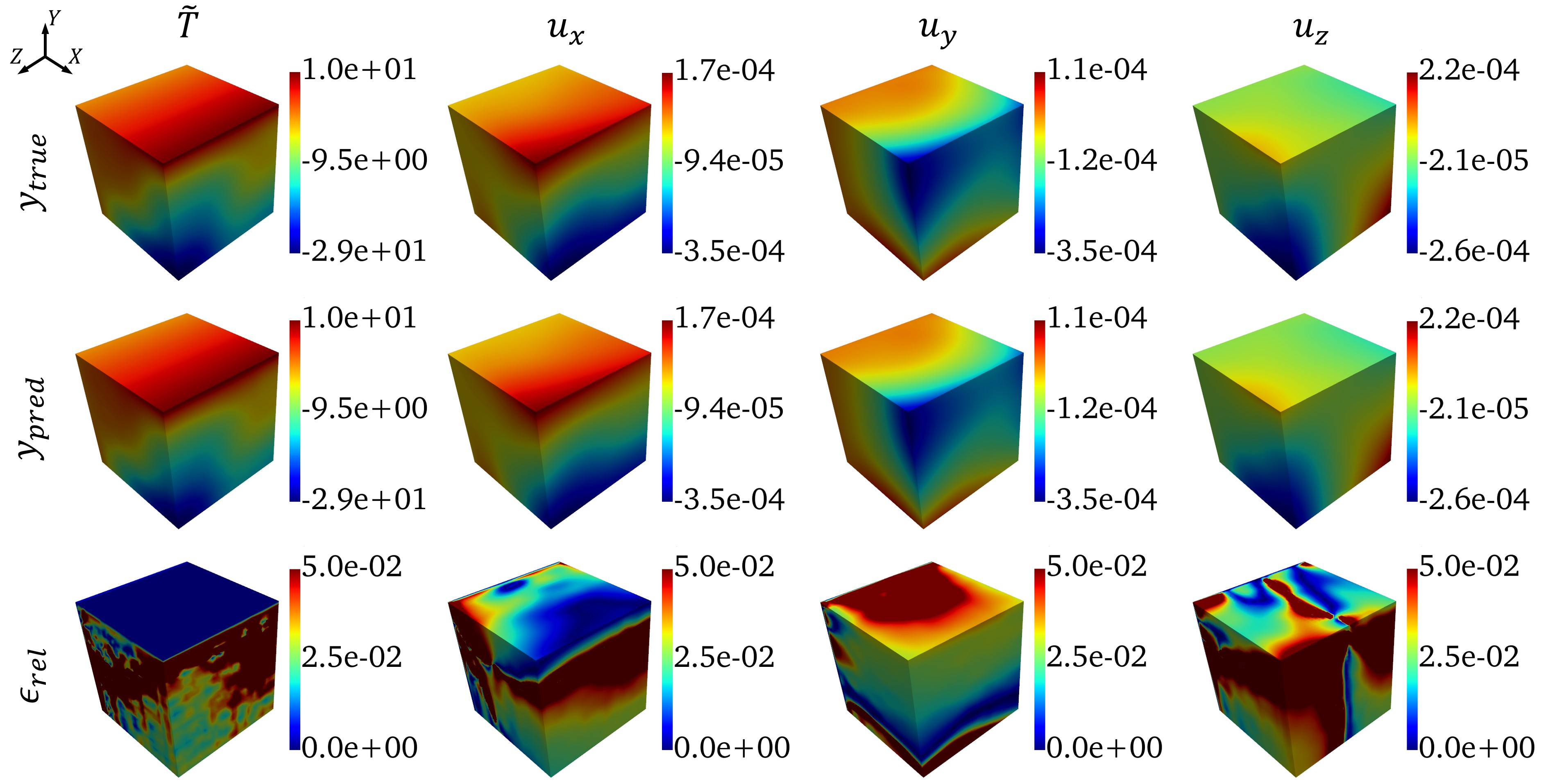}
\caption{Solution obtained using the fully coupled FEM solver ($y_{true}$) and I-FENN ($y_{pred}$) for the featured load case at the 86th time step for the 3D cube example.}
\label{fig:cube_physics}
\end{figure}

The trained model is incorporated into the I-FENN framework as described in Section~\ref{section:methodology}. The I-FENN results for the median load case are presented in Fig.~\ref{fig:cube_median}. In addition, a featured load case is chosen based on the highest thermal load integral, representing the maximum amount of heat change introduced to the body. Throughout all the testing load cases, the selected featured load case exhibited the maximum absolute temperature change  ($\widetilde{T}$) and maximum absolute strain trace. The I-FENN results for the featured load case are presented in Fig.~\ref{fig:cube_physics}, with maximum response values detected at the 86th time step.

In both figures (Fig.~\ref{fig:cube_median} and Fig.~\ref{fig:cube_physics}), the coupled FEM results are considered as the true values ($y_{true}$), which are plotted in the top row, followed by the I-FENN response ($y_{pred}$) in the second row. The third row presents the relative error ($\epsilon_{rel}$) capped at 5\%. The first column depicts the temperature changes ($\widetilde{T}$), followed by the displacement components ($u_x, u_y, \text{and } u_z$). The I-FENN results demonstrate strong agreement with the FEM results. Overall, the results indicate excellent consistency, particularly for zones with extreme response values, which exhibit relative errors well below 5\%. The relative error zones exceeding 5\% are attributed to very small response values close to zero, which magnifies the relative error values. A consistent behavior of I-FENN is found in the results for the 10th and 90th error-percentile load cases, depicted in Fig.~\ref{fig:cube_10th} and Fig.~\ref{fig:cube_90th}, respectively.  

\subsubsection{Computational savings and scalability}

The network was trained using a coarse mesh of $(11 \times 11 \times 11)$ nodes, and then accurately predicted the response for testing on a fine mesh of $(31 \times 31 \times 31)$ nodes. A complementary analysis is provided in Appendix~\ref{section:numerical_examples:cube:appendix:fine}, where Fig.~\ref{fig:cube_median_fine75} confirms that the framework maintains its accuracy on further mesh refinement. To assess performance and scalability, the framework's efficiency in computing responses on finer meshes is examined. The results are tabulated in Table~\ref{tab:fem_ifenn_times}, showing the computational time of the I-FENN framework and the fully coupled FEM solver for different domain sizes (cells per direction).

\begin{table}[hbt!]
\renewcommand{\arraystretch}{1.2}
\centering
\caption{Comparison of FEM and I-FENN computation times in minutes across various domain sizes.}
\begin{tabular}{cccc}
\hline
\makecell {Domain size \\ $[$Cells$]$} &  \makecell {FEM time \\ $[$Minutes$]$} & \makecell {I-FENN time \\ $[$Minutes$]$}  & Savings \\
\hline
$30 \times 30 \times 30$ & 5.1 & 3.3 & 35\% \\
$45 \times 45 \times 45$ & 20.0 & 12.5 & 38\% \\
$60 \times 60 \times 60$ & 56.9 & 34.2 & 40\% \\
$75 \times 75 \times 75$ & 134.0 & 75.8 & 43\% \\
\hline
\end{tabular}
\label{tab:fem_ifenn_times}
\end{table}

For all tested domain sizes, the I-FENN framework exceeded the performance of the fully coupled FEM solver, showing a consistent computational efficiency. To better visualize the results, computational times and savings across different domain sizes are plotted in Fig.~\ref{fig:cube_median}. In addition to the demonstrated computational savings, this plot highlights the scalability of the I-FENN framework as the domain is refined. This scalability is a critical feature for a framework designed to improve performance in multiphysics simulations, ensuring its applicability to more complex and computationally demanding problems in the future. It is important to highlight that the reported computational times of I-FENN exclude the training phase of the DeepONet. For just a few simulations, on a coarse mesh, the overall time (including DeepONet training and I-FENN execution) can exceed that of running a fully coupled FEM solver. However, thanks to I-FENN's demonstrated scalability, when performing many simulations for different load cases, especially on fine meshes, the combined cost becomes significantly lower than that of the traditional FEM approach.  

\begin{figure}[hbt!]
\centering
\includegraphics[width=.75\textwidth]{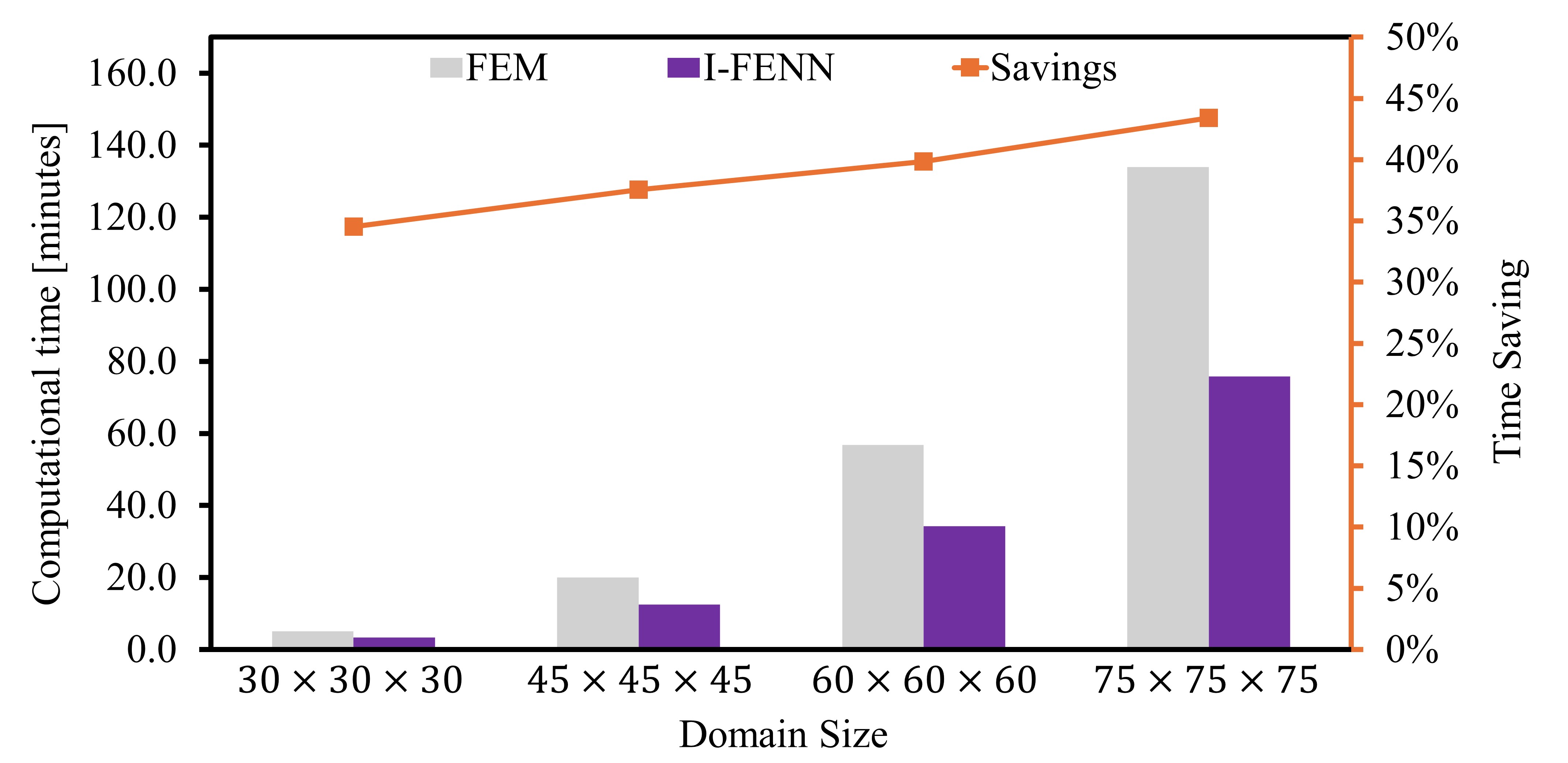}
\caption{Computation time comparison and scalability of I-FENN across domain sizes.}
\label{fig:cube_savings}
\end{figure}

I-FENN outperforming the fully coupled FEM is a particularly noteworthy observation, given that the deal.II library, written in C++, is highly optimized for performance and designed with high-performance computing (HPC) compatibility in mind. In contrast, the I-FENN framework is still under active development, with numerous performance optimizations yet to be implemented.

\subsubsection{I-FENN stability}
\label{section:numerical_examples:cube:stability}

During the initial stages of developing the network, it is evident that there was an issue with the output stability. There was an error accumulation pattern through time steps,  resulting in higher errors towards the end of the simulation. To address this issue, a modified framework design was established to investigate the root cause of the error. 

In this approach, true values (computed using the fully coupled FEM solver) were incorporated into the framework. The simulation starts with using true strain trace and temperature values as inputs to the DeepONet and the mechanical FEM solver, respectively. After step number 33, the DeepONet output replaces the true temperature and is sent to the FEM solver. At step number 66, the strain trace output from the mechanical solver replaces the true strain and is used as input to the DeepONet. 

\begin{figure}[hbt!]
\centering
\includegraphics[width=1.0\textwidth]{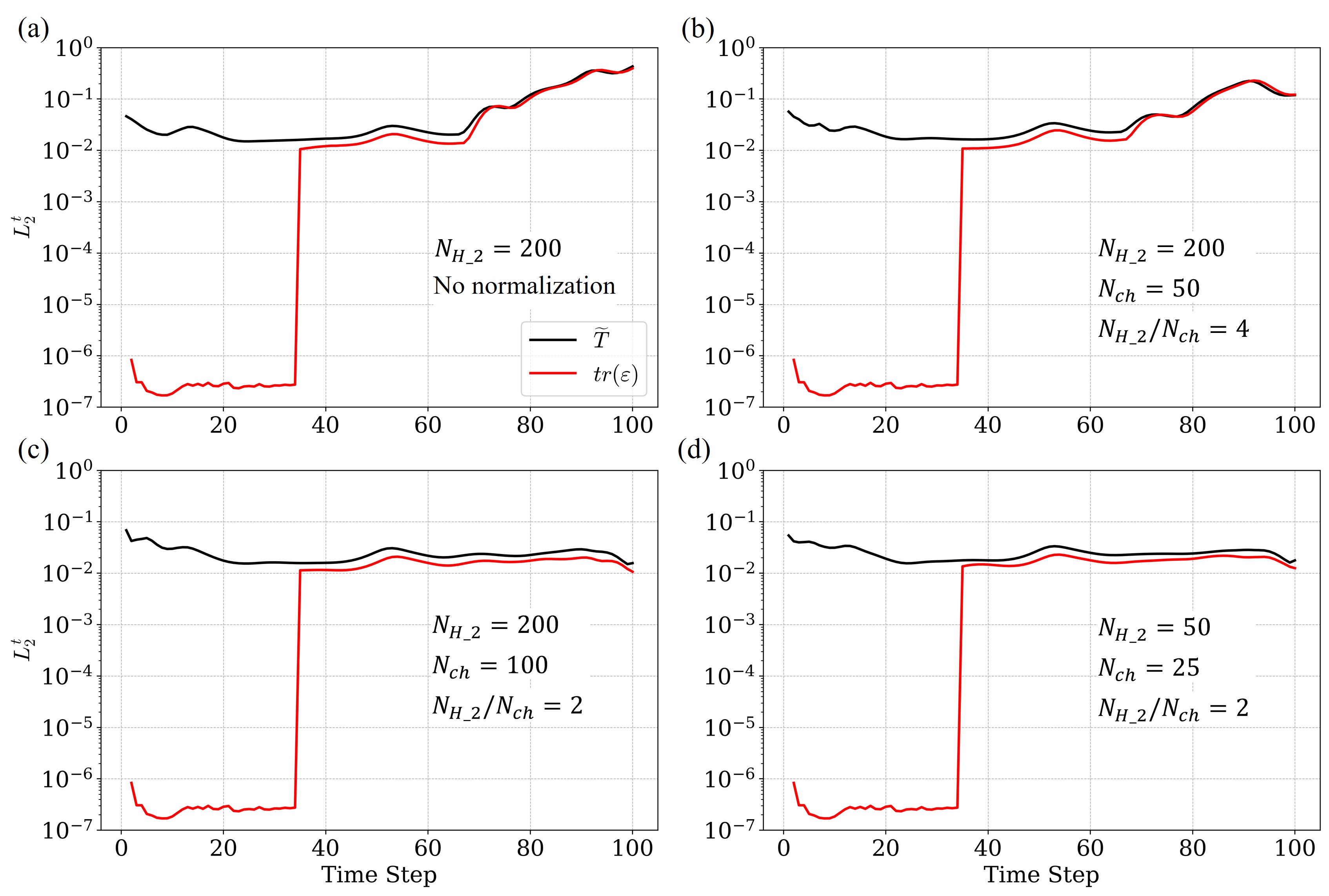}
\caption{I-FENN framework stability: showing normalized error $L_2^t$ vs. time steps across different setups of the DeepONet model: a) $N_{H\_2}=200$ without group normalization, b) $N_{H\_2}=200$ with group normalization using $N_{ch}=50$, c) $N_{H\_2}=200$ with $N_{ch}=100$, and d) $N_{H\_2}=50$ with $N_{ch}=25$. $N_{H\_2}$ is the GRU hidden layer size for the strain branch, and $N_{ch}$ is the number of normalization channels. All tests are conducted for the reference model's 90th error-percentile load case. True temperature values are used till step 33, and true strain values are used till step 66.}
\label{fig:cube_stability_comparison}
\end{figure}

This testing approach was applied to different networks with different architectures, with results summarized in Fig.~\ref{fig:cube_stability_comparison}. The figure depicts the normalized error norm per time step $L_2^t$ for strain trace and temperature values across time steps as the simulation progresses incrementally from one step to another. Four design options are tested for the strain branch: a) GRU hidden output of size 200 ($N_{H\_2}=200$) without group normalization, b) $N_{H\_2}=200$ with group normalization channels of $N_{ch}=50$, c) $N_{H\_2}=200$ with $N_{ch}=100$, and d) $N_{H\_2}=50$ with $N_{ch}=25$. The reader is referred to Section~\ref{section:deeponet_for_ifenn:detailed} for more details on the adopted group normalization approach. 

For the first design option with no normalization, the error values of strain trace and temperature show different patterns, which can be described along three phases. Phase one starts from step 1 to step 33. The strain trace relative error is very low, given that this is a response of the mechanical FEM solver using true temperature inputs. In the first stage, also, the DeepONet output (temperature), shows a relatively reasonable error which is dependent on the network training. 

Moving to the second phase between steps 33 and 66, only one parameter is changed. The temperature predictions from DeepONet are used as inputs to the mechanical FEM solver, resulting in a corresponding stable increase in the strain trace error. The third stage starts at the 66th step by using the FEM mechanical strain trace as input to the DeepONet. Therefore, it can be concluded that the DeepONet model is susceptible to minor changes in the strain trace inputs.

To overcome this sensitivity to changes in the strain input, the group normalization approach is introduced in the other three design options. Design options (c) and (d) show enhanced stability in the network performance. Option (d) provided an optimized network size and, hence, is selected for further implementation (refer to Table~\ref{tab:cube_deeponet_arch}). Finally, an additional study on the 3D cube example is conducted to investigate the impact of not enforcing boundary conditions. For brevity, the detailed analysis and corresponding results are presented in Appendix~\ref{section:numerical_examples:cube:appendix:bcs}.

\subsection{Thermoelasticity example: 3D Thick-walled tube with thermal surface load}
\label{section:numerical_examples:tube}
\subsubsection{Problem setup}
\label{section:numerical_examples:tube:setup}
\begin{figure}[hbt!]
\centering
\includegraphics[width=.5\textwidth]{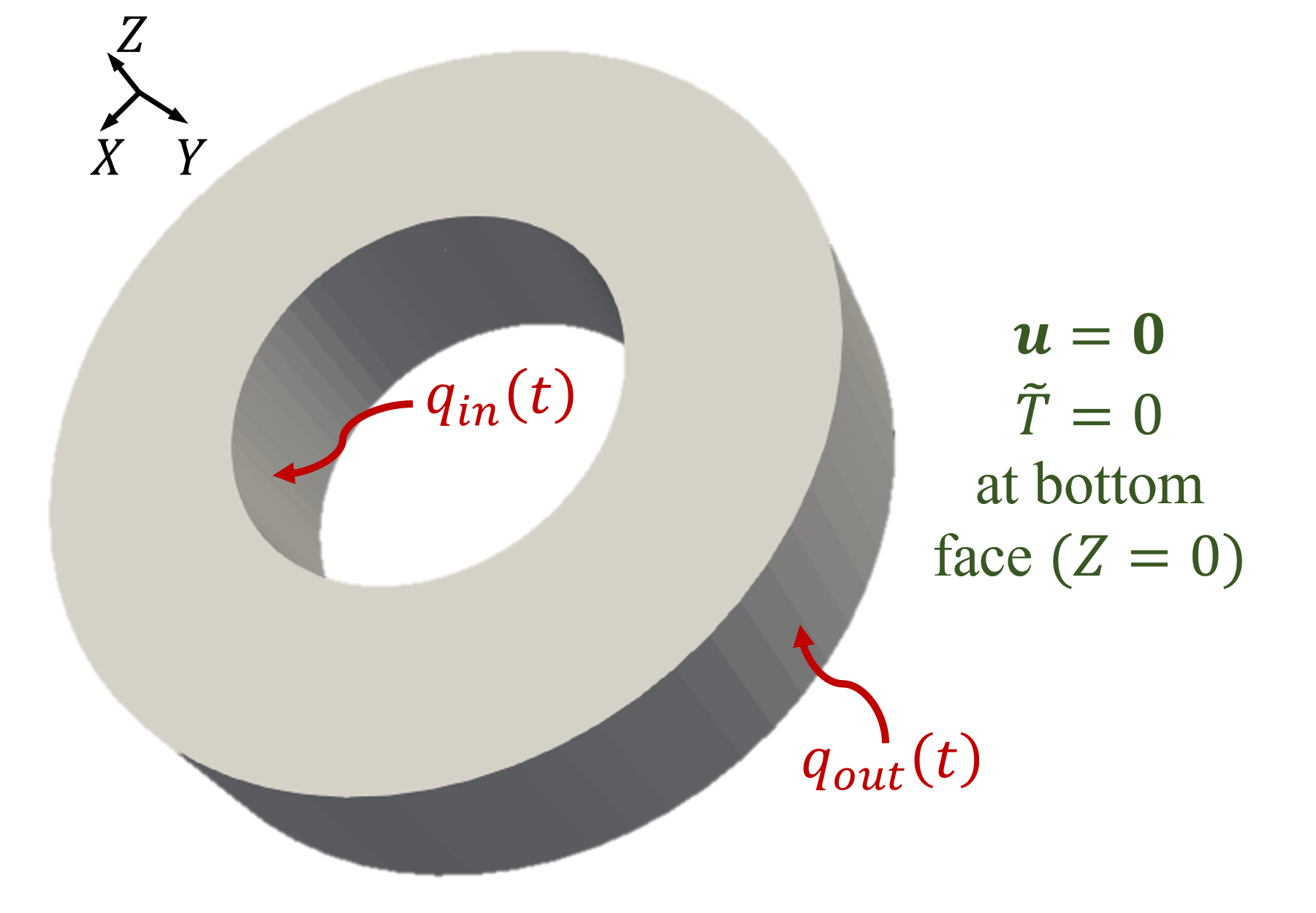}
\caption{Schematic representation of the geometry and boundary conditions
of the 3D tube problem.}
\label{fig:tube_setup}
\end{figure}

The second example is a 3D thick-walled tube, as depicted in Fig.~\ref{fig:tube_setup}, with inner and outer radii of 1.0 m and 2.0 m, respectively. The tube has a length of 1.0 m along its longitudinal axis (Z-axis). Zero essential boundary conditions are imposed on the base plane ($z=0$). Material properties are the same as those selected for example 1 in Section~\ref{section:numerical_examples:cube:setup}.  However, for this example, no body loads were incorporated. Instead, boundary flux values are applied on the inner and outer walls of the tube. Both inner flux $(q_{in})$ and outer flux $(q_{out})$ values are defined as functions of time and angular location $(\theta = \arctan(y/x))$. Flux functions are defined as $q_{in}(\theta, t)=q_{in-0}+q_{in-1}\sin{(\omega_tt)}\sin{(\omega_{in-r}\theta)}$, and $q_{out}(\theta,t)=q_{out-0}+q_{out-1}\sin{(\omega_tt)}\sin{(\omega_{out-r}\theta)}$.

\subsubsection{Training and testing results}
\label{section:numerical_examples:tube:training}
The parameters  $q_{in-0}$, $q_{in-1}$, $q_{out-0}$, $q_{out-1}$, $\omega_{in-r}$, $\omega_{out-r}$, and $\omega_t$ were randomly sampled to generate $1000$ unique loading cases. Each case is simulated for $60000$ seconds, divided into 120 equal increments. A subset of $900$ load cases was split into training and validation sets with a ratio of $4:1$, while the remaining $100$ load cases were used for testing. Comprehensive parametric studies, detailed in Appendix ~\ref{section:numerical_examples:tube:appendix}, are conducted to optimize the model's hyperparameters. Based on these studies, the hyperparameters listed in Table~\ref{tab:tube_deeponet_arch} are selected for training the DeepONet model employed in I-FENN.

\begin{table}[hbt!]
\renewcommand{\arraystretch}{1.2}
\centering
\caption{Hyperparameters of the DeepONet model trained for the 3D tube example}
\begin{tabular}{ccccccc}
\hline 
& Input Size & $N_{GRU}$ & $N_H$ & $N_{ch}$& $N_{FC}$ & $D_{out}$ \\
\hline
Branch \# 1 &$N_l=$128 & 2 & 64 & -  & 1 & 64 \\
Branch \# 2 &$N_b=$512 & 2 & 64  & 32 & 1 & 64  \\
Trunk       &$N_d=$3   & - & 256& -  & 4& 128\\
\hline
\end{tabular}
\label{tab:tube_deeponet_arch}
\end{table}

\begin{figure}[hbt!]
\centering
\includegraphics[width=1.0\textwidth]{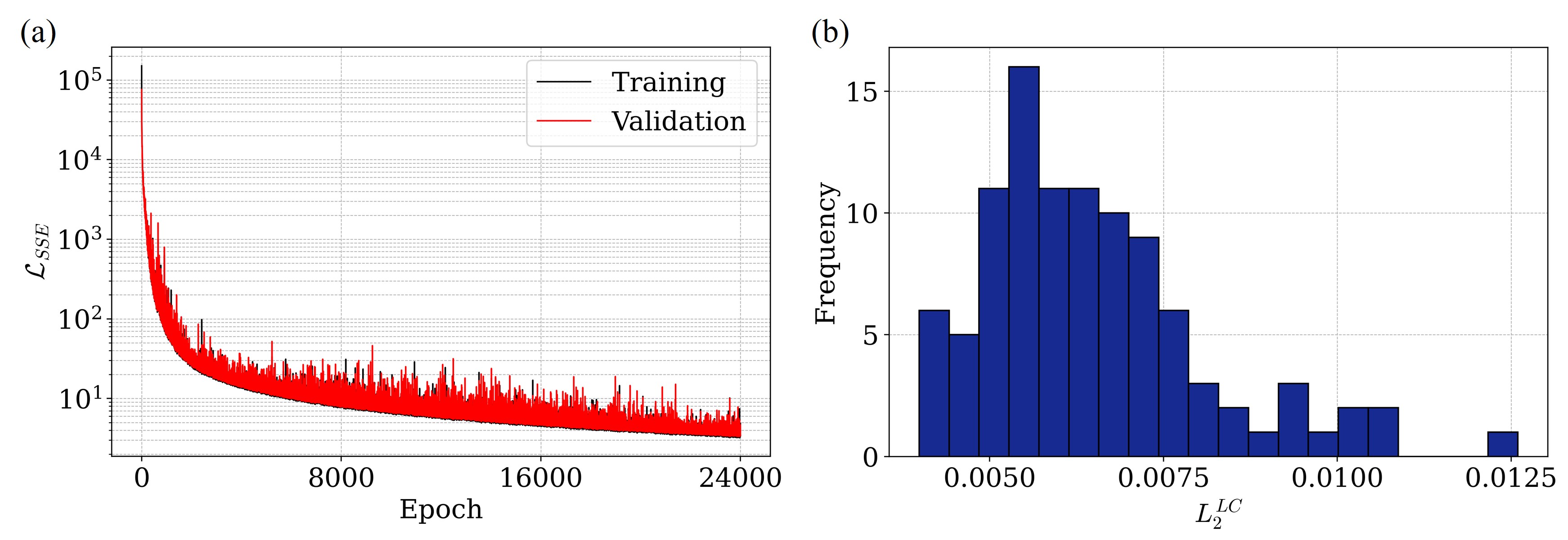}
\caption{a) DeepONet model training loss $\mathcal{L}_{SEE}$ for the 3D tube example, b) Histogram of the DeepONet model testing loss $L_2^{LC}$ for different load cases of the 3D tube example.}
\label{fig:tube_training_testing_loss}
\end{figure}

A total of 128 equally spaced points are selected for surface load input, with 64 points per surface. Within the whole domain, 512 structurally spaced points are selected to represent the strain trace field. A total number of $5184$ nodes is utilized for training the network ($N_n=5184$). A sum of square error loss function $\mathcal{L}_{\text{SSE}}$ is utilized for calculating training loss for a total number of $24,000$ epochs, requiring 12 hours and 35 minutes for training. Training vs validation loss curves are provided in Fig.~\ref{fig:tube_training_testing_loss}a, showing a stable descending learning curve with no signs of overfitting. 

A histogram of testing error is depicted in Fig.~\ref{fig:tube_training_testing_loss}b, showing a roughly normal error distribution with slight skew towards the left. The median, 10th, and 90th error-percentile load cases are selected for I-FENN implementation. The median load case is covered in the following subsection, while the 10th and 90th error-percentile load cases are provided in Appendix~\ref{section:numerical_examples:tube:appendix:10-90}.

\subsubsection{I-FENN results}
\label{section:numerical_examples:tube:results}
\begin{figure}[hbt!]
\centering
\includegraphics[width=1.0\textwidth]{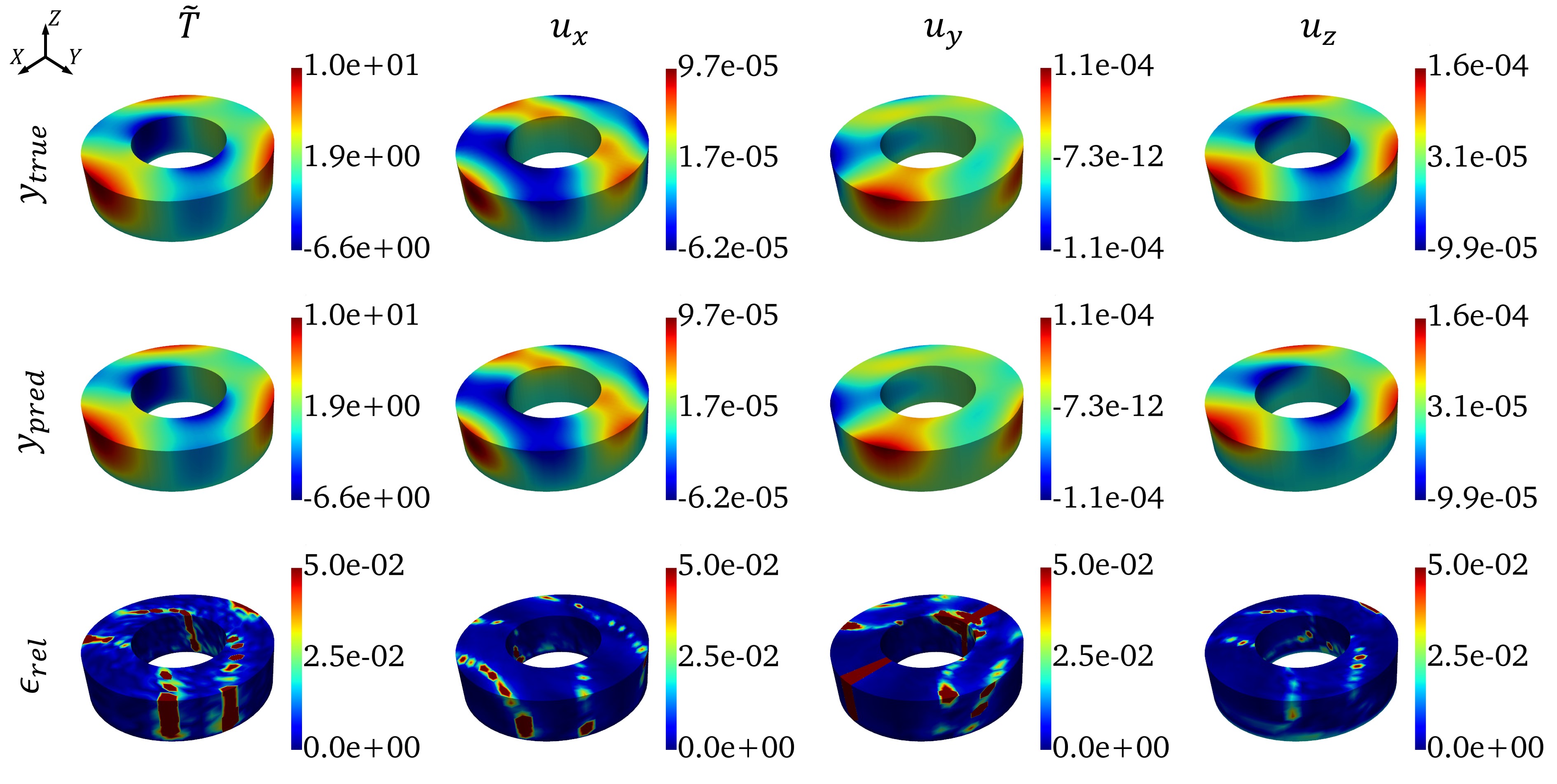}
\caption{Solution obtained using the fully coupled FEM solver ($y_{true}$) and I-FENN ($y_{pred}$) for the median load case at the 120th time step for the 3D tube example.}
\label{fig:tube_median}
\end{figure}

I-FENN implementation is done as described in Section~\ref{section:methodology}, incorporating the trained DeepONet model. I-FENN results and the fully coupled FEM response for the median load case are depicted in Fig.~\ref{fig:tube_median} along with the relative error values. The coupled FEM results are plotted in the top row, as the true values ($y_{true}$), followed by the I-FENN response ($y_{pred}$) in the second row, and the relative error ($\epsilon_{rel}$) capped at 5\% at the third row. All I-FENN response fields show a good agreement with the fully coupled FEM response, with relative error values much lower than 5\% at the zones with high response values. Only zones with response values close to zero exhibited relative error exceeding 5\%. 

\begin{figure}[hbt!]
\centering
\includegraphics[width=1.0\textwidth]{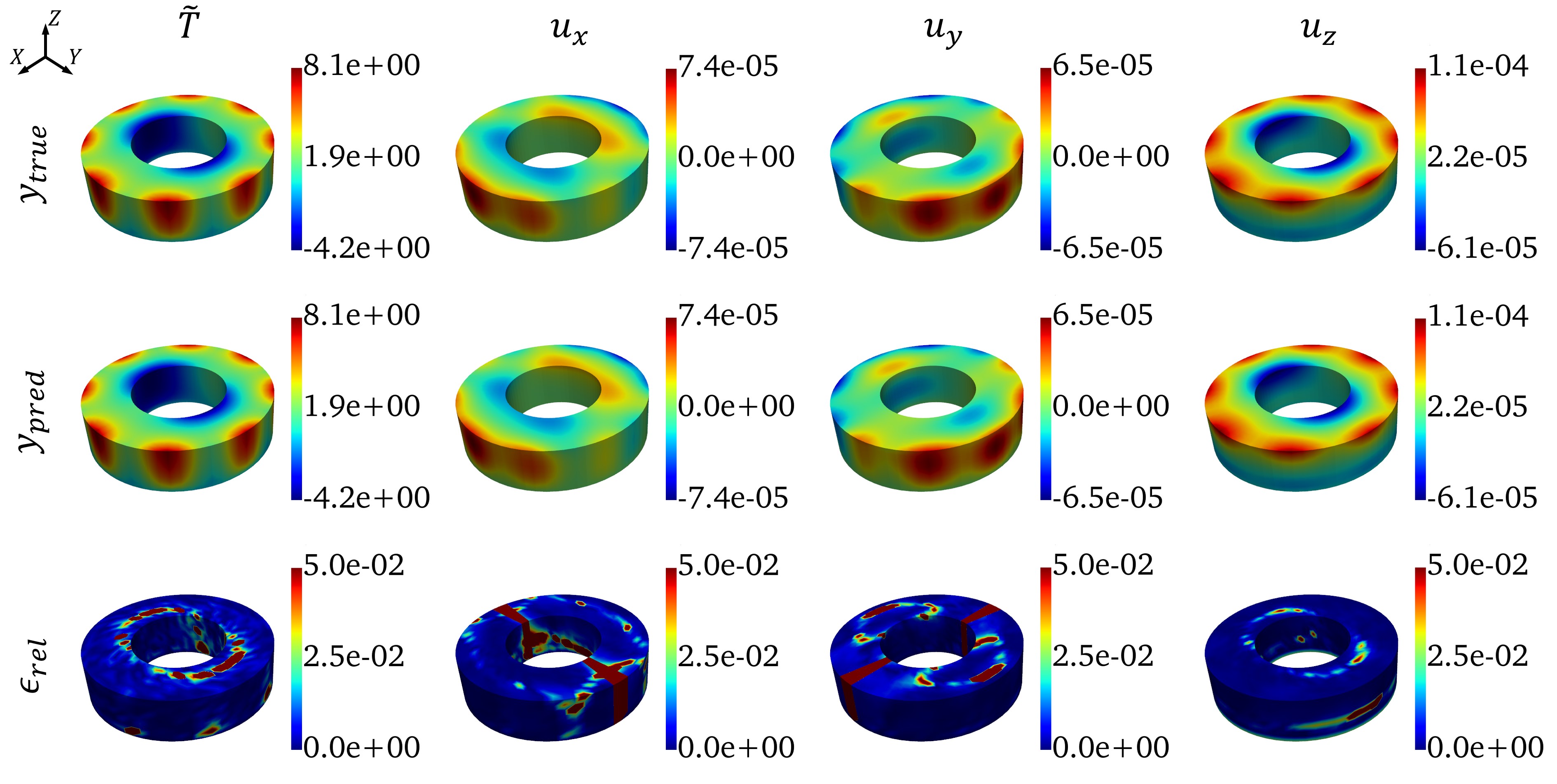}
\caption{Solution obtained using the fully coupled FEM solver ($y_{true}$) and I-FENN ($y_{pred}$) for the featured load case at the 90th time step for the 3D tube example.}
\label{fig:tube_physics}
\end{figure}

In addition to the median load case, a featured load case is chosen based on the highest discrepancy in the spatial frequency between inner and outer boundary fluxes. Figure.~\ref{fig:tube_physics} presents the results of the featured load case at the 90th time step, showing the ability of I-FENN to capture extreme response spacial discrepancies with high accuracy. The I-FENN framework continues to demonstrate consistent and robust performance across this example and the previous one (Section~\ref{section:numerical_examples:cube}). The current example further highlights the versatility of the I-FENN approach, showcasing its ability to handle flux values that vary both spatially and temporally within a more complex geometry than that of the previous example.

\subsubsection{I-FENN vs. simulation substitution}

As discussed in the literature review (Section~\ref{section:introduction:literature_review}), a possible approach for machine learning in computational mechanics is simulation substitution, where the whole response is predicted using a neural network as a surrogate model. To test this approach,  a DeepONet model is trained for predicting all components as a surrogate model. The surrogate model has only one branch for load, as the strain trace will not exist as an output from the mechanical solver. For the surrogate model, $N_c$ is set as four to represent the four response components ($\widetilde{T}, u_x, u_y, \text{and } u_z$). The surrogate model hyperparameters (listed in Table~\ref{tab:tube_deeponet_surrogate_arch}) are calibrated to provide a roughly close number of trainable parameters as those of I-FENN integrated DeepONet.

\begin{table}[hbt!]
\renewcommand{\arraystretch}{1.2}
\centering
\caption{Hyperparameters of the surrogate DeepONet model trained for simulation substitution for the 3D tube example}
\begin{tabular}{ccccccc}
\hline
& Input Size & $N_{GRU}$ & $N_H$ & $N_{ch}$& $N_{FC}$ & $D_{out}$ \\
\hline
Branch \# 1 &$N_l=$128 & 2 & 84 & -  & 1 & 84 \\
Trunk       &$N_d=$3   & - & 332& -  & 4& 84\\
\hline
\end{tabular}
\label{tab:tube_deeponet_surrogate_arch}
\end{table}

The DeepONet for I-FENN and the surrogate model have a number of trainable parameters of  372,032  and 375,740, respectively.  Labeled pairs of input/output data were used for training the model, with four output components per node. Then, the surrogate model is trained using the same parameters used for training the DeepONet for I-FENN, such as training/testing load cases, the loss function, the learning rate, and the number of epochs.

\begin{figure}[hbt!]
\centering
\includegraphics[width=1.0\textwidth]{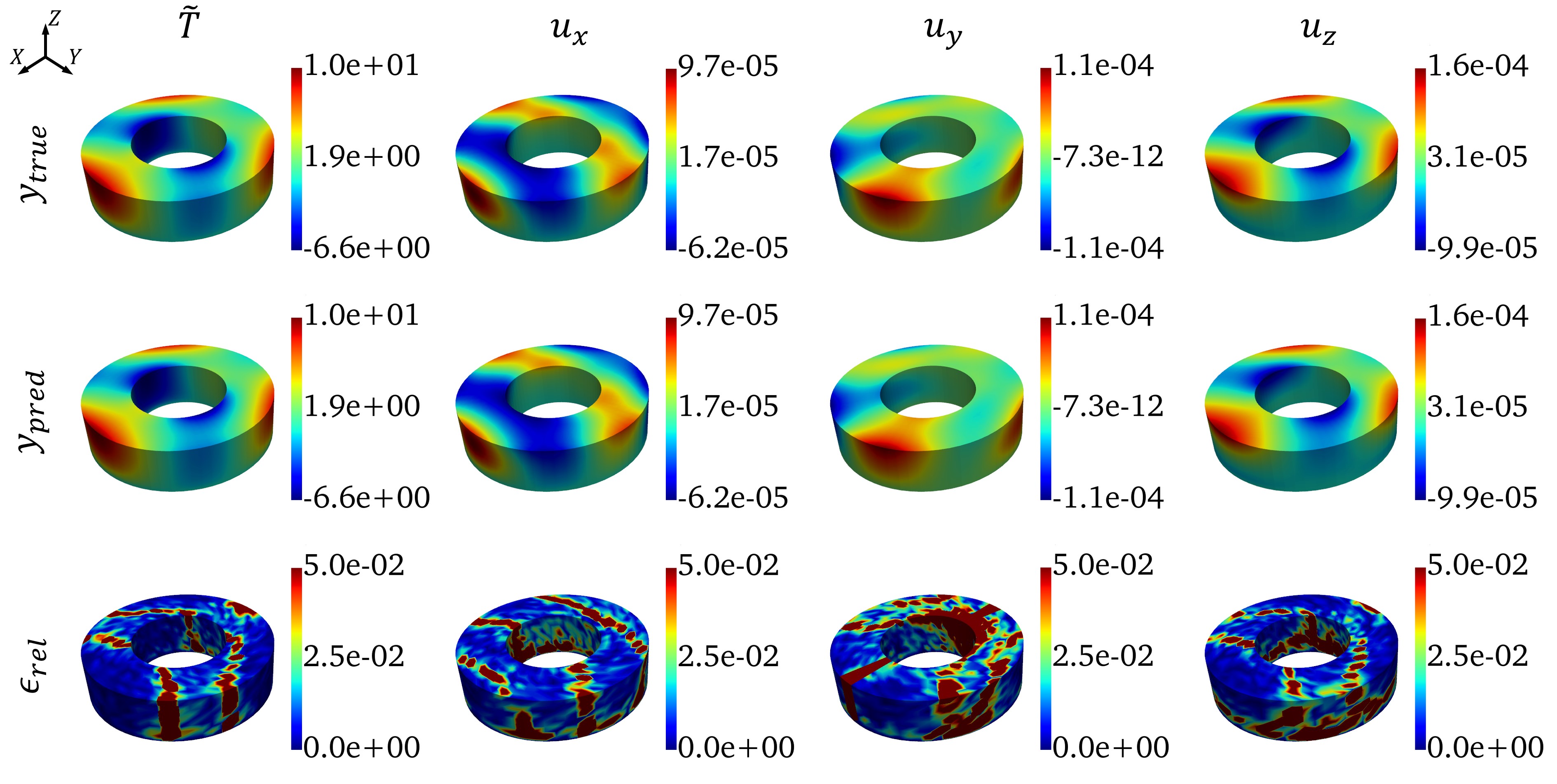}
\caption{Solution obtained using the fully coupled FEM solver ($y_{true}$) and the surrogate model ($y_{pred}$) for the median load case at the 120th time step for the 3D tube example.}
\label{fig:tube_median_surrogate}
\end{figure}

\begin{figure}[hbt!]
\centering
\includegraphics[width=1.0\textwidth]{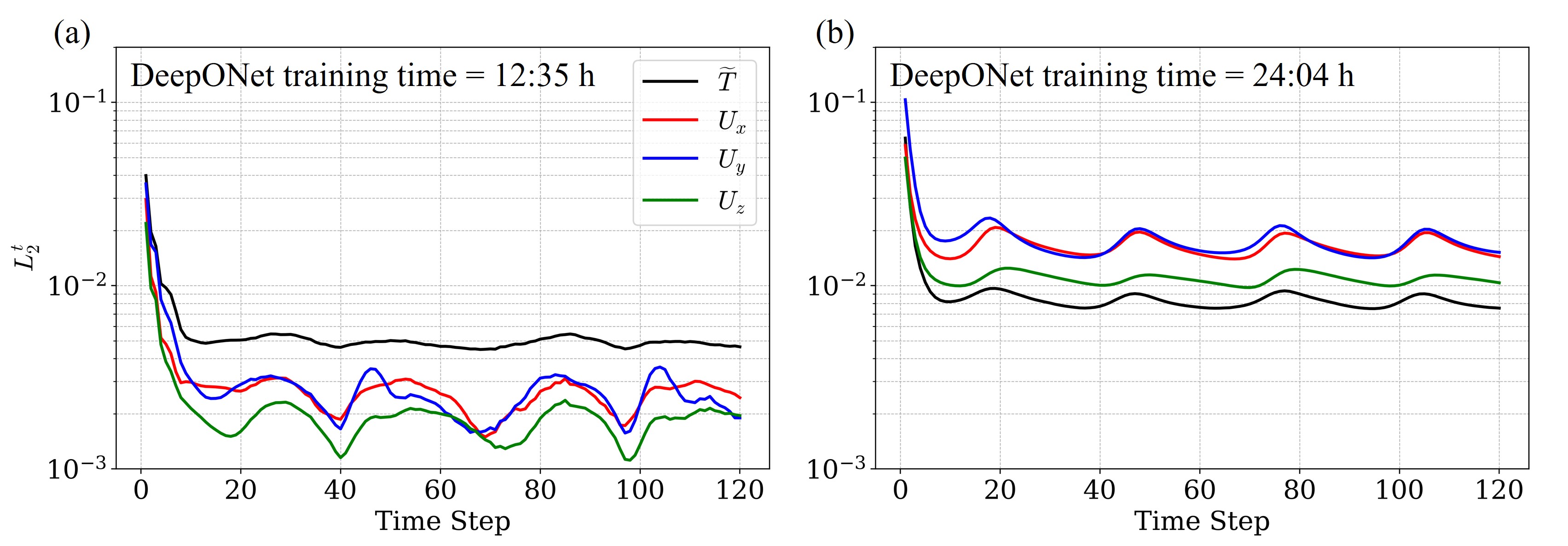}
\caption{Normalized error $L_2^t$ for all components of the median load case applied to the 3D tube example, using: a) I-FENN, b) simulation substitution (surrogate model).}
\label{fig:tube_ifenn_vs_surrogate_median}
\end{figure}

The trained surrogate model results for the median load case at the last time step are depicted in Fig.~\ref{fig:tube_median_surrogate}. Noting that the median load case here refers to the same load case detected from the frequency distribution reported in Fig.~\ref{fig:tube_training_testing_loss}b. Compared to I-FENN results (Fig.~\ref{fig:tube_median}), the surrogate model exhibits higher error values across all components, with larger zones exceeding 5\% relative error. For a more comprehensive assessment, the testing error $L_2^t$ values for I-FENN and the surrogate model for all time steps and response components are plotted in Fig.~\ref{fig:tube_ifenn_vs_surrogate_median}. Error plots show that I-FENN errors are almost one order of magnitude less than those of the surrogate model, especially for the displacement components. It is worth noting that the higher accuracy in displacement components is related to the mechanical solver integrated within the I-FENN framework, highlighting the accuracy edge provided by I-FENN compared to surrogate models.

Compared to I-FENN, the surrogate model requires four times larger output training datasets, making it considerably more demanding, particularly for more complex or high-dimensional problems. In addition to higher memory requirements, training the surrogate model took twice as long as training the I-FENN integrated DeepONet. Additionally, the surrogate model functions as a black box, offering neither interpretability nor flexibility, and lacks the capability for deeper analysis, such as the stability study presented in Section~\ref{section:numerical_examples:cube:stability}. While the surrogate model may be acceptable in specific contexts, the superior accuracy, interpretability, and adaptability of the I-FENN framework make it a significantly more appropriate choice for multiphysics simulations.

Finally, it is essential to comment on the relatively early prediction errors, shown in Fig.\ref{fig:tube_ifenn_vs_surrogate_median}. These errors are primarily concentrated within the initial steps and can be attributed to the transient phase introduced by the GRU network. This transient phase arises from the sensitivity of the GRU to its initial hidden states, which are typically initialized to random values or zeros \cite{mohajerin_multistep_2019}. To mitigate this issue, some researchers employ a technique named as the washout technique, where the first few prediction time steps are excluded from training, allowing the network to update the hidden states and washout the influence of the arbitrary initial conditions \cite{bonassi_recurrent_2022, bianchi_recurrent_2017, bonassi_stability_2021} However, the washout approach introduces its own drawbacks, including unreliable early predictions and potential training instability \cite{mohajerin_multistep_2019}.  In our approach, the entire sequence is used during training, and although moderate errors appear in the early steps, they remain within acceptable bounds, as discussed in Appendix~\ref{section:numerical_examples:tube:appendix:early_stages}. Importantly, as shown in Fig.~\ref{fig:tube_ifenn_vs_surrogate_median}, these initial errors did not propagate or degrade performance in subsequent predictions.

\subsection{Poroelasticity example: 2D Excavation problem with hydraulic boundary flux (Dewatering)}
\label{section:numerical_examples:excav}
\subsubsection{Problem setup}
\label{section:numerical_examples:excav:setup}
\begin{figure}[hbt!]
\centering
\includegraphics[width=.65\textwidth]{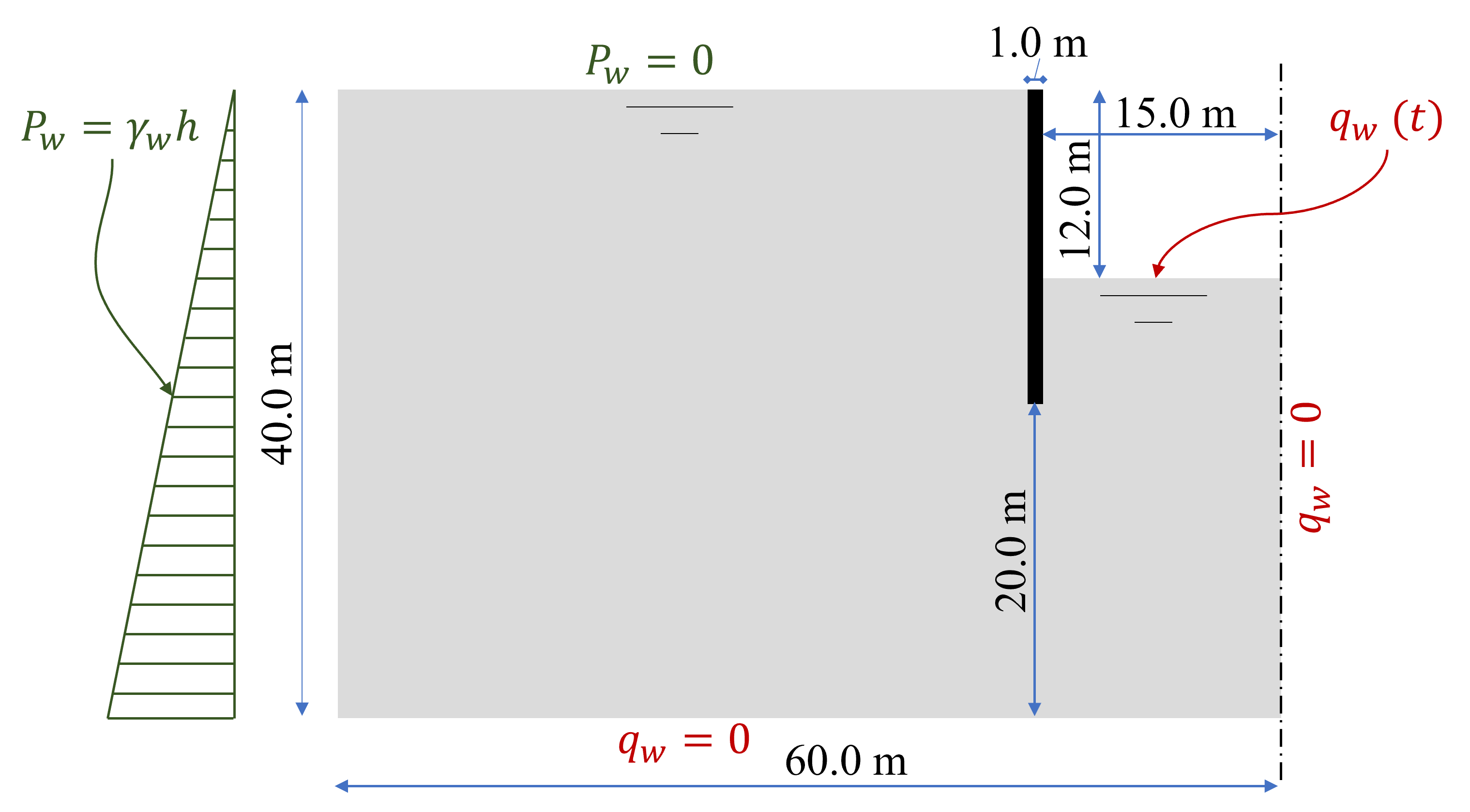}
\caption{Schematic representation of the geometry and boundary conditions
of the 2D poroelasticity excavation problem.}
\label{fig:excav_setup}
\end{figure}

Previous examples have demonstrated I-FENN's performance and accuracy in thermoelasticity problems. In this example, we extend our validation set to poroelasticity problems. We propose a poroelasticity problem setup inspired by geotechnical engineering problems in literature \cite{potts_finite_2001, li_finite_2004}. Figure~\ref{fig:excav_setup} depicts an excavation setup where a dewatering system is applied to reduce the water level within the excavated part. 

The porous medium properties are set as $\lambda=8.375$ MPa, $\mu=5.58$ MPa, $K_s=55556$ MPa, $K_f=2200$ MPa, $\phi = 0.4$, $\mathbf{K}_H = I \times 0.0001$ m/s. A fluctuating water flow $q_W (t)$ is assumed at the excavated part water level. For this purpose, Gaussian random histories are generated to represent different scenarios of dewatering flux. The simulation is done over a time span of $3\times10^{6}$ seconds, discretized over 60 equally spaced increments.

\subsubsection{Training and testing results}
\label{section:numerical_examples:excav:training}
A total of 700 dewatering flux histories are utilized as follows: 600 load cases for training and validation with ratio of 4:1, and 100 load cases for testing. A model was trained with the hyperparameters listed in Table~\ref{tab:excav_deeponet_arch}. The input to the load branch varying in time, while spatially it is a scalar field representing a uniform flux value along the excavated surface. The input to the strain branch is a vector 584 strain trace points uniformly distributed along the 2D mesh. The model is trained to predict output for $N_n = 2309$ Nodes, with trunk input size $N_d$ representing the nodes' coordinates. Notably, no enforcement of boundary conditions was applied for this model. Further details on the training stabilization are provided in Appendix~\ref{section:numerical_examples:excav:appendix:stabilization}.

\begin{figure}[hbt!]
\centering
\includegraphics[width=1.0\textwidth]{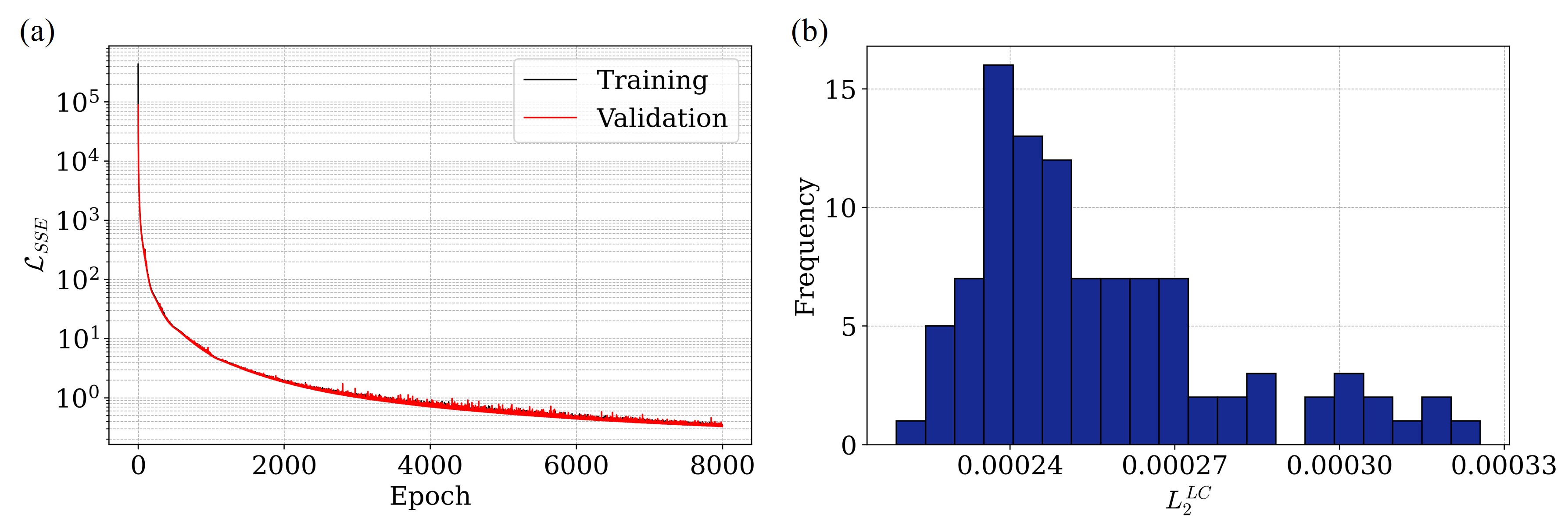}
\caption{a) DeepONet model training loss $\mathcal{L}_{SEE}$ for the 2D excavation example, b) Histogram of the DeepONet model testing loss $L_2^{LC}$ for different load cases of the 2D excavation example.}
\label{fig:excav_training_testing_loss}
\end{figure}

A total of 8000 epochs are employed, requiring 18 minutes, for model training. Training and validation loss curves ($\mathcal{L}_{\text{SSE}}$) are depicted in Fig.~\ref{fig:excav_training_testing_loss}a. Both curves show enhanced learning, with the curves descending through epochs with no significant signs of overfitting. The testing loss distribution, shown in Fig.~\ref{fig:excav_training_testing_loss}b, indicates relatively low error values overall, with an apparent left skew. This suggests that the majority of cases exhibit minimal error, while a smaller number of instances have higher loss values. The load cases with the median, 10th, and 90th error-percentile loss values are selected for I-FENN implementation. The following section will provide the results for the median load case, while the 10th and 90th error-percentile ones are presented in Appendix~\ref{section:numerical_examples:excav:appendix:10-90}. 

\begin{table}[hbt!]
\renewcommand{\arraystretch}{1.2}
\centering
\caption{Hyperparameters of the DeepONet model trained for the 2D excavation example}
\begin{tabular}{ccccccc}
\hline
& Input Size & $N_{GRU}$ & $N_H$ & $N_{ch}$& $N_{FC}$ & $D_{out}$ \\
\hline
Branch \# 1 &$N_l=$1   & 2 & 64 & -  & 1 & 64  \\
Branch \# 2 &$N_b=$584 & 2 & 64 & 32 & 1 & 64  \\
Trunk       &$N_d=$2   & - & 256& -  & 4& 128  \\
\hline
\end{tabular}
\label{tab:excav_deeponet_arch}
\end{table}

\subsubsection{I-FENN results}
\label{section:numerical_examples:excav:results}
\begin{figure}[hbt!]
\centering
\includegraphics[width=1.0\textwidth]{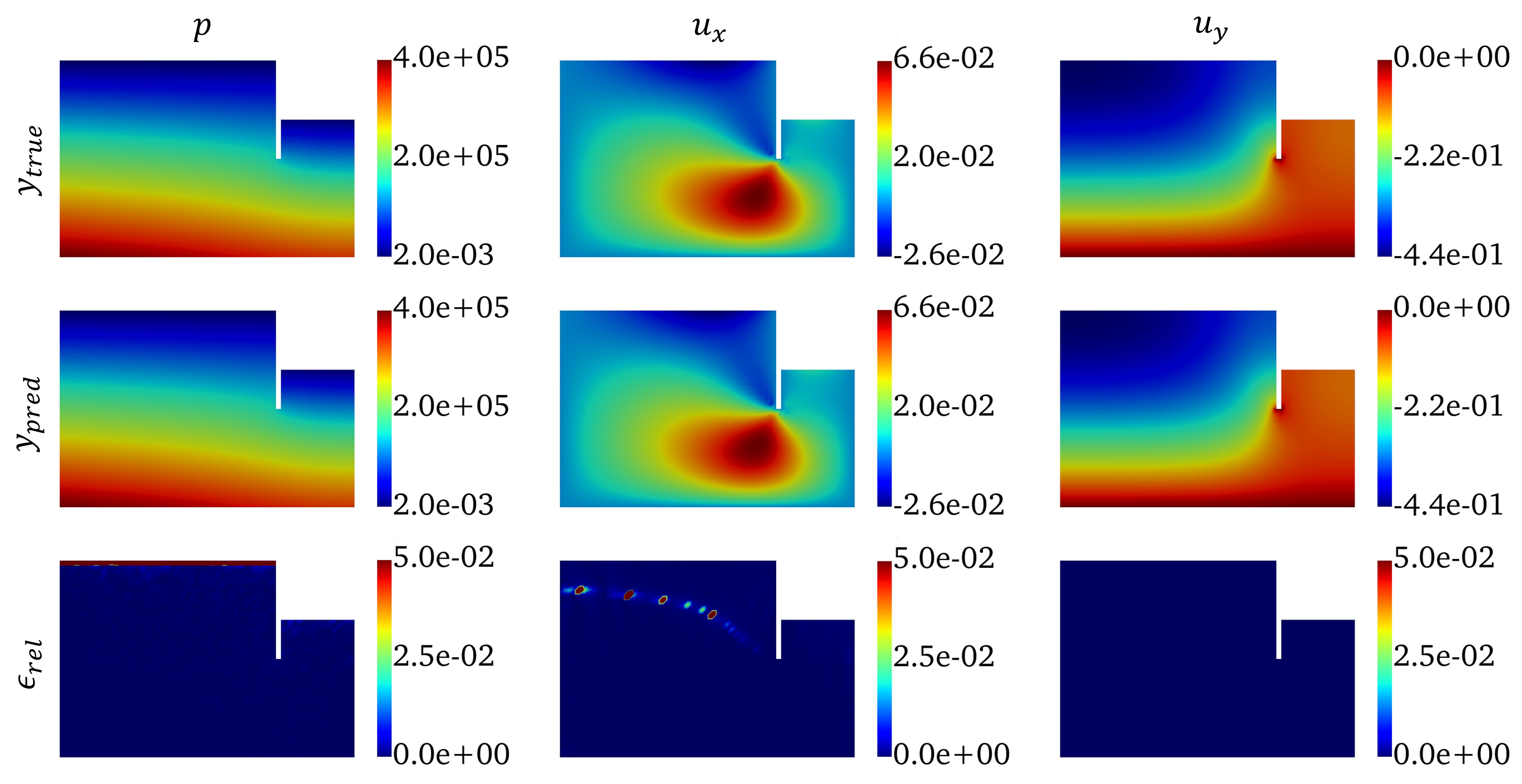}
\caption{Solution obtained using the fully coupled FEM solver ($y_{true}$) and I-FENN ($y_{pred}$) for the median load case at the 60th time step for the 2D excavation example.}
\label{fig:excav_median}
\end{figure}

\begin{figure}[hbt!]
\centering
\includegraphics[width=1.0\textwidth]{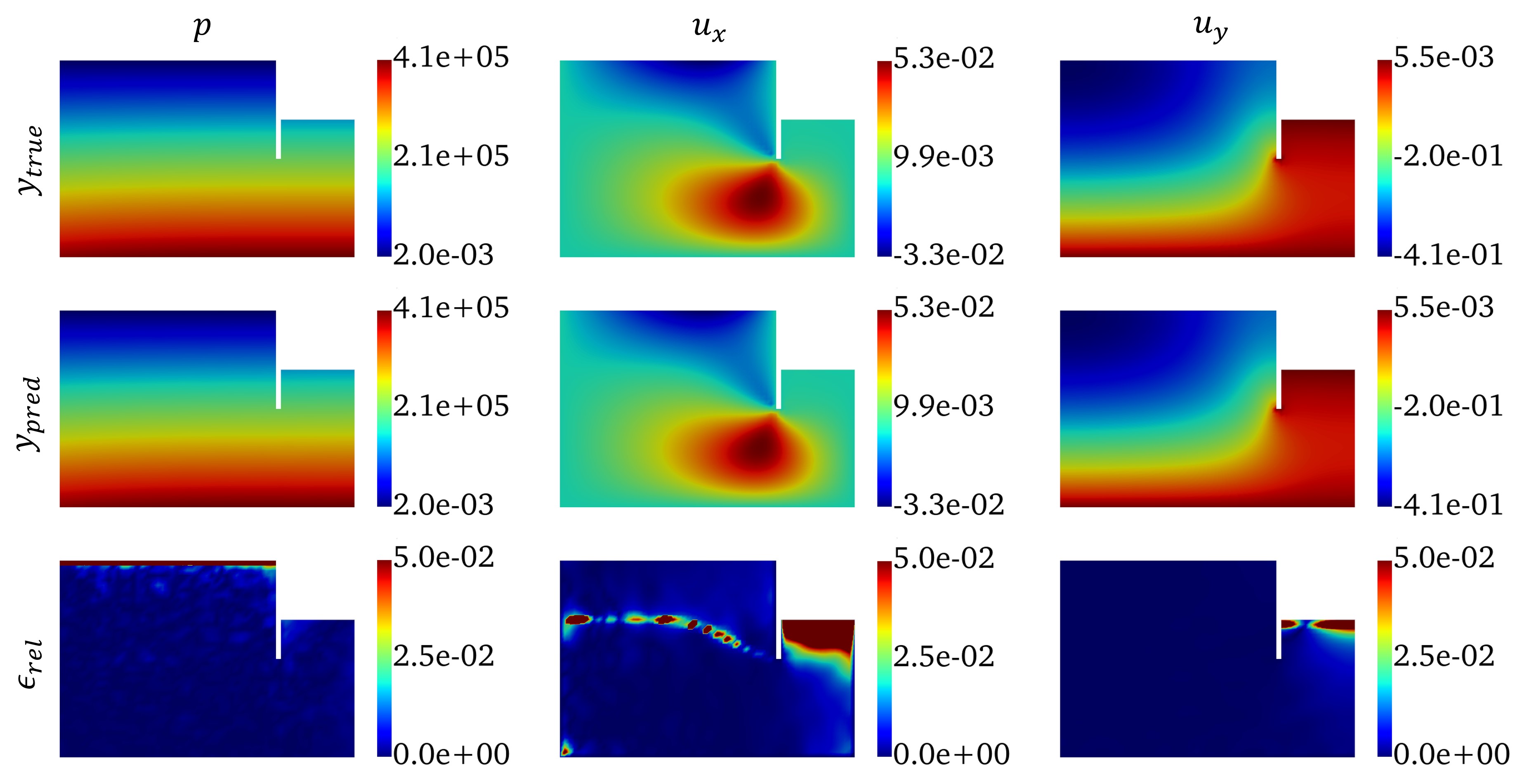}
\caption{Solution obtained using the fully coupled FEM solver ($y_{true}$) and I-FENN ($y_{pred}$) for the load case with maximum detected pressure at the 1st time step for the 2D excavation example.}
\label{fig:excav_physics_max}
\end{figure}

\begin{figure}[hbt!]
\centering
\includegraphics[width=1.0\textwidth]{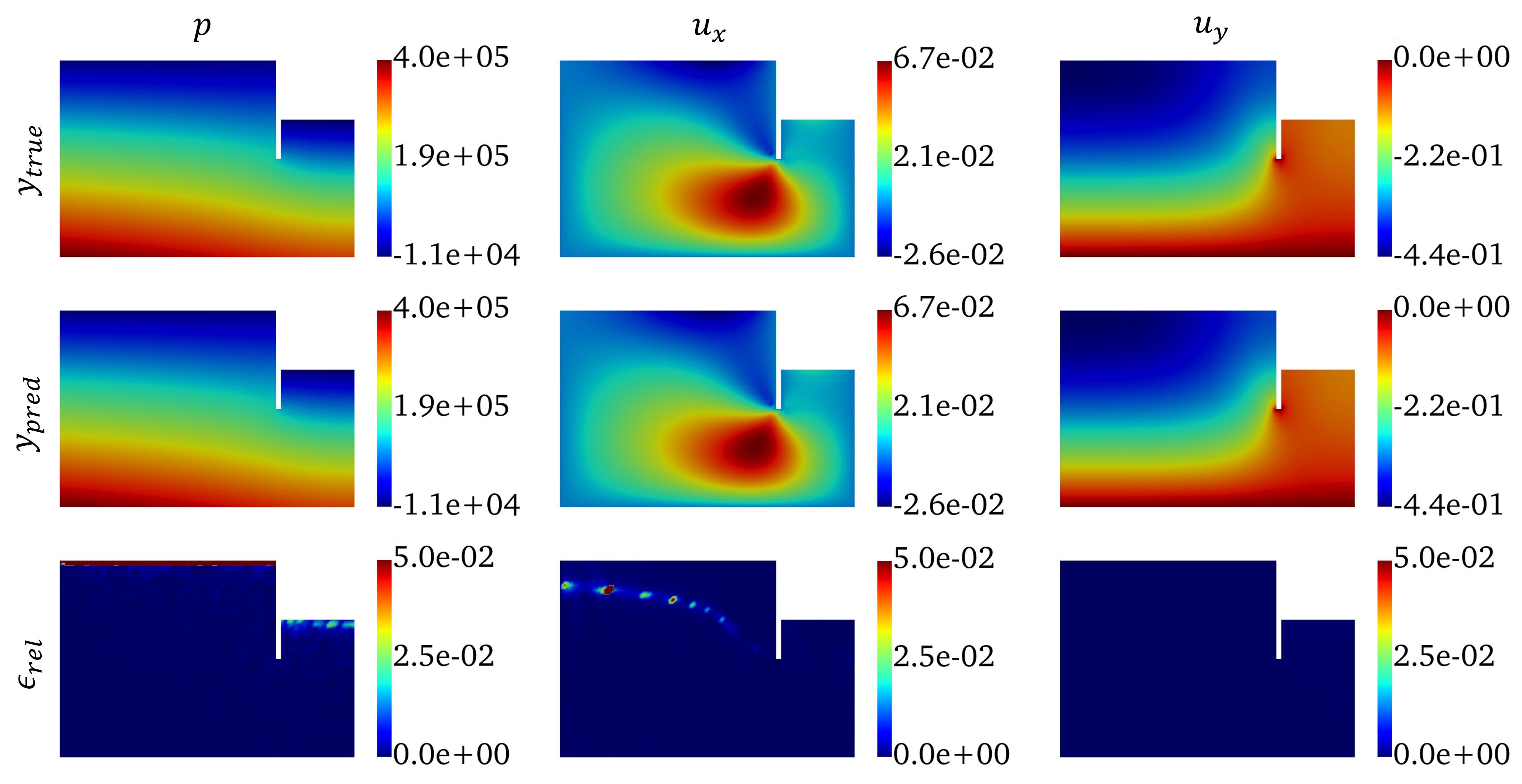}
\caption{Solution obtained using the fully coupled FEM solver ($y_{true}$) and I-FENN ($y_{pred}$) for the load case with minimum detected pressure at the 54th time step for the 2D excavation example.}
\label{fig:excav_physics_min}
\end{figure}

Figure~\ref{fig:excav_median} depicts the I-FENN and the fully coupled FEM solver outputs for the median load case at the last time step. The top row displays the coupled FEM results as the ground truth values ($y_{true}$), the second row shows the I-FENN predictions ($y_{pred}$), and the third row presents the relative error ($\epsilon_{rel}$), which is limited to a maximum of 5\%. In addition to the median load case, a featured load case is chosen for I-FENN implementation. The selected featured load case exhibited the highest and lowest recorded pressure values among all the testing load cases. Results at time steps with maximum and minimum pressure values are depicted in  Fig.~\ref{fig:excav_physics_max} and Fig.~\ref{fig:excav_physics_min}, respectively. Overall, all I-FENN results show a good match with the fully coupled FEM results, with relative errors much lower than 5\%. Only zones with response values close to zero show high relative error exceeding 5\%. 

\begin{figure}[hbt!]
\centering
\includegraphics[width=1.0\textwidth]{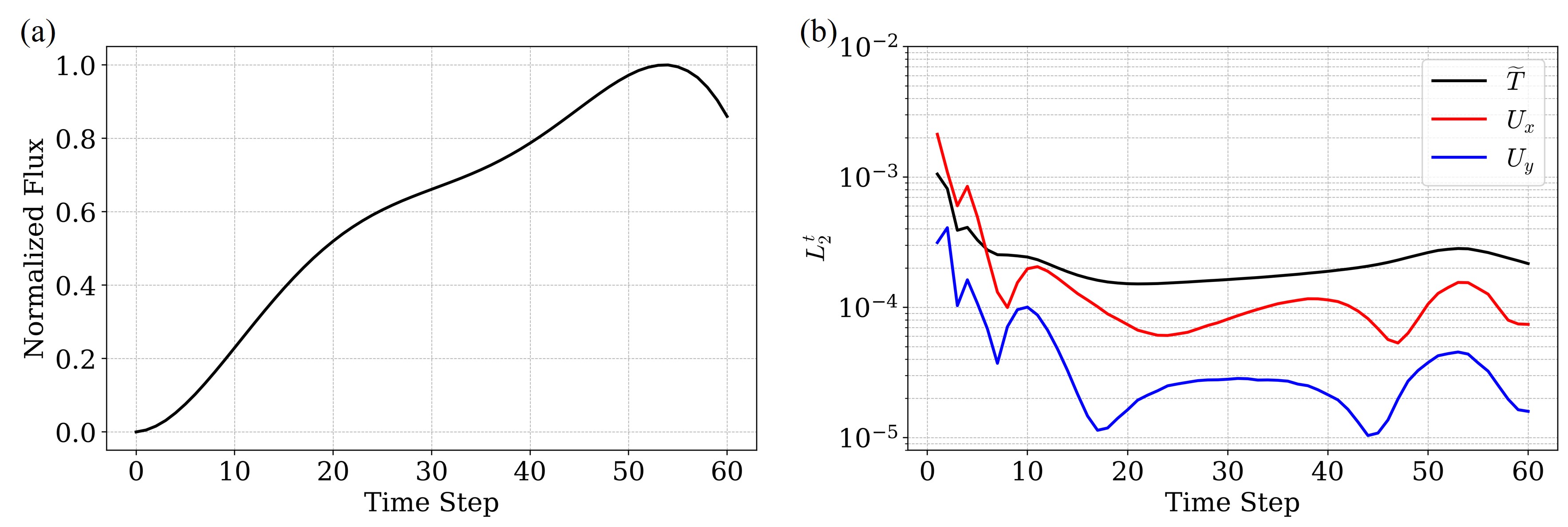}
\caption{a) Normalized flux values, b) normalized error $L_2^t$ per component using I-FENN for the load case with maximum/minimum detected pressure values.}
\label{fig:excav_physics_flux_and_error}
\end{figure}

Figure~\ref{fig:excav_physics_flux_and_error} presents the normalized dewatering flux and the corresponding error for all time steps of the featured load case. The dewatering flux (shown in Fig.~\ref{fig:excav_physics_flux_and_error}a) is almost monotonic with low frequency, enabling the system to exhibit the extreme response values. Error curves in Fig.~\ref{fig:excav_physics_flux_and_error}b show small error values across all time steps, with displacement errors one to two orders of magnitude less than the pressure errors. This pattern is consistent with I-FENN errors depicted in Fig.~\ref{fig:tube_ifenn_vs_surrogate_median}a. The displacement field, computed using the mechanical FEM solver, shows lower errors than the coupled field predicted by DeepONet. This highlights the effectiveness of the I-FENN approach in maintaining high accuracy for the mechanical field (predicted using FEM), even when the coupled field predictions (using DeepONet) are comparatively less accurate. These observations offer basis for future research on understanding how errors scale in I-FENN simulations for multiphysics problems.

\section{Summary and conclusions}
\label{section:summary_conclusion}
In this paper, we present the I-FENN framework for multiphysics problems using DeepONets for thermoelasticity and poroelasticity simulations. A GRU-enabled DeepONet architecture is introduced for integration within the I-FENN framework, handling sequence inputs of time history excitations. The integrated I-FENN framework is introduced in a versatile approach, suitable for deployment on high-performance computing (HPC) systems with support for parallel processing. Three distinct examples covering thermoelasticity and poroelasticity problems are introduced with different domain and surface loading conditions. Based on the analysis and discussions, the key findings can be summarized as follows:
\begin{itemize}
    \item Integration of the GRU-enabled DeepONets within the I-FENN framework provided high flexibility in terms of modeling different geometries and loading conditions in different multiphysics problems.
    \item The I-FENN accuracy for different problems was less than 5\% of relative error for all non-trivial (non-zero) points in the domain.
    \item I-FENN predictions for mechanical field displacements demonstrated superior accuracy, even in the presence of higher errors in the coupled field (temperature or pressure predicted by DeepONet). This underscores the I-FENN's robustness, as mechanical equilibrium is rigorously enforced through FEM, ensuring minimal error in the mechanical field despite inaccuracies in the coupled field. These observations open up new avenues for research to investigate the error and scalability of I-FENN simulations for large-scale multiphysics problems. 
    \item I-FENN simulations required significantly less training time and training data and yielded significantly lower errors than simulation substitution (surrogate model) approaches.
    \item The I-FENN framework shows the ability to accurately predict responses for finer meshes despite being trained on a coarse mesh.
    \item I-FENN demonstrated superior performance compared to the fully coupled finite element solver, achieving up to 40\% savings in computational cost when applied to finer meshes, highlighting its scalability and efficiency.
\end{itemize}

\section*{Acknowledgments}
This work was partially supported by the Sand Hazards and Opportunities for Resilience, Energy, and Sustainability (SHORES) Center, funded by Tamkeen under the NYUAD Research Institute Award CG013. The authors also acknowledge the support of the NYUAD Center for Research Computing, which provided resources, services, and staff expertise that contributed to this work.

\section*{Data Availability}
The data and source code supporting the findings of this study will be made publicly available upon publication of the article.

\appendix
\numberwithin{equation}{section}
\counterwithin{figure}{section}
\counterwithin{table}{section}

\makeatletter
\renewcommand{\theHsection}{\Alph{section}}
\makeatother

\section{Mathematical formulation for multiphysics problems}
\label{section:appendix:math}

\subsection{Thermoelasticity}
\label{section:appendix:math:thermoelasticity}

\subsubsection{Strong Form}
\label{section:appendix:math:thermoelasticity:strong}
\begin{figure}[hbt!]
\centering
\includegraphics[width=.4\textwidth]{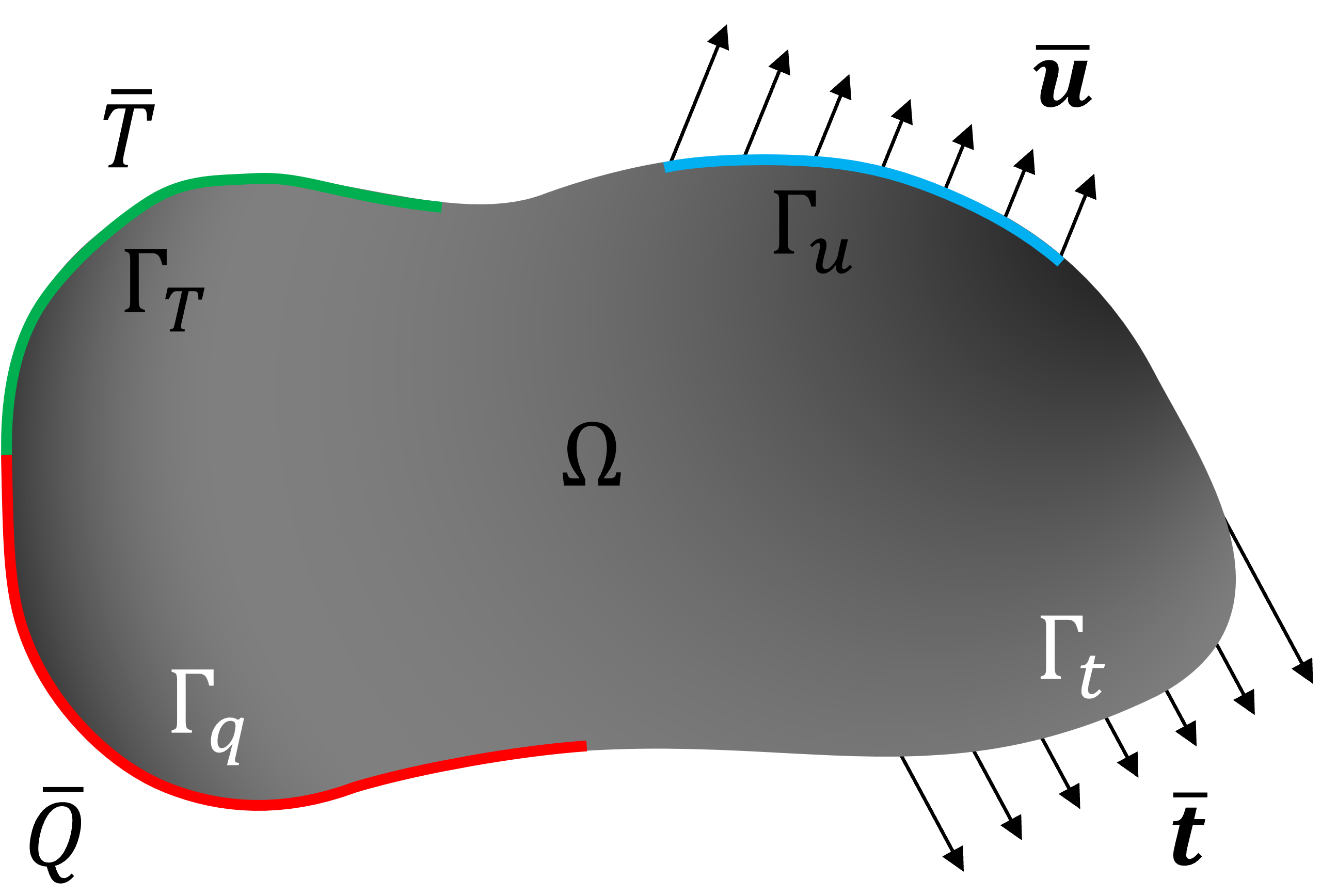}
\caption{Schematic representation of a body with prescribed boundary conditions for a thermoelasticity problem.}
\label{fig:schematic_thermo}
\end{figure}

 The governing system of differential equations can be defined as follows \cite{gurtin_linear_1973 , abeyaratne_continuum_1998 , truesdell_linear_1973}:

\begin{align}
    \boldsymbol{\nabla}\cdot \boldsymbol{\sigma} + \boldsymbol{b} &= \boldsymbol{0}, \quad  \boldsymbol{x}\in\Omega,
    \label{eq:th_s_blm}
    \\
    \rho T_{o} \dot{\eta} + \boldsymbol{\nabla}\cdot \boldsymbol{q} - r &= 0, \quad  \boldsymbol{x}\in\Omega, 
    \label{eq:th_s_energy}
\end{align}

\noindent
where $\boldsymbol{\sigma}$ denotes the Cauchy stress tensor, $\boldsymbol{b}$ is the mechanical body force vector, $\rho$ is the material mass density, $T_{o}$ is the reference temperature, $\eta$ is the entropy per unit mass, $\boldsymbol{q}$ is the heat flux vector, and $r$ is the heat body source or sink. $\Omega$ denotes the physical domain (depicted in  Fig.~\ref{fig:schematic_thermo}), while $\boldsymbol{\nabla}$ is the gradient operator and $\boldsymbol{\nabla}\cdot$ is the divergence operator.

The boundary conditions are defined as follows:
\begin{equation}\label{eq:thermo-strong-bc}
    \begin{aligned}
        \boldsymbol{u} &= \boldsymbol{\overline{u}}, \quad x \in \Gamma_u, 
        \\
        \boldsymbol{t} &= \boldsymbol{\overline{t}}, \quad x \in \Gamma_t,
        \\
        T &= \overline{T}, \quad x \in \Gamma_T, 
        \\
        Q &= \overline{Q}, \quad x \in \Gamma_q 
    \end{aligned}
\end{equation}

\noindent
where the domain boundary ($\Gamma$) is defined as $\Gamma=\Gamma_t \cup \Gamma_u=\Gamma_T \cup \Gamma_q$. Symbols $\boldsymbol{u}$ and $T$ represent the displacement and temperature fields, with their boundary conditions imposed on boundary parts $\Gamma_u$ and $\Gamma_T$, respectively. Traction ($\boldsymbol{t}$) is defined as $\boldsymbol{t} = \boldsymbol{\sigma}\cdot\boldsymbol{n}$ on the traction boundary part $\Gamma_t$, where  $\boldsymbol{n}$ represents the normal unit vector at the boundary $\Gamma$. Finally, the flux boundary condition is defined as $Q=\boldsymbol{q}\cdot\boldsymbol{n}$ on the flux boundary part $\Gamma_q$.

The heat flux vector $\boldsymbol{q}$ can be defined using Fourier's law as \cite{gurtin_linear_1973 , abeyaratne_continuum_1998 , truesdell_linear_1973}:

\begin{equation}
\begin{aligned}
    \boldsymbol{q} = -k \boldsymbol{\nabla} T,
    \label{th_flux}
\end{aligned}
\end{equation}

\noindent
where $k$ is the thermal conductivity of the material. Thermoelasticity constitutive laws for stress ($\boldsymbol{\sigma}$) and entropy ($\eta$) are expressed as \cite{gurtin_linear_1973 , abeyaratne_continuum_1998 , truesdell_linear_1973}:
\begin{align}
    \boldsymbol{\sigma} &= \boldsymbol{\hat{C}} : (\boldsymbol{\varepsilon} - \alpha\boldsymbol{I} (T-T_0))
    \label{eq:th_sigma}
    \\
    \rho T_{o} \dot{\eta} &= \rho C_{\varepsilon} \dot{T} + \alpha\boldsymbol{I}:\boldsymbol{\hat{C}} : \boldsymbol{\dot{\varepsilon}}    
    \label{eq:th_entropy}
\end{align}

\noindent
where $C_{\varepsilon}$ is the specific heat per unit mass at constant strain, and $\boldsymbol{\hat{C}}$ is the elasticity tensor defined as $ \hat{C}_{ijkl} = \lambda \delta_{ij} \delta_{kl} + \mu \left(\delta_{ik} \delta_{jl} + \delta_{il} \delta_{jk}\right)$ for isotropic linear materials. While $\boldsymbol{\varepsilon}$ is the strain defined as $\boldsymbol{\varepsilon} = \boldsymbol{\nabla}^s \boldsymbol{u} =\frac{1}{2}(\boldsymbol{\nabla} \boldsymbol{u}+\boldsymbol{\nabla} \boldsymbol{u}^T)$ for small deformations. Parameters $\mu$ and $\lambda$ are Lamé constants, and $\alpha$ is the coefficient of thermal expansion.

Substituting Eq.~\eqref{eq:th_entropy} in Eq.~\eqref{eq:th_s_energy} yields:

\begin{equation}\label{eq:th_s_energy_2}
\begin{aligned}
    \rho C_{\varepsilon} \dot{T} +\alpha\boldsymbol{I}:\boldsymbol{\hat{C}} : \boldsymbol{\dot{\varepsilon}} +  \boldsymbol{\nabla}\cdot \boldsymbol{q} - r &= 0
\end{aligned}
\end{equation}

\subsubsection{Weak Form}
\label{section:appendix:math:thermoelasticity:weak}

Using the balance of linear momentum equation in \eqref{eq:th_s_blm} and the stress constitutive relation \eqref{eq:th_sigma}, the mechanical weak form can be written as:

\begin{align} \label{eq:th_w_blm}
    \int_{\Omega} \nabla^{s} \boldsymbol{\widehat{w}} : (\boldsymbol{\hat{C}} : \boldsymbol{\varepsilon}) \, d\Omega  \;- \int_{\Omega} \nabla^{s} \boldsymbol{\widehat{w}} : (\boldsymbol{\hat{C}} : \alpha (T-T_0)\boldsymbol{\mathit{I}}) \, d\Omega \nonumber 
    \\
    = \int_{\Gamma_t} \boldsymbol{\widehat{w}} \cdot \boldsymbol{\overline{t}} \, d\Gamma 
    \;+ \int_{\Omega} \boldsymbol{\widehat{w}} \cdot \boldsymbol{b} \, d\Omega 
    \quad \forall \, \boldsymbol{\widehat{w}} \in \mathcal{W}_u
\end{align}

\noindent
where $\widehat{\boldsymbol{w}}$ is the displacement test function, and $\mathcal{W}_u$ denotes the displacement function space. Substituting the  Fourier definition of flux \eqref{th_flux}  into the energy equation \eqref{eq:th_s_energy_2}, the thermal weak form can be written as:  

\begin{align}\label{eq:th_w_heat}
    \int_{\Omega} \widehat{T} \rho C_{\varepsilon} \dot{T} \, d\Omega 
    \;+ \int_{\Omega} \nabla \widehat{T} \cdot \left( k \nabla T \right) \, d\Omega 
    \;+ \int_{\Omega} \widehat{T} \left( \alpha \boldsymbol{\mathit{I}}: \, \boldsymbol{\hat{C}} : \dot{\boldsymbol{\varepsilon}} \right) T_0  \, d\Omega \nonumber 
    \\
    = - \int_{\Gamma_q} \widehat{T} \overline{Q} \, d\Gamma 
    \;+ \int_{\Omega} \widehat{T} r \, d\Omega 
    \quad \forall \, \widehat{T} \in \mathcal{W}_T
\end{align}
\noindent
where  $\widehat{T}$ is the temperature test function, and  $\mathcal{W}_T$ is the temperature function space. Using the definition of the elasticity tensor for isotropic linear materials $ (\hat{C}_{ijkl} = \lambda \delta_{ij} \delta_{kl} + \mu (\delta_{ik} \delta_{jl} + \delta_{il} \delta_{jk}))$, the weak form of  the energy equation (Eq.~\eqref{eq:th_w_heat}) can be written as

\begin{align}\label{eq:th_w_heat_2}
    \int_{\Omega} \widehat{T} \rho C_{\varepsilon} \dot{T} \, d\Omega 
    \;+ \int_{\Omega} \nabla \widehat{T} \cdot \left( k \nabla T \right) \, d\Omega 
    \;+ \int_{\Omega} \widehat{T} \alpha (n_{dim}\lambda + 2\mu) \text{tr} \left(\dot{\boldsymbol{\varepsilon}}  \right) T_0  \, d\Omega \nonumber 
    \\
    = - \int_{\Gamma_q} \widehat{T} \overline{Q} \, d\Gamma 
    \;+ \int_{\Omega} \widehat{T} r \, d\Omega 
    \quad \forall \, \widehat{T} \in \mathcal{W}_T
\end{align}
\noindent
where $n_{dim}$ is the number of spatial dimensions.

\subsection{Poroelasticity}
\label{section:appendix:math:poroelasticity}

\subsubsection{Strong Form}
\label{section:appendix:math:poroelasticity:strong}

\begin{figure}[hbt!]
\centering
\includegraphics[width=.4\textwidth]{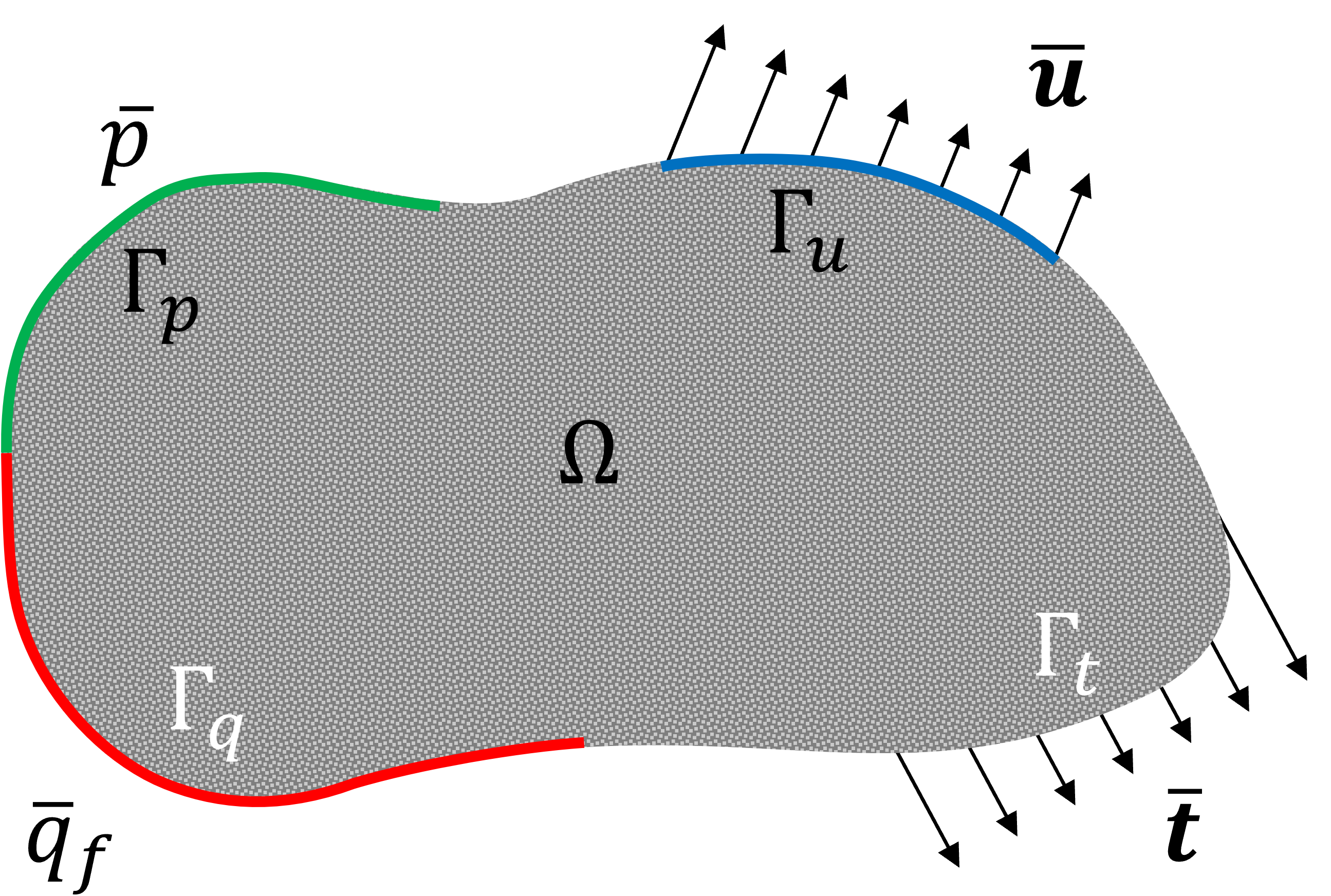}
\caption{Schematic representation of a body with prescribed boundary conditions for a poroelasticity problem.}
\label{fig:schematic_poro}
\end{figure}

For a representative elementary volume composed of solid grains and pores filled with a fluid, the governing equations can be written as \cite{potts_finite_1999, coussy_poromechanics_2004,cheng_poroelasticity_2016, yi_consistent_2020, chen_thermodynamically_2022}:  
\begin{align}
	\label{eq:poro_st_blm}
	\nabla \cdot \boldsymbol{\sigma}_t +\boldsymbol{b} &=\boldsymbol{0}
    \\
    \label{eq:poro_st_cont}
	\dot{\zeta}+\nabla \cdot \boldsymbol{v}_{f} &= Q_f
\end{align}

\noindent
where $\boldsymbol{\sigma}_t$ is the total Cauchy stress due to deformation and fluid pressure ($p$), $\zeta$ is the increment of fluid content,  $\boldsymbol{v}_{f}$ is the fluid velocity vector, and $Q_f$ represents any fluid source or sink. 
Domain boundary is defined as  $\Gamma=\Gamma_u \cup \Gamma_t=\Gamma_p \cup \Gamma_q$ (see  Fig.~\ref{fig:schematic_poro}), with boundary conditions described as follows:
\begin{equation}\label{eq:thermo-strong-bc}
    \begin{aligned}
        \boldsymbol{u} &= \boldsymbol{\overline{u}}, \quad x \in \Gamma_u, 
        \\
        \boldsymbol{t} &= \boldsymbol{\overline{t}}, \quad x \in \Gamma_t,
        \\
        p &= \overline{p}, \quad x \in \Gamma_p, 
        \\
        q_f &= \overline{q}_f, \quad x \in \Gamma_q 
    \end{aligned}
\end{equation}

\noindent
where $\boldsymbol{\overline{u}}$ and $\boldsymbol{\overline{t}}$ are the displacement and traction boundary conditions defined over boundaries $ \Gamma_u $ and  $ \Gamma_t $, respectively.   $\overline{p}$ is the fluid pressure over boundary $ \Gamma_p $, while  $ \overline{q}_f $ is the fluid flux over boundary $\Gamma_q$, defined as $ \overline{q}_f = \boldsymbol{v}_{f} \cdot \boldsymbol{n} $.

Assuming fluid flow is controlled by Darcy's law, fluid velocity  and increment can be defined as \cite{potts_finite_1999, cheng_poroelasticity_2016, yi_consistent_2020, chen_thermodynamically_2022}:
\begin{align}
    \label{eq:fluid_velocity}
    \boldsymbol{v}_{f} &= -\boldsymbol{K}_H \cdot \nabla H_f
    \\
    \label{eq:fluid_increment}
	\zeta&= \dfrac{p}{M} + \alpha^\prime \text{tr}(\boldsymbol{\varepsilon})
\end{align}
\noindent
where $\boldsymbol{K}_H$ is the Hydraulic conductivity tensor defined as $\boldsymbol{K}_H = \gamma_f \boldsymbol{K}_I / \mu_f  $. Given that, $\boldsymbol{K}_I$ is the permeability tensor, $\gamma_f$ is the weight density, and $\mu_f$ is the fluid viscosity.  $H_f$ is fluid total head defined as $ H_f = {p} / {\gamma_f} + \boldsymbol{x} \cdot \boldsymbol{i}_g$, where  $\boldsymbol{x}$ is the coordinate vector, and  $\boldsymbol{i}_g$ is the unit vector parallel to gravity, but in the opposite direction. 

Finally, $\alpha^\prime$ and  $M$ are the Biot's coefficient and modulus, respectively, defined as \cite{potts_finite_1999, cheng_poroelasticity_2016, yi_consistent_2020, chen_thermodynamically_2022}: 
\begin{align}
	\label{biot_eqn}
	\alpha^\prime = 1 - \dfrac{K_b}{K_s}
    \\
    \dfrac{1}{M} = \dfrac{\phi}{K_f} + \dfrac{\alpha^\prime-\phi}{K_s}
\end{align}
\noindent
where  $\phi$ is porosity of the medium, $K_s$ is the solid grain bulk modulus, $K_f$ is the fluid bulk modulus, and $K_b$ is the bulk moduli of the solid skeleton defined as \cite{cheng_poroelasticity_2016}:
\begin{align}
	K_b = \lambda + \dfrac{2\mu}{3}
	\label{bulk_moduli}
\end{align}
Finally, the total stress  $\boldsymbol{\sigma}_t$ is computed as a function of the solid effective stress $\boldsymbol{\sigma}^\prime$ and fluid pressure $\boldsymbol{\sigma}_f$, as follows:
 \begin{align}
    \label{poro_sigma}
    \boldsymbol{\sigma}_t &= \boldsymbol{\sigma}^\prime + \boldsymbol{\sigma}_f \notag
    \\
    &=\boldsymbol{\hat{C}} : \boldsymbol{\varepsilon} - \alpha^\prime p \boldsymbol{\mathit{I}}
\end{align}

Substituting Eq.~\eqref{poro_sigma} into Eq.~\eqref{eq:poro_st_blm}, the linear momentum equilibrium equation for the two-phase mixture can be written as:
\begin{align}
	\label{eq:poro_st_blm_2}
	\nabla \cdot (\boldsymbol{\hat{C}} : \boldsymbol{\varepsilon} - \alpha^\prime p \boldsymbol{\mathit{I}})+\boldsymbol{b} &= \boldsymbol{0}
\end{align}

Incorporating definitions in Eq.~\eqref{eq:fluid_velocity} and Eq.~\eqref{eq:fluid_increment} into Eq.~\eqref{eq:poro_st_cont}, the fluid flow continuity equation can be written as:
\begin{align}
	\dfrac{\dot{p}}{M} + \alpha^\prime \text{tr}(\dot{\boldsymbol{\varepsilon}})  - \nabla \cdot \left( \dfrac{ \boldsymbol{K}_I}{\mu_f} \cdot (\nabla p + \gamma_f \boldsymbol{i}_g) \right) &= Q_f
	\label{eq:poro_st_cont_2}
\end{align}

\subsubsection{Weak Form and FEM}
\label{section:appendix:math:poroelasticity:weak}

Given the final definitions, the linear momentum equilibrium equation Eq.~\eqref{eq:poro_st_blm_2} and the fluid flow continuity equation Eq.~\eqref{eq:poro_st_cont_2}, their corresponding weak forms can be respectively listed as:
\begin{align} 
    \label{poro_w_blm}
    &\int_{\Omega} \nabla^{s} \boldsymbol{\widehat{w}} : \boldsymbol{\hat{C}} : \boldsymbol{\varepsilon} \, d\Omega  \;- \int_{\Omega} \nabla^{s} \boldsymbol{\widehat{w}} : (\alpha^\prime p \boldsymbol{\mathit{I}}) \, d\Omega \nonumber 
    \\
    &= \int_{\Gamma_t} \boldsymbol{\widehat{w}} \cdot \boldsymbol{\overline{t}} \, d\Gamma 
    \;+ \int_{\Omega} \boldsymbol{\widehat{w}} \cdot \boldsymbol{b} \, d\Omega 
    \quad \forall \, \boldsymbol{\widehat{w}} \in \mathcal{W}_u
\end{align}

\begin{align}
    \label{poro_w_cont}
    &\int_{\Omega} \widehat{p} \dfrac{\dot{p}}{M} \, d\Omega 
    \;+ \int_{\Omega} \widehat{p} \alpha^\prime \text{tr}(\dot{\boldsymbol{\varepsilon}})  \, d\Omega
    \;+ \int_{\Omega} \nabla \widehat{p} \cdot \left(\dfrac{ \boldsymbol{K}_I}{\mu_f} \cdot \nabla p \right) \, d\Omega 
     \nonumber 
    \\
    &= -\int_{\Omega} \nabla \widehat{p} \cdot \left(\dfrac{ \boldsymbol{K}_I}{\mu_f} \cdot \gamma_f \boldsymbol{i}_g \right) \, d\Omega - \int_{\Gamma_q} \widehat{p} \;\overline{q}_f \, d\Gamma 
    \;+ \int_{\Omega} \widehat{p} \; Q_f \, d\Omega 
    \quad \forall \, \widehat{p} \in \mathcal{W}_p
\end{align}

\noindent
where $\widehat{\boldsymbol{w}}$ is the displacement test function,  $\mathcal{W}_u$ is the displacement function space,  $\widehat{p}$ is the fluid pressure test function, and  $\mathcal{W}_p$ is the fluid pressure function space. 

\section{Additional results and parametric studies}
\label{section:appendix:numerical_examples}

This section presents additional results and parametric studies conducted for the three examples in Section~\ref{section:numerical_examples}, offering further insights that may be of interest to the reader.

\subsection{Thermoelasticity example: 3D Cube with thermal body load}
\label{section:numerical_examples:cube:appendix}

\subsubsection{I-FENN results for the 10th and 90th error-percentile load cases}
\label{section:numerical_examples:cube:appendix:10-90}

\begin{figure}[hbt!]
\centering
\includegraphics[width=1.0\textwidth]{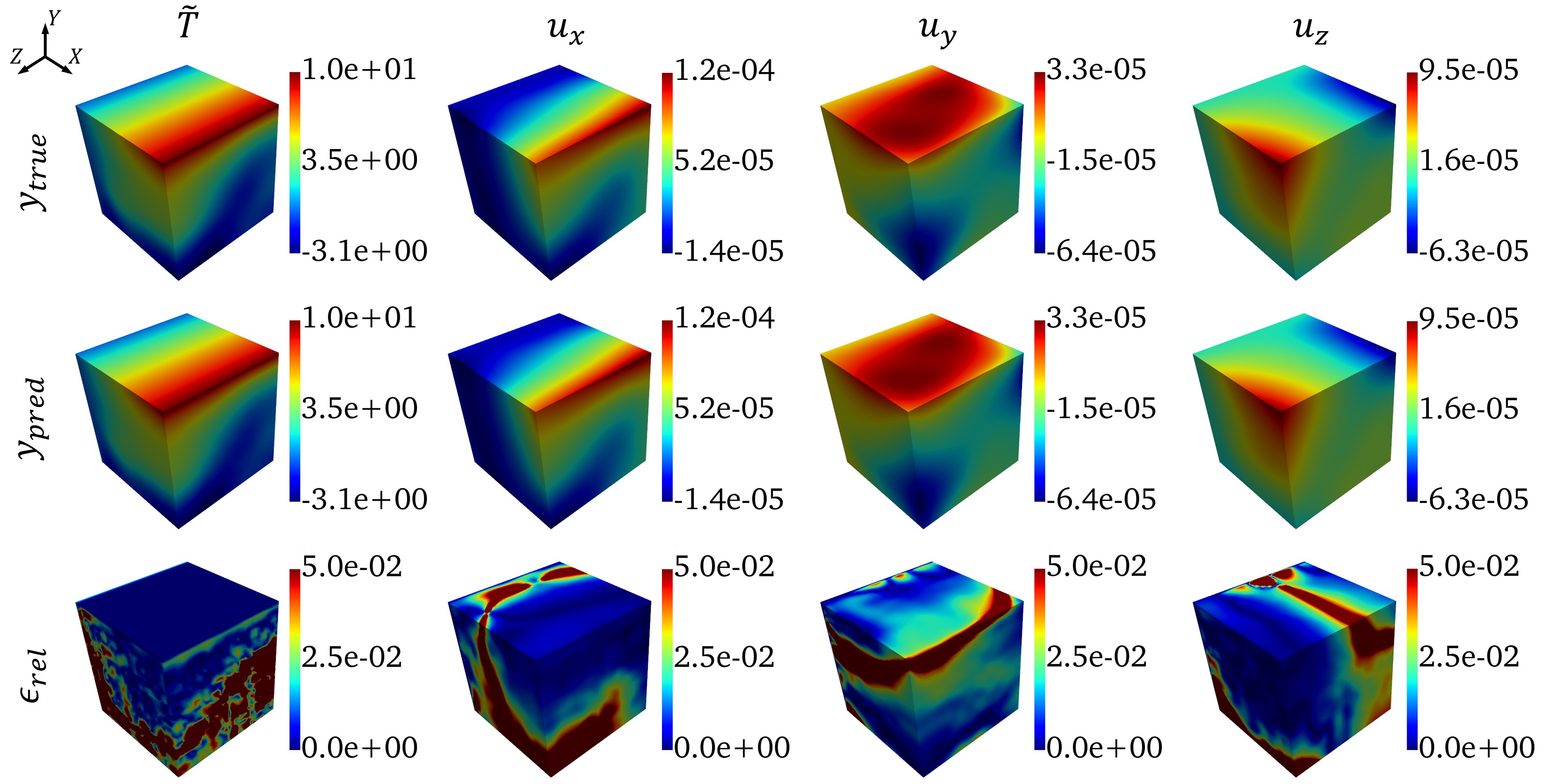}
\caption{Solution obtained using the fully coupled FEM solver ($y_{true}$) and I-FENN ($y_{pred}$) for the 10th error-percentile load case at the 100th time step for the 3D cube example.}
\label{fig:cube_10th}
\end{figure}

\begin{figure}[hbt!]
\centering
\includegraphics[width=1.0\textwidth]{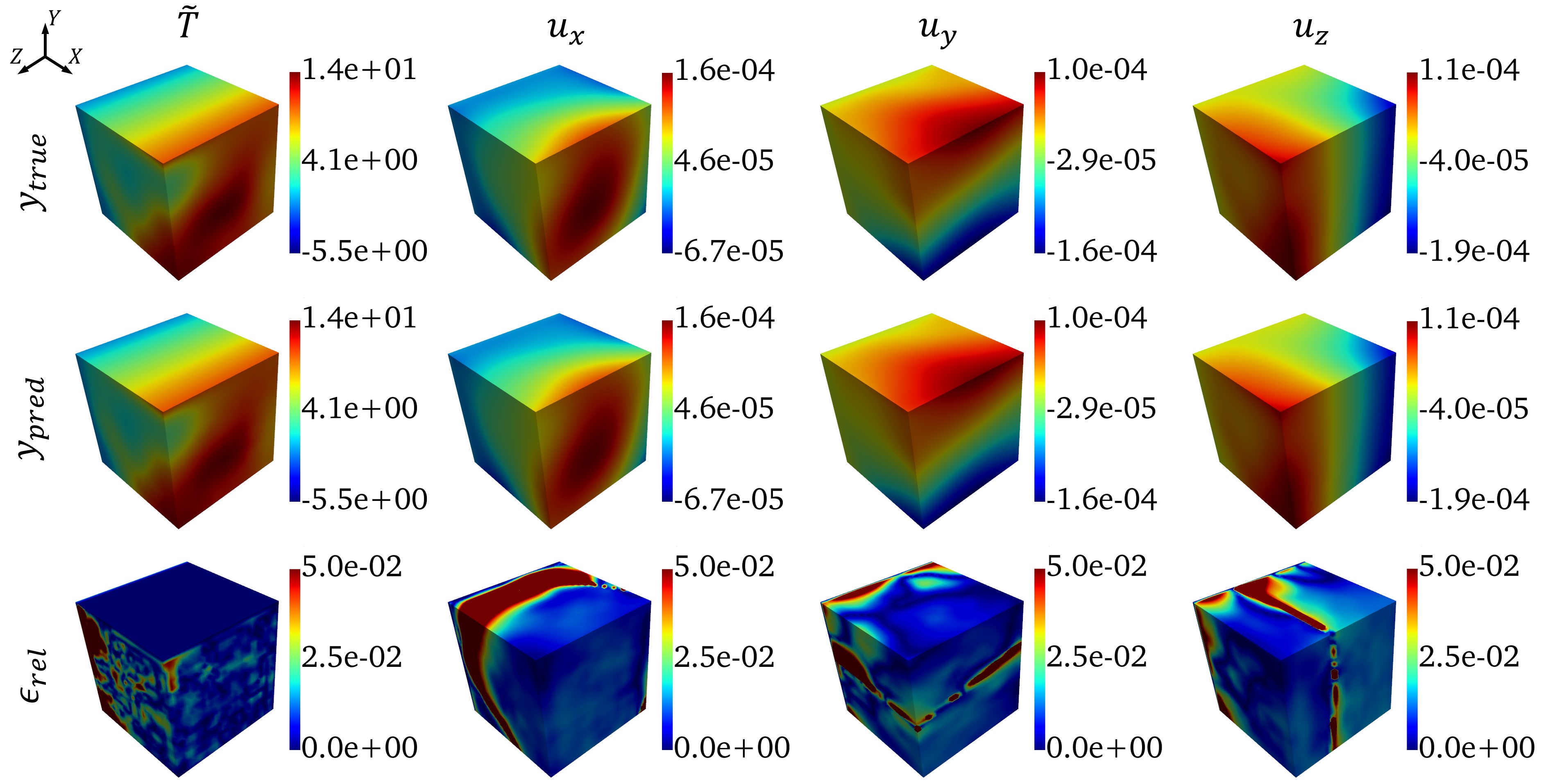}
\caption{Solution obtained using the fully coupled FEM solver ($y_{true}$) and I-FENN ($y_{pred}$) for the 90th error-percentile load case at the 100th time step for the 3D cube example.}
\label{fig:cube_90th}
\end{figure}

I-FENN results for the 10th and 90th error-percentile load cases are depicted in Fig.~\ref{fig:cube_10th} and Fig.~\ref{fig:cube_90th}, respectively. The results show consistent accuracy compared to the median load case results presented in Fig.~\ref{fig:cube_median}. 

\subsubsection{I-FENN results for the median load case considering a finer mesh}
\label{section:numerical_examples:cube:appendix:fine}
\begin{figure}[hbt!]
\centering
\includegraphics[width=1.0\textwidth]{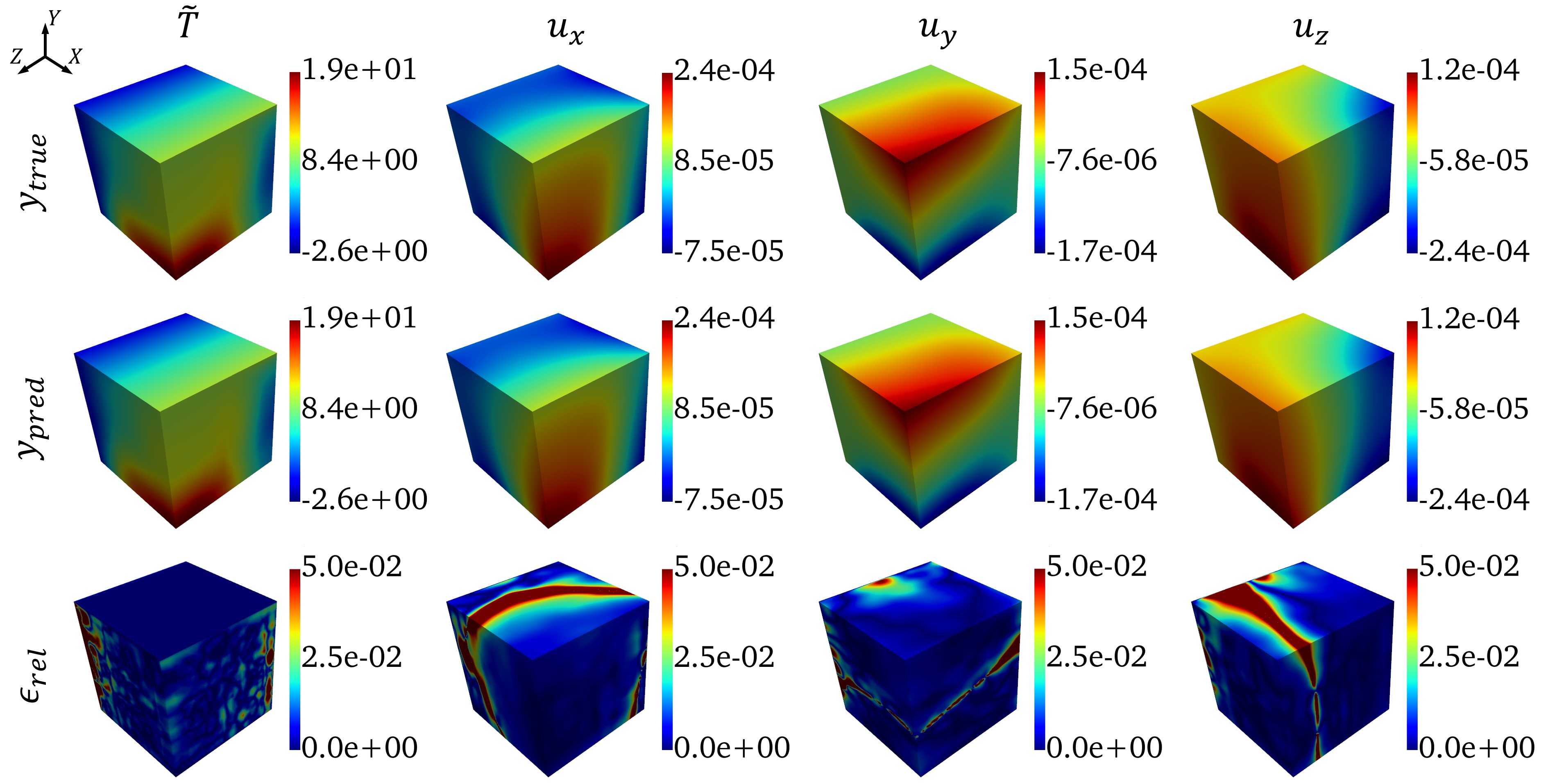}
\caption{Solution obtained using the fully coupled FEM solver ($y_{true}$) and I-FENN ($y_{pred}$) for the median load case at the 100th time step for a fine mesh ($75 \times 75 \times 75$ cells) of the 3D cube example.}
\label{fig:cube_median_fine75}
\end{figure}

The I-FENN results for the median load case applied on a fine mesh of $75 \times 75 \times 75$ cells are presented in Fig.~\ref{fig:cube_median_fine75}. The error limits and distributions closely match that of the coarser mesh of $30 \times 30 \times 30$ cells (see Fig.~\ref{fig:cube_median}). This observation further confirms the framework's ability to preserve accuracy under mesh refinement.

\subsubsection{I-FENN results for the median load case without enforcing the boundary conditions}
\label{section:numerical_examples:cube:appendix:bcs}

\begin{figure}[hbt!]
\centering
\includegraphics[width=1.0\textwidth]{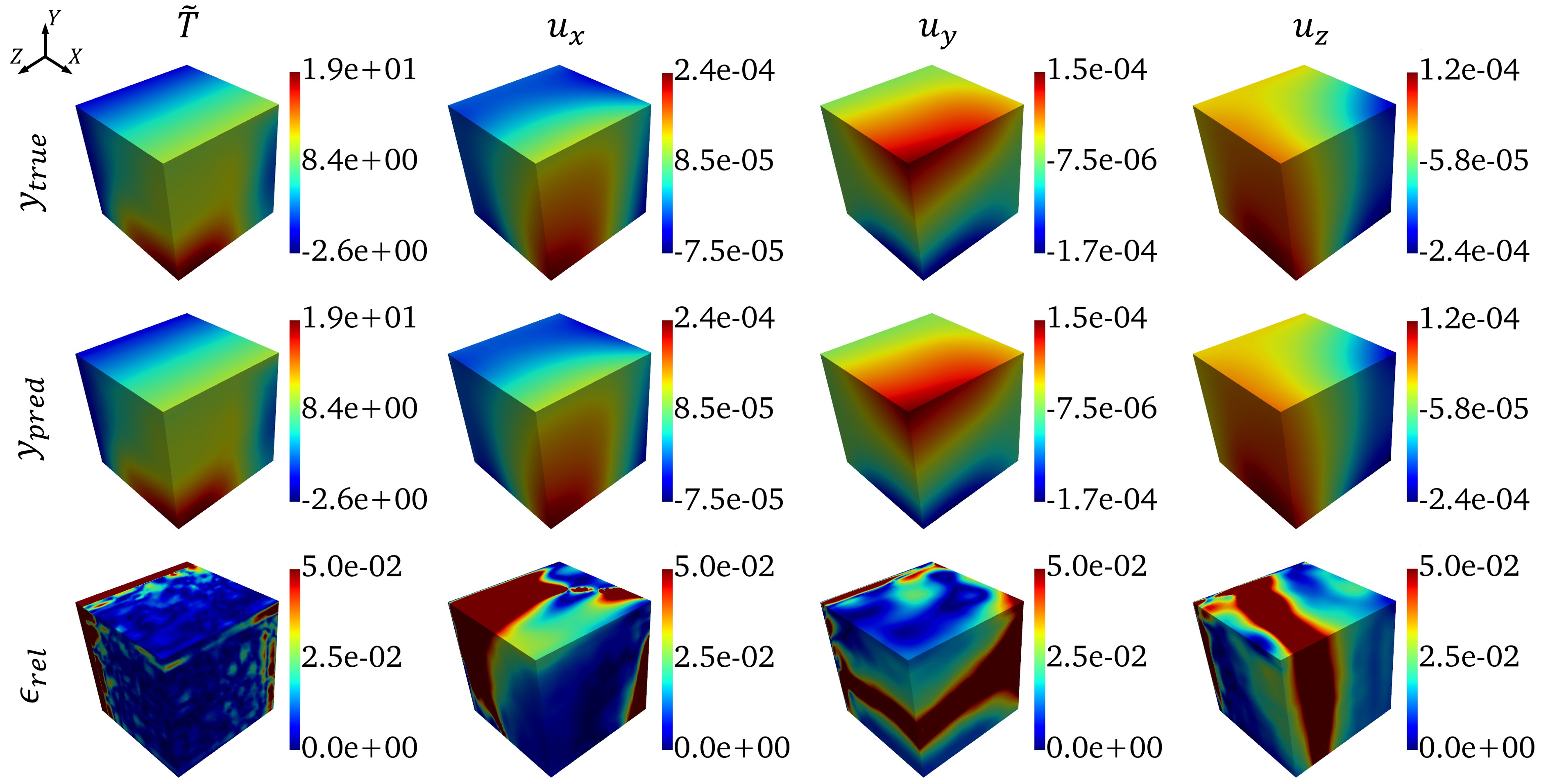}
\caption{Solution obtained using the fully coupled FEM solver ($y_{true}$) and  I-FENN ($y_{pred}$) with DeepONet trained without enforcing BCs, for the median load case at the 100th time step for the 3D cube example.}
\label{fig:cube_median_no_bc}
\end{figure}

To verify the validity of the boundary conditions enforcement approach, presented in Section~\ref{section:deeponet_for_ifenn:bcs}, a DeepONet model is trained without enforcing boundary conditions. Afterward, the trained model is integrated into the I-FENN framework and tested for the median load case, identified in Section~\ref{section:numerical_examples:cube:training}. Figure~\ref{fig:cube_median_no_bc} presents the I-FENN results for the tested configuration in which temperature boundary conditions were not enforced through the DeepONet. The results show a good agreement with the true values; however, the zones with relative error exceeding 5\% are much larger than those with enforced boundary conditions (see Fig.~\ref{fig:cube_median}). This observation underscores the added value of enforcing boundary conditions and suggests that, despite the potential complexity, they should be applied whenever feasible to improve accuracy.

\subsection{Thermoelasticity example: 3D Thick-walled tube with thermal surface load}
\label{section:numerical_examples:tube:appendix}

\subsubsection{Error in early stages}
\label{section:numerical_examples:tube:appendix:early_stages}
\begin{figure}[hbt!]
\centering
\includegraphics[width=1.0\textwidth]{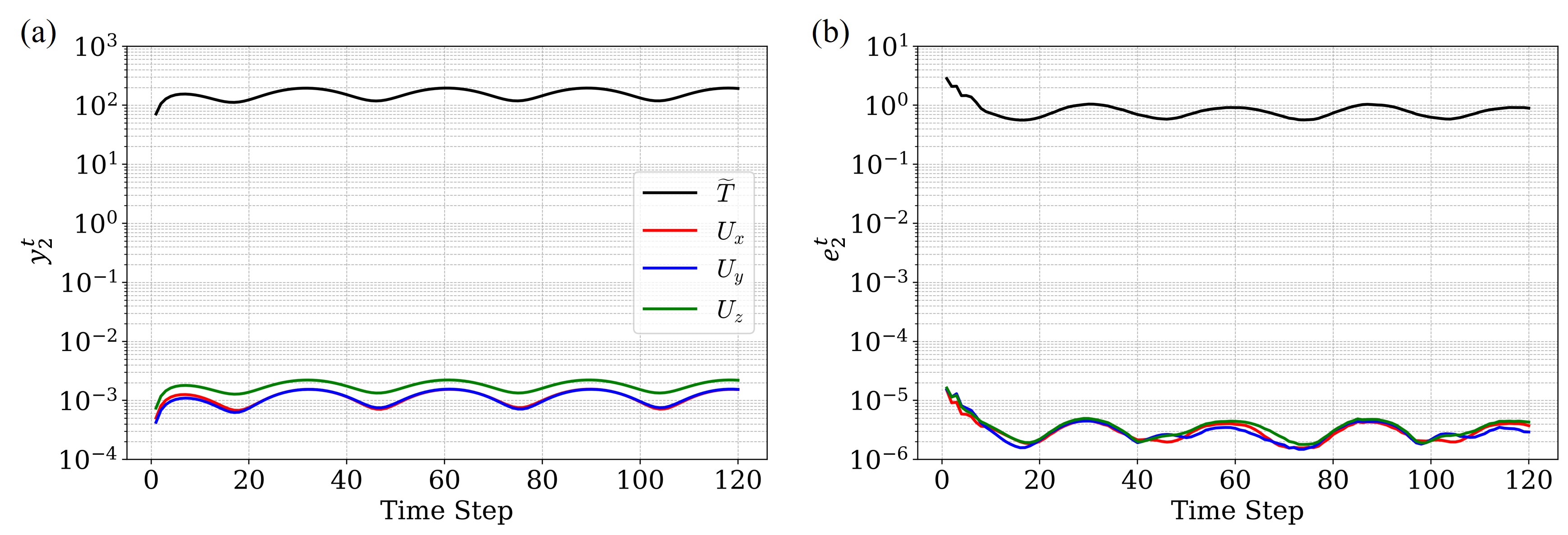}
\caption{a) Norm of true response $y_2^t$ and b) error norm $e_2^t$ for all components of the median load case applied to the 3D tube example using I-FENN.}
\label{fig:tube_early_stages}
\end{figure}

Error plots in Fig.~\ref{fig:tube_ifenn_vs_surrogate_median} revealed a relatively high error in the early stages of the simulation. To investigate the cause, we plot the norms of true values ($y_2^t = \left\| y_{true} \right\|_2$) and norms of errors ($e_2^t = \left\| y_{true} -y_{pred} \right\|_2$), as shown in Fig.~\ref{fig:tube_early_stages}. In the initial time steps, the true response norm $y_2^t$ is relatively small, while the error norm $e_2^t$ is relatively high compared to the later time steps. These two factors are participating in inflating the relative error $L_2^t$ in the early stages of the simulation (see Fig.~\ref{fig:tube_ifenn_vs_surrogate_median}), since relative error $L_2^t$ is computed as a ratio and becomes more sensitive when the denominator (true norm $y_2^t$) is small. 

\begin{figure}[hbt!]
\centering
\includegraphics[width=1.0\textwidth]{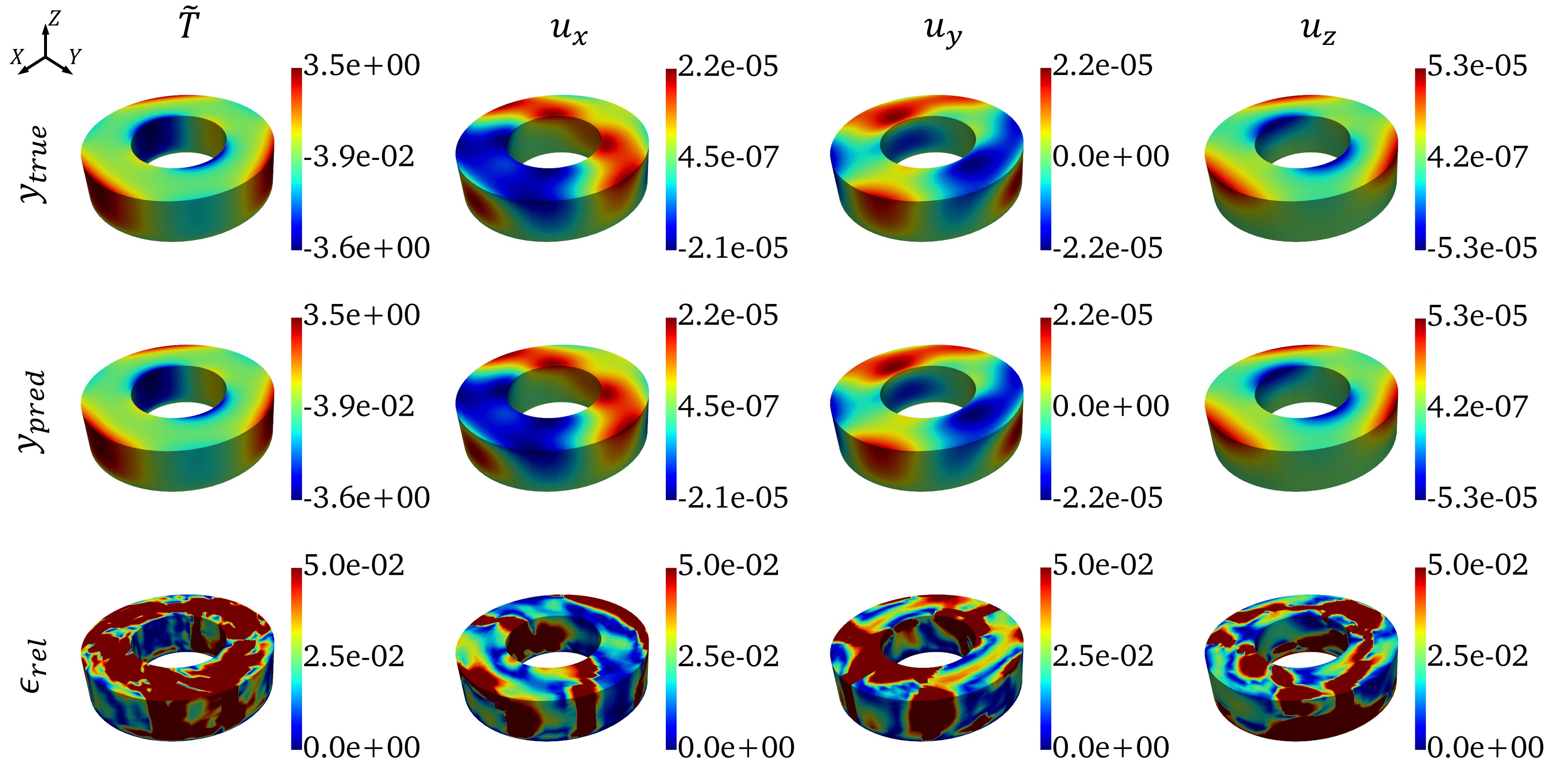}
\caption{Solution obtained using the fully coupled FEM solver ($y_{true}$) and I-FENN ($y_{pred}$) for the median load case at the 1st time step for the 3D tube example.}
\label{fig:tube_median_t1}
\end{figure}

\begin{figure}[hbt!]
\centering
\includegraphics[width=1.0\textwidth]{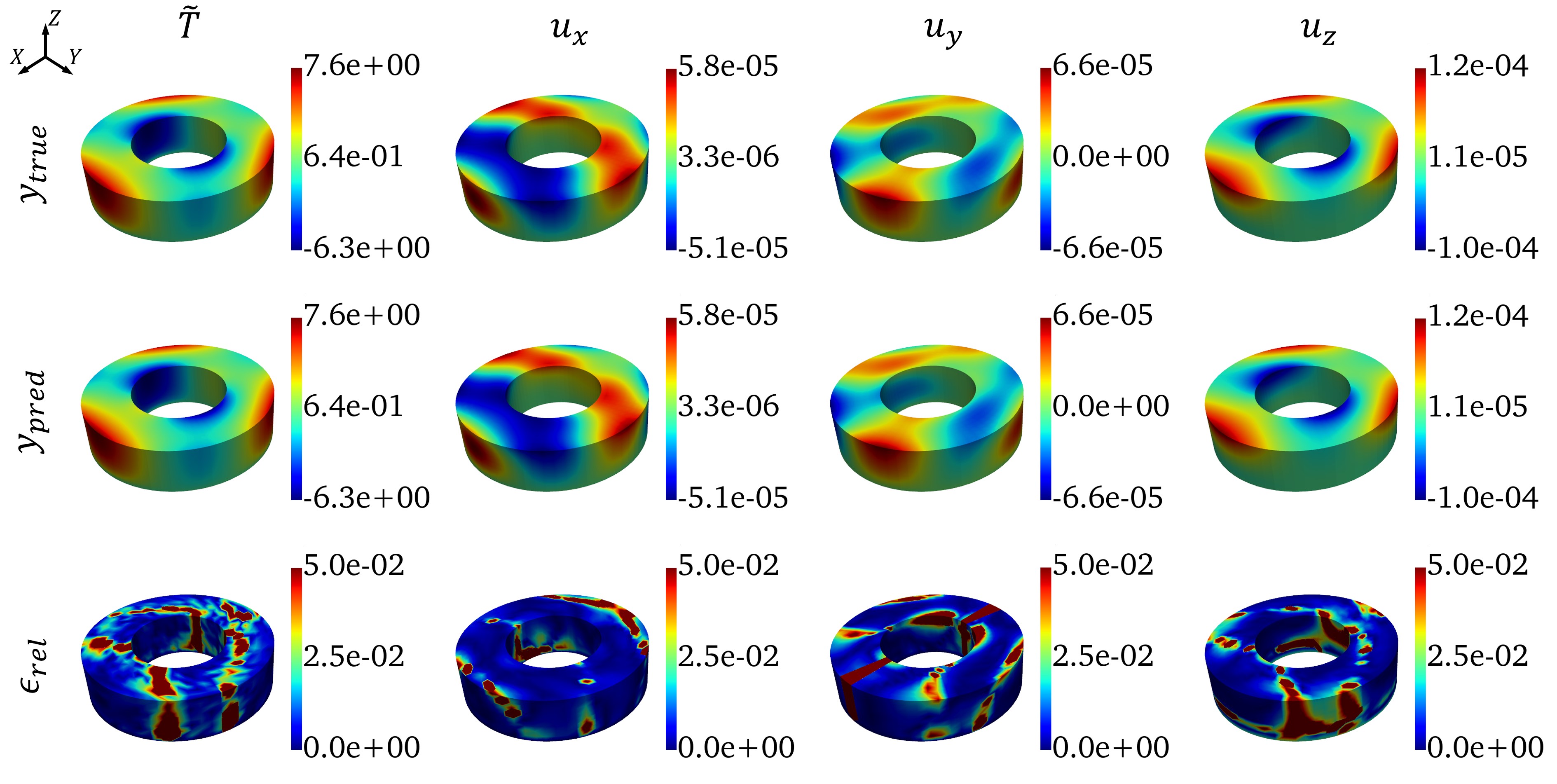}
\caption{Solution obtained using the fully coupled FEM solver ($y_{true}$) and I-FENN ($y_{pred}$) for the median load case at the 5th time step for the 3D tube example.}
\label{fig:tube_median_t5}
\end{figure}

To investigate the extent of the error in initial time steps, I-FENN results for the median load case applied to the 3D tube are plotted at the 1st and 5th time steps in Fig.~\ref{fig:tube_median_t1} and Fig.~\ref{fig:tube_median_t5}, respectively. It can be seen that at the 1st time step, the relative error values $\epsilon_{rel}$ at zones with maximum response are significantly lower than 5\%. However, areas with response values close to zero show large boundaries exceeding 5\%. Moving to the 5th time step Fig.~\ref{fig:tube_median_t5}, the results show a consistent behavior to that shown at the 120th time step (see Fig.~\ref{fig:tube_median}), indicating a stabilized behavior. In conclusion, the figures introduced in this section (Fig.~\ref{fig:tube_early_stages}, Fig.~\ref{fig:tube_median_t1}, and Fig.~\ref{fig:tube_median_t5}) suggest that despite having initial high $L_2^t$ at early time steps, the discrepancies are mainly concentrated in areas with response close to zero. Moreover, after almost 5 time steps (5\% of the whole simulation duration), the effects of early-stage errors introduced by the GRU appear to diminish significantly.

\subsubsection{Training dataset size}
\label{section:numerical_examples:tube:appendix:data_size}
\begin{figure}[hbt!]
\centering
\includegraphics[width=1.0\textwidth]{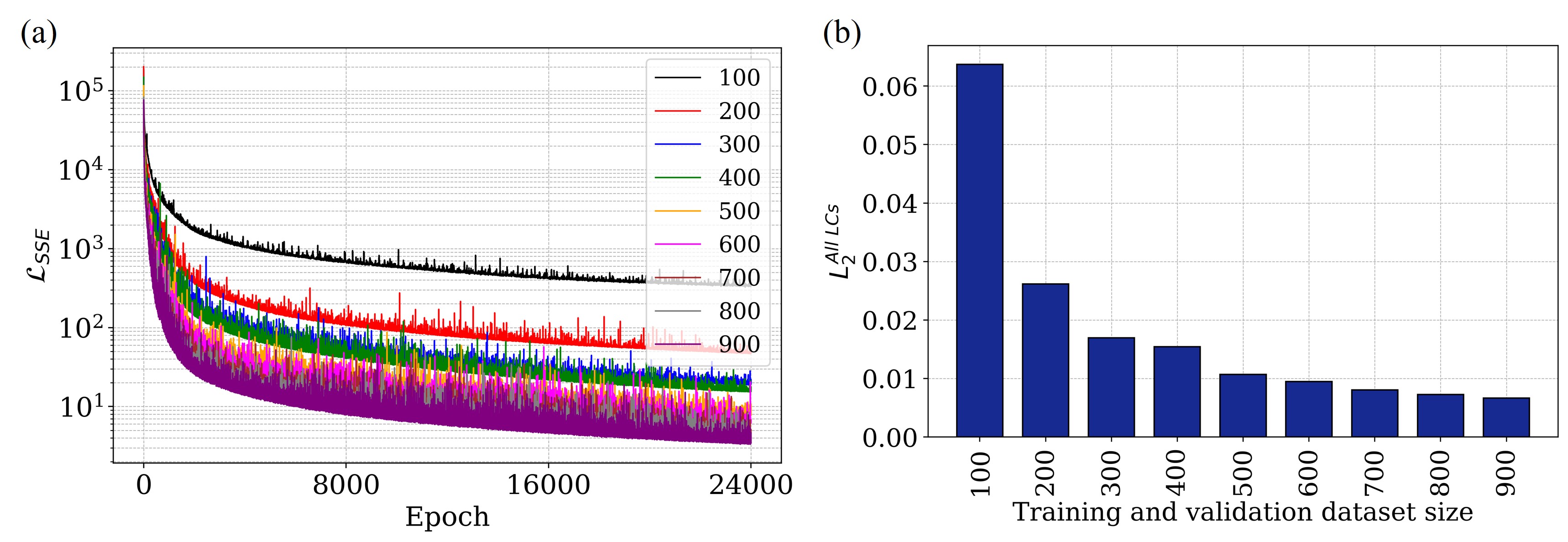}
\caption{DeepONet model performance considering different dataset sizes for the 3D tube example: a) validation loss $\mathcal{L}_{SEE}$, b) testing loss $L_2^{All\ LCs}$.}
\label{fig:tube_dataset_size_validation_testing}
\end{figure}

One of the important parametric studies carried out during the development of the current model is the dataset size and its effect on the model's generalizability. In this study, different dataset sizes (load case count) are considered for training the model, ranging from 100 to 900 load cases. A consistent set of 100 load cases is used for testing all the trained models. The validation loss during training and the final testing loss values are presented in Fig.~\ref{fig:tube_dataset_size_validation_testing}. Both plots exhibit a consistent trend: error values decrease as the dataset size increases, confirming that increased dataset size enhances the model's learning. 

All models are trained for the same number of epochs, which raises the question of whether the improved accuracy is primarily due to the increased number of load cases seen per epoch, rather than the diversity of those cases. To address this, a complementary study is conducted. In this study, the model trained with 200 load cases is trained four times the number of epochs used for the model with 800 load cases, and the model with 400 load cases is trained for twice as many epochs as the model trained with 800 load cases. 

\begin{figure}[hbt!]
\centering
\includegraphics[width=1.0\textwidth]{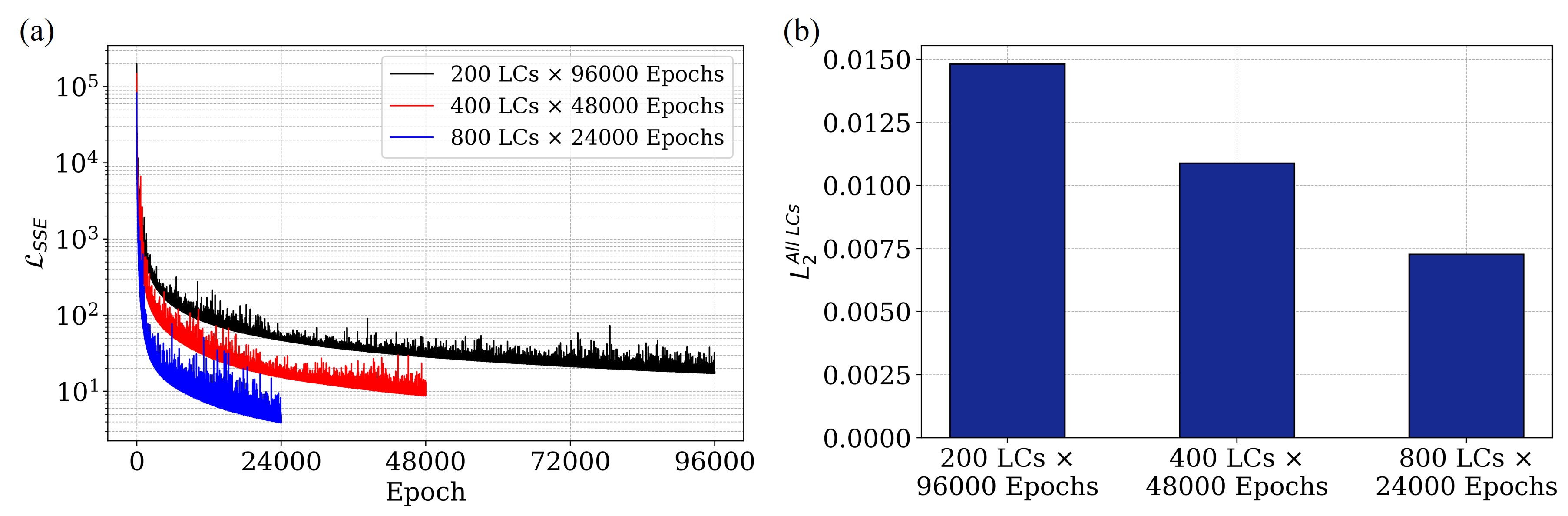}
\caption{DeepONet model performance considering different dataset sizes and number of epochs for the 3D tube example: a) validation loss $\mathcal{L}_{SEE}$, b) testing loss $L_2^{All\ LCs}$.}
\label{fig:tube_dataset_size_epochs_validation_testing}
\end{figure}

The validation loss during training and the final testing loss values for this study are presented in Fig.~\ref{fig:tube_dataset_size_epochs_validation_testing}. The results indicate that increasing the number of training epochs alone does not lead to the same improvement in model accuracy as increasing the dataset size. This suggests that the model benefits from the increasing load cases in the training data more than additional exposure to the same data, confirming the data diversity and its role in the learning process. 

\subsubsection{Training stabilization}
\label{section:numerical_examples:tube:appendix:stabilization}
\begin{figure}[hbt!]
\centering
\includegraphics[width=1.0\textwidth]{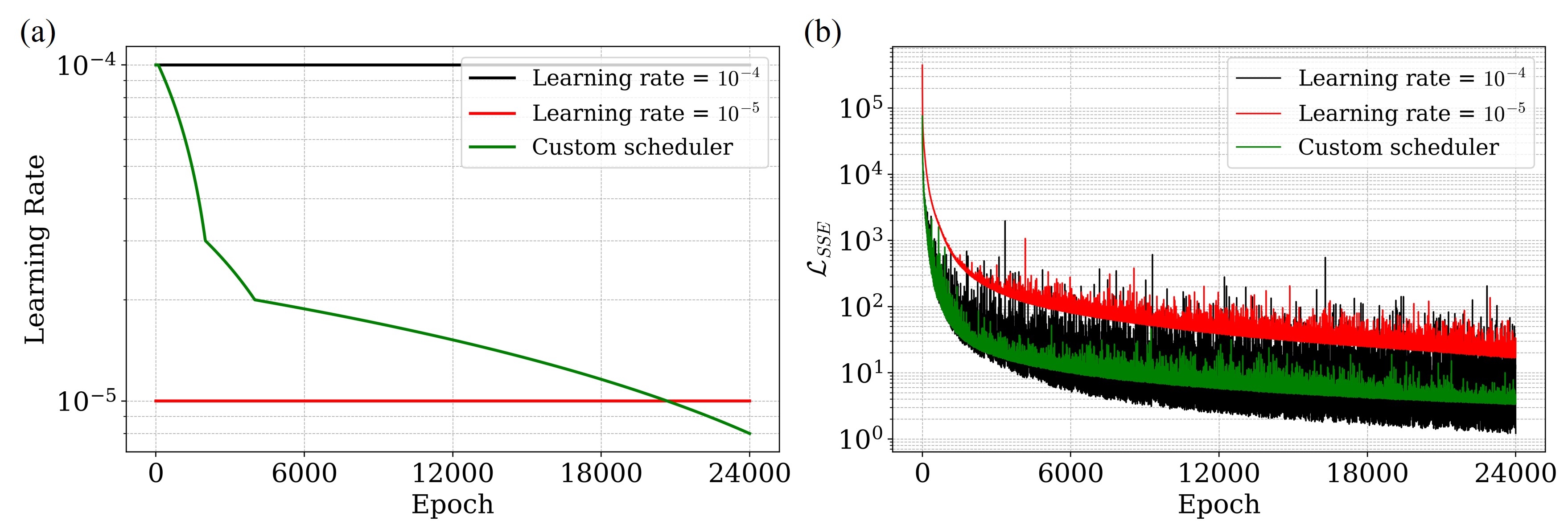}
\caption{Comparison between different learning rates investigated during training the DeepONet for the 3D tube example: a) learning rates, b) validation loss $\mathcal{L}_{SEE}$.}
\label{fig:tube_stabilization_lr_validation}
\end{figure}

One of the challenges that existed in the early stages of training the model was the training stability, where the loss curves showed high fluctuation along the epochs. To tackle this issue, a custom learning scheduler is adopted to stabilize training without impacting the training accuracy or the need for increased training epochs. Three different cases of learning rate options and their corresponding validation loss curves are shown in Fig.~\ref{fig:tube_stabilization_lr_validation}.  The results show that using a custom scheduler enhanced the learning stability, showing less noise than the high learning rate model and more accuracy than the low learning rate model. The custom scheduler option is adopted in developing the DeepONet model used for I-FENN integration.

\subsubsection{Normalization}
\label{section:numerical_examples:tube:appendix:normalization}
Various normalization strategies were evaluated, and three key approaches are highlighted here: 
\begin{itemize}
    \item (a) min-max normalization to scale the data between 0 and 1, expressed as:
    \begin{align}
        x_{\text{normalized}} = \frac{x - x_{\min}}{x_{\max} - x_{\min}}
    \end{align}
    \item (b) min-max normalization to scale the data between -1 and 1, expressed as:
    \begin{align}
       x_{\text{normalized}} = 2 \left( \frac{x - x_{\min}}{x_{\max} - x_{\min}} \right) - 1
    \end{align}
    \item (c) standardization using the mean and standard deviation of the data, expressed as:
    \begin{align}
       x_{\text{normalized}} = \frac{x - \mu_x}{\sigma_x}
    \end{align}
\end{itemize}

\begin{figure}[hbt!]
\centering
\includegraphics[width=1.0\textwidth]{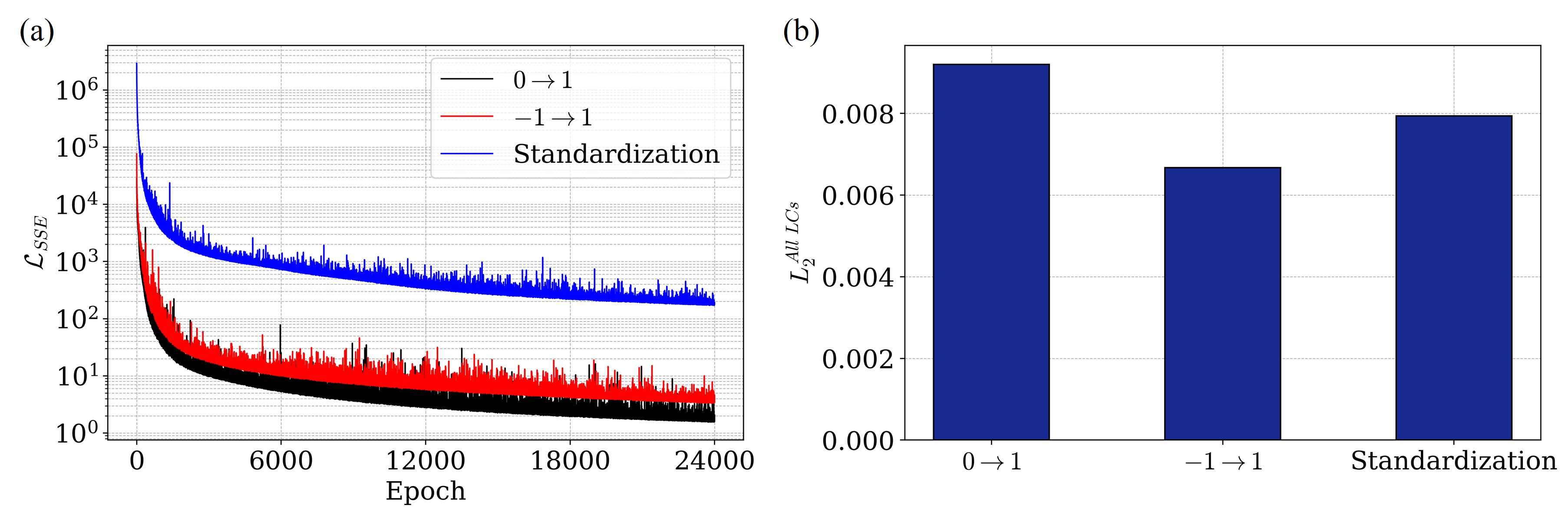}
\caption{DeepONet model performance considering different data normalization/standardization options: a) validation loss $\mathcal{L}_{SEE}$, b) testing loss $L_2^{All\ LCs}$.}
\label{fig:tube_normalization_training_testing}
\end{figure}

The validation losses vs epochs using different normalization techniques are provided in Fig.~\ref{fig:tube_normalization_training_testing}a, showcasing training stability for all options. However, accuracy can not be inferred using loss values as they are computed during the training phase with different normalization techniques, which inherently shift the data range, affecting normalized error computation. To address this discrepancy, during the testing phase, all outputs are denormalized before computing the loss metric $ L_2^{All\ LCs}$ described in Section~\ref{section:deeponet_for_ifenn:training_testing}.  Given that the lowest testing error is given by normalizing between -1 and 1 (refer to Fig.~\ref{fig:tube_normalization_training_testing}b), this approach is adopted for the DeepONet trained for I-FENN implementation for the 3D tube example. 

\subsubsection{Loss function}
\label{section:numerical_examples:tube:appendix:loss_function}
\begin{figure}[hbt!]
\centering
\includegraphics[width=1.0\textwidth]{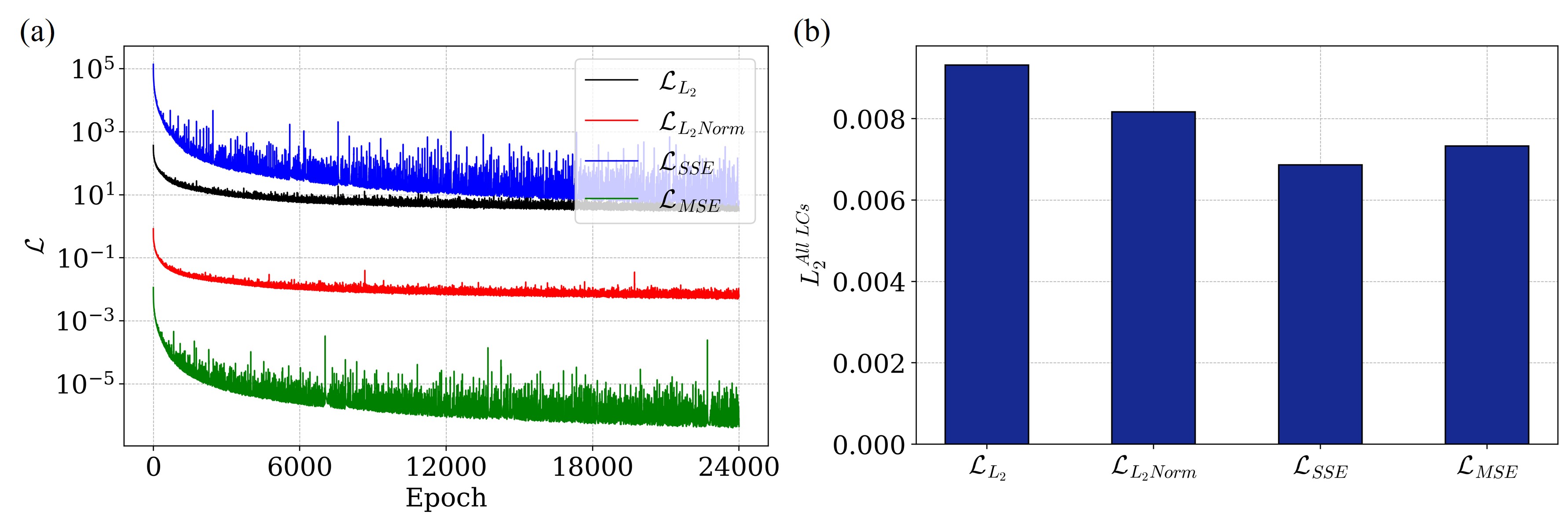}
\caption{DeepONet model performance on initial trials considering different training loss functions: a) validation loss $\mathcal{L}$, b) testing loss $L_2^{All\ LCs}$.}
\label{fig:tube_loss_function_training_testing_init}
\end{figure}

Different loss functions were tested in the initial stage of developing the DeepONet model. The reader is referred to Section~\ref{section:deeponet_for_ifenn:training_testing} for definitions of those loss functions. Figure~\ref{fig:tube_loss_function_training_testing_init} shows the validation and testing loss using different loss functions during the early stages of model training trials. The loss function $\mathcal{L}_{\text{SSE}}$ showed the least testing error and, hence, was selected for developing the final model. 

\begin{figure}[hbt!]
\centering
\includegraphics[width=1.0\textwidth]{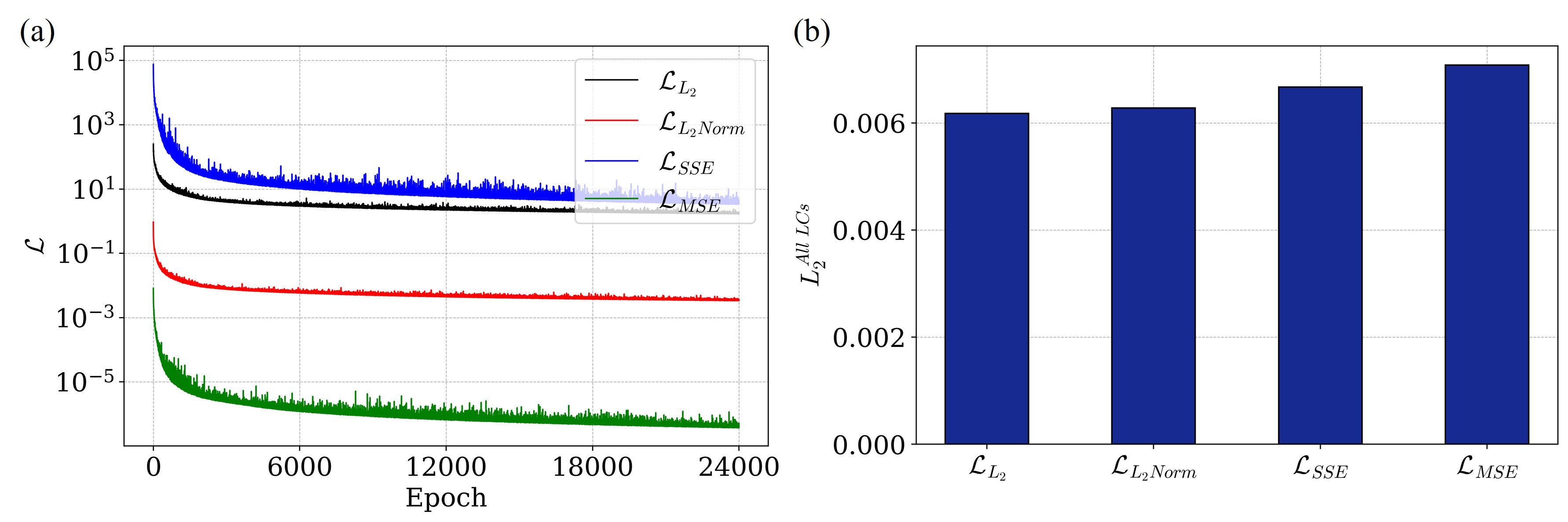}
\caption{Latest DeepONet model performance considering different training loss functions: a) validation loss $\mathcal{L}$, b) testing loss $L_2^{All\ LCs}$.}
\label{fig:tube_loss_function_training_testing}
\end{figure}

During the development of the final model, and due to different parametric studies, many hyperparameters were changed, including learning rate, batch size, and dataset size. On refreshing the comparison between different loss functions for the latest training setup, we find that the previous conclusion holds no more, as shown in Fig.~\ref{fig:tube_loss_function_training_testing}. For the latest setup, the best-performing loss function is $\mathcal{L}_{L_2}$ compared to $\mathcal{L}_{\text{SSE}}$ in the initial training trials. This finding highlights the dynamic nature of NNs and how changes in hyperparameters can affect the effectiveness of other training hyperparameters.

\subsubsection{Branch and trunk size}
\label{section:numerical_examples:tube:appendix:branch_trunk_size}
\begin{figure}[hbt!]
\centering
\includegraphics[width=1.0\textwidth]{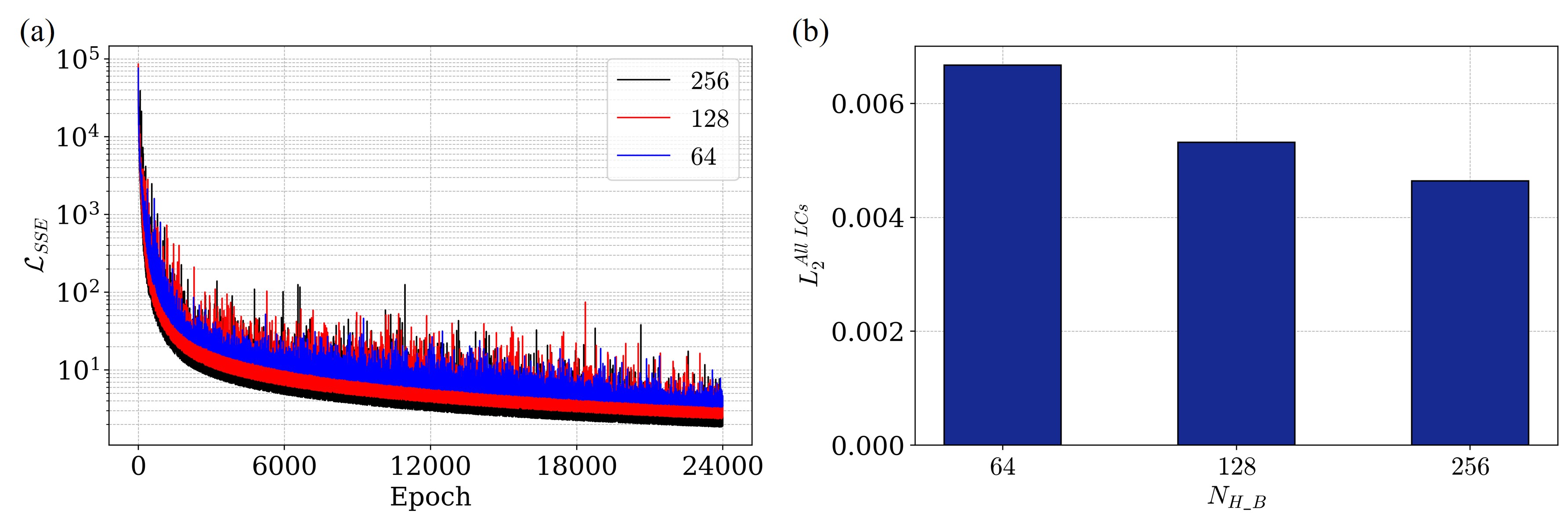}
\caption{Effect of the branch hidden layer size $N_{H\_B}$ on the DeepONet model: a) validation loss $\mathcal{L}_{SEE}$, b) testing loss $L_2^{All\ LCs}$.}
\label{fig:tube_branch_validation_testing}
\end{figure}

\begin{figure}[hbt!]
\centering
\includegraphics[width=1.0\textwidth]{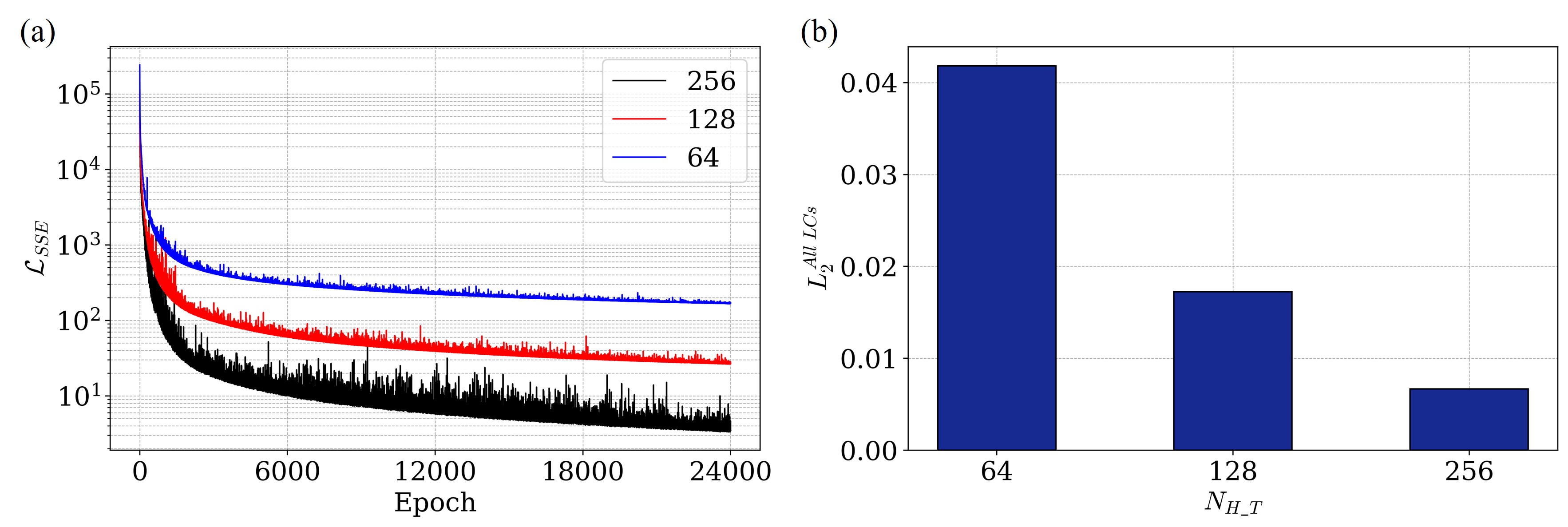}
\caption{Effect of the trunk hidden layer size $N_{H\_T}$ on the DeepONet model: a) validation loss $\mathcal{L}_{SEE}$, b) testing loss $L_2^{All\ LCs}$.}
\label{fig:tube_trunk_validation_testing}
\end{figure}

Among the parametric studies conducted for hyperparameter selection, different values for the hidden layer size $(N_H)$ for the branch and trunk were investigated. Fig.~\ref{fig:tube_branch_validation_testing} shows the effect of branch hidden size on validation and testing errors. Alternatively, Fig.~\ref{fig:tube_trunk_validation_testing} depicts the effect of branch hidden size on validation and testing errors. Results show that no significant accuracy improvement is detected on increasing the branch hidden size. However, increasing the trunk hidden size showed a substantial increase in accuracy, highlighting the complexity of geometry being embedded through the trunk. Given these results, hidden layer sizes of 64 and 256 are selected for the branch and trunk, respectively, as listed in Table~\ref{tab:tube_deeponet_arch}.

\subsubsection{I-FENN results for the 10th and 90th error-percentile load cases}
\label{section:numerical_examples:tube:appendix:10-90}

\begin{figure}[hbt!]
\centering
\includegraphics[width=1.0\textwidth]{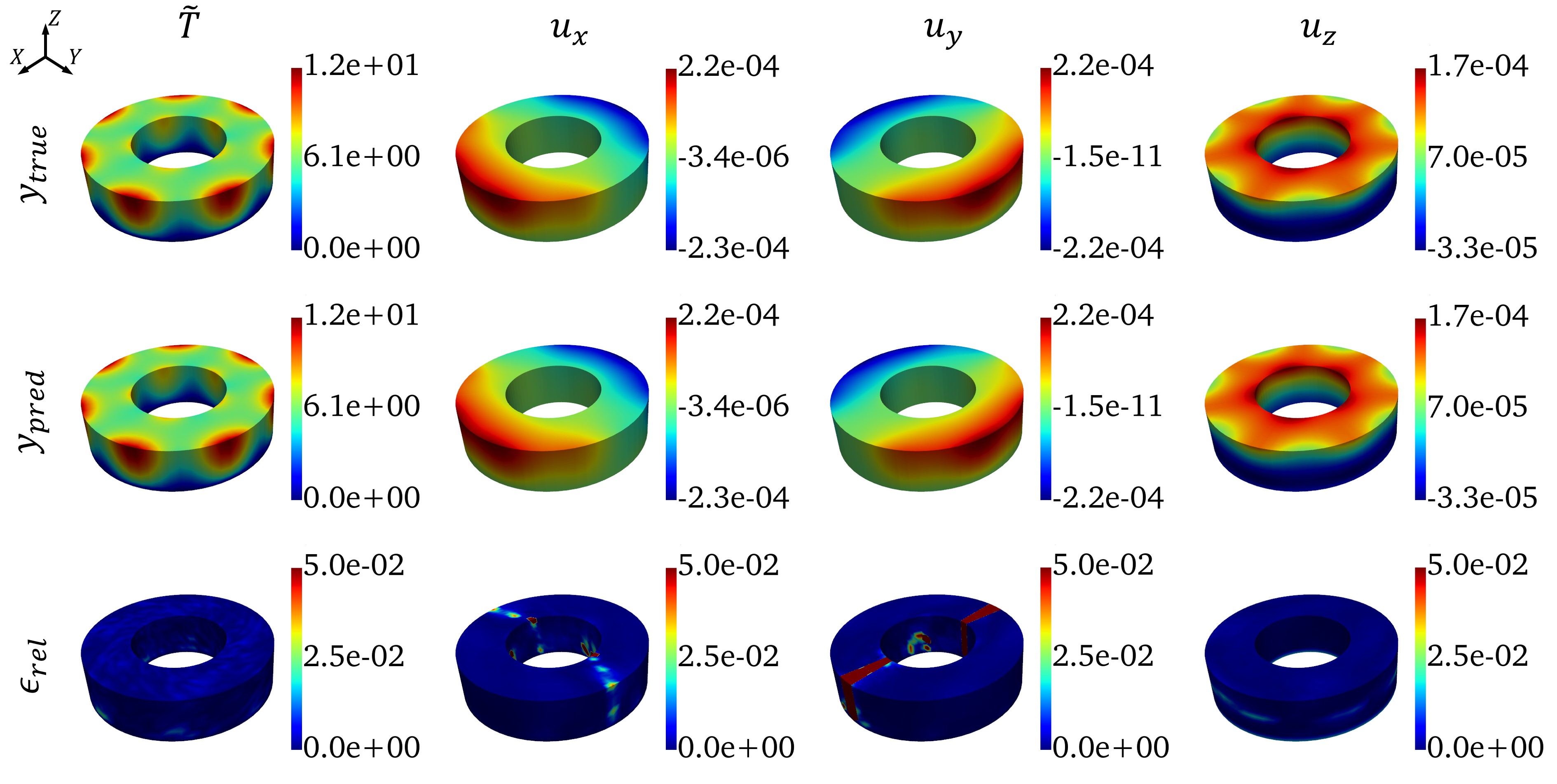}
\caption{Solution obtained using the fully coupled FEM solver ($y_{true}$) and I-FENN ($y_{pred}$) for the 10th error-percentile load case at the 120th time step for the 3D tube example.}
\label{fig:tube_10th}
\end{figure}

\begin{figure}[hbt!]
\centering
\includegraphics[width=1.0\textwidth]{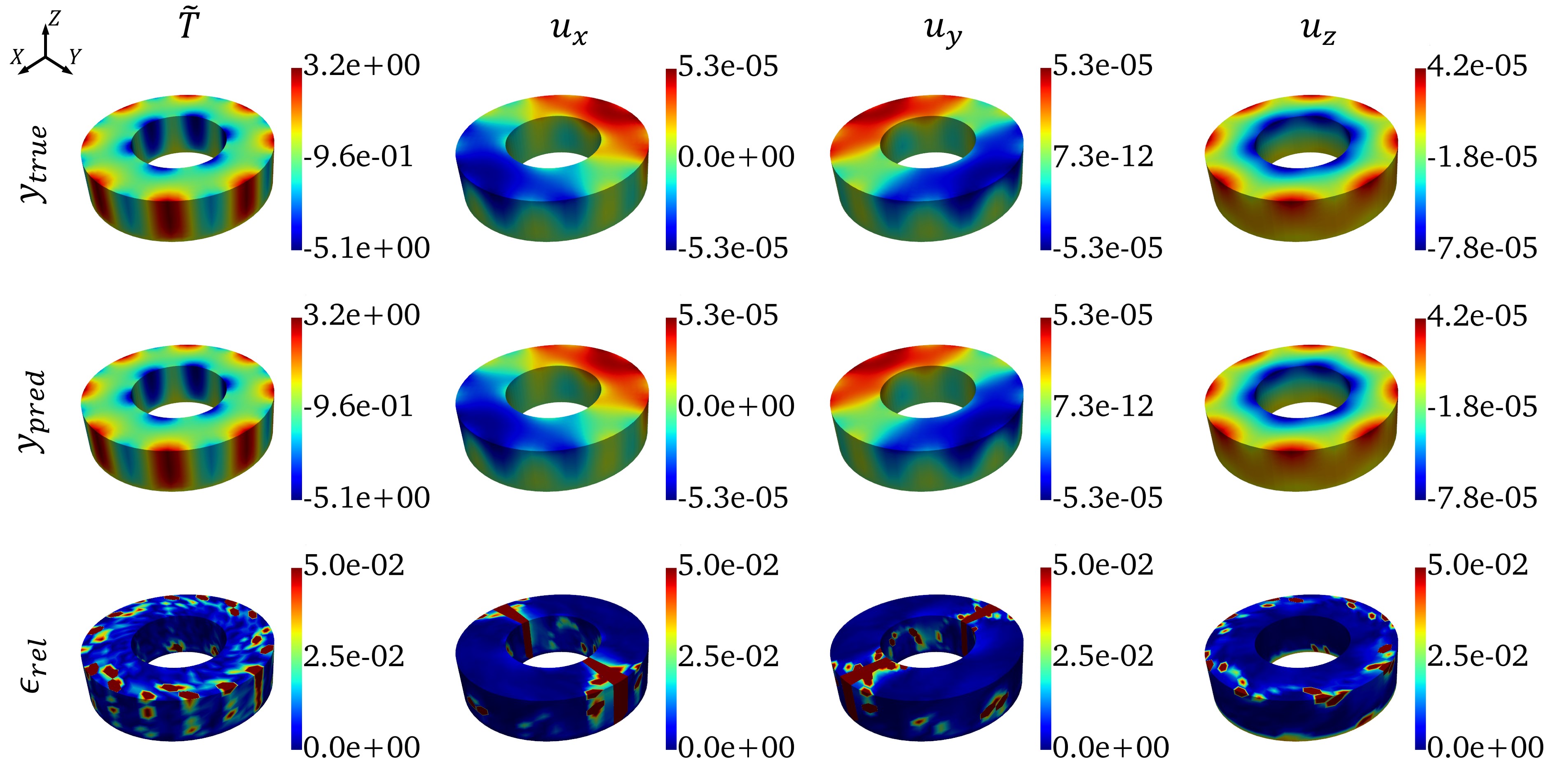}
\caption{Solution obtained using the fully coupled FEM solver ($y_{true}$) and I-FENN ($y_{pred}$) for the 90th error-percentile load case at the 120th time step for the 3D tube example.}
\label{fig:tube_90th}
\end{figure}

I-FENN results for the 10th and 90th error-percentile load cases are depicted in  Fig.~\ref{fig:tube_10th} and Fig.~\ref{fig:tube_90th}, respectively. The accuracy behavior of the framework is consistent with the median load case results shown in Fig.~\ref{fig:tube_median}, with much less error for the 10th error-percentile load case. Results demonstrate the reliability of the I-FENN framework across varying loading conditions. 

\subsection{Poroelasticity example: 2D Excavation problem with hydraulic boundary flux (Dewatering)}
\label{section:numerical_examples:excav:appendix}

\subsubsection{Training stabilization}
\label{section:numerical_examples:excav:appendix:stabilization}
\begin{figure}[hbt!]
\centering
\includegraphics[width=1.0\textwidth]{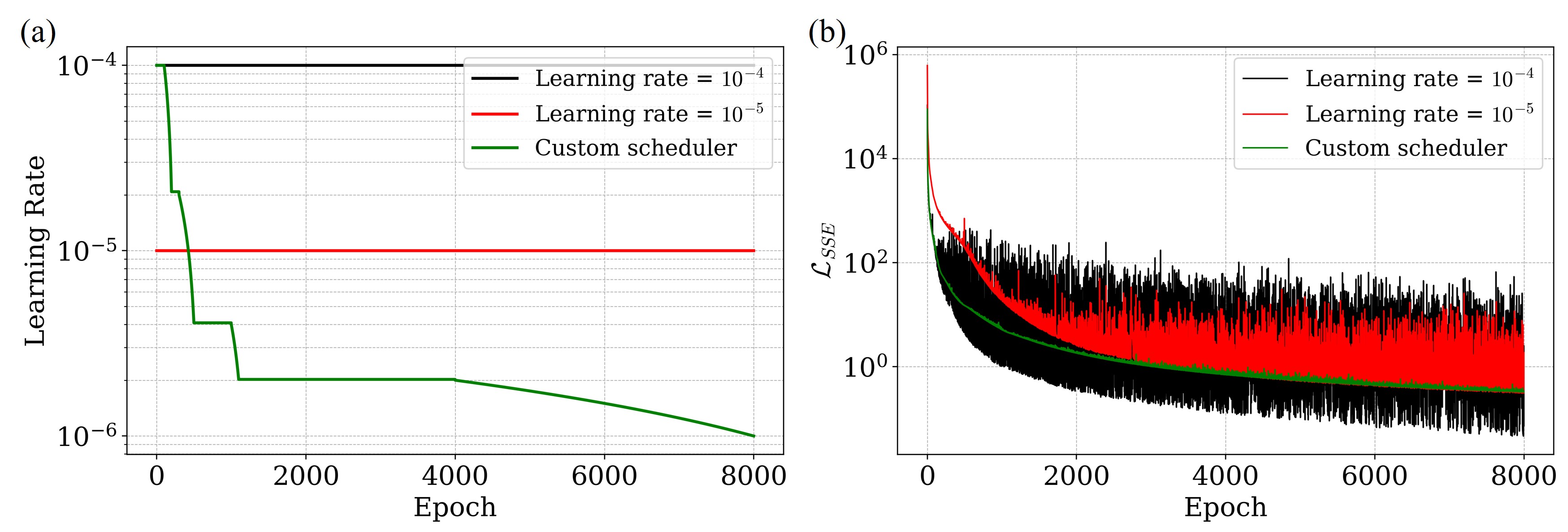}
\caption{Comparison between different learning rates investigated during training the DeepONet for the 2D excavation example: a) learning rates, b) validation loss $\mathcal{L}_{SEE}$.}
\label{fig:excav_stabilization_lr_validation}
\end{figure}

Training stability was one of the main challenges encountered during the early stages of model development, as the loss curves exhibited significant fluctuations across epochs. A custom learning rate scheduler was implemented, aiming to stabilize training without compromising accuracy or requiring additional training epochs. Three different learning rate choices with their corresponding validation loss curves are depicted in Fig.~\ref{fig:excav_stabilization_lr_validation}. The results highlight the importance of using a training scheduler that allows broad exploration in the early stages through a high learning rate, followed by a gradual reduction to enable precise fine-tuning as training progresses. The custom scheduler learning rate shown in Fig.~\ref{fig:excav_stabilization_lr_validation}a is the adopted scheduler for the DeepONet model used in the I-FENN implementation, with results in Section~\ref{section:numerical_examples:excav:results}

\subsubsection{I-FENN results for the 10th and 90th error-percentile load cases}
\label{section:numerical_examples:excav:appendix:10-90}
\begin{figure}[hbt!]
\centering
\includegraphics[width=1.0\textwidth]{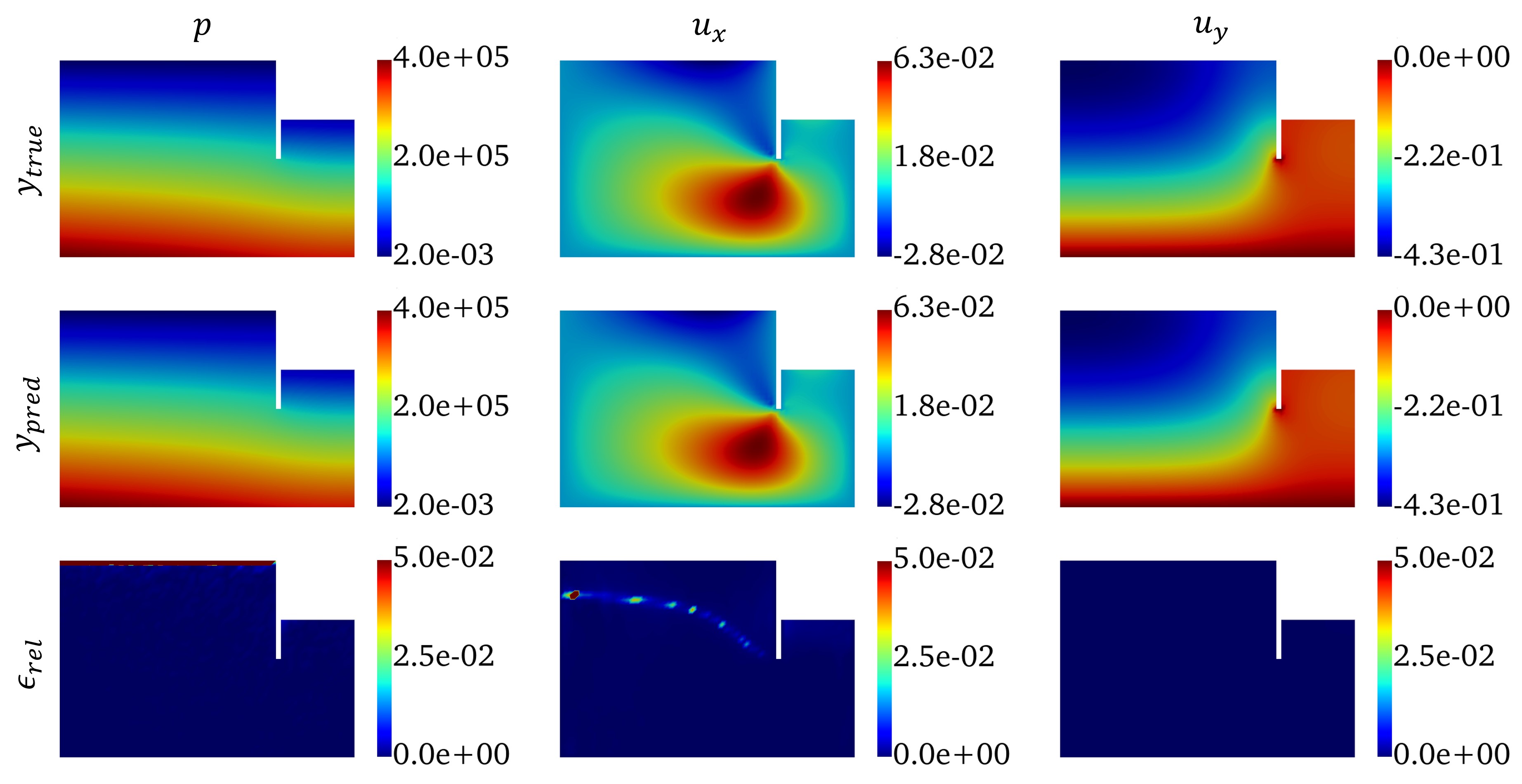}
\caption{Solution obtained using the fully coupled FEM solver ($y_{true}$) and I-FENN ($y_{pred}$) for the 10th error-percentile load case at the 60th time step for the 2D excavation example.}
\label{fig:excav_10th}
\end{figure}

\begin{figure}[hbt!]
\centering
\includegraphics[width=1.0\textwidth]{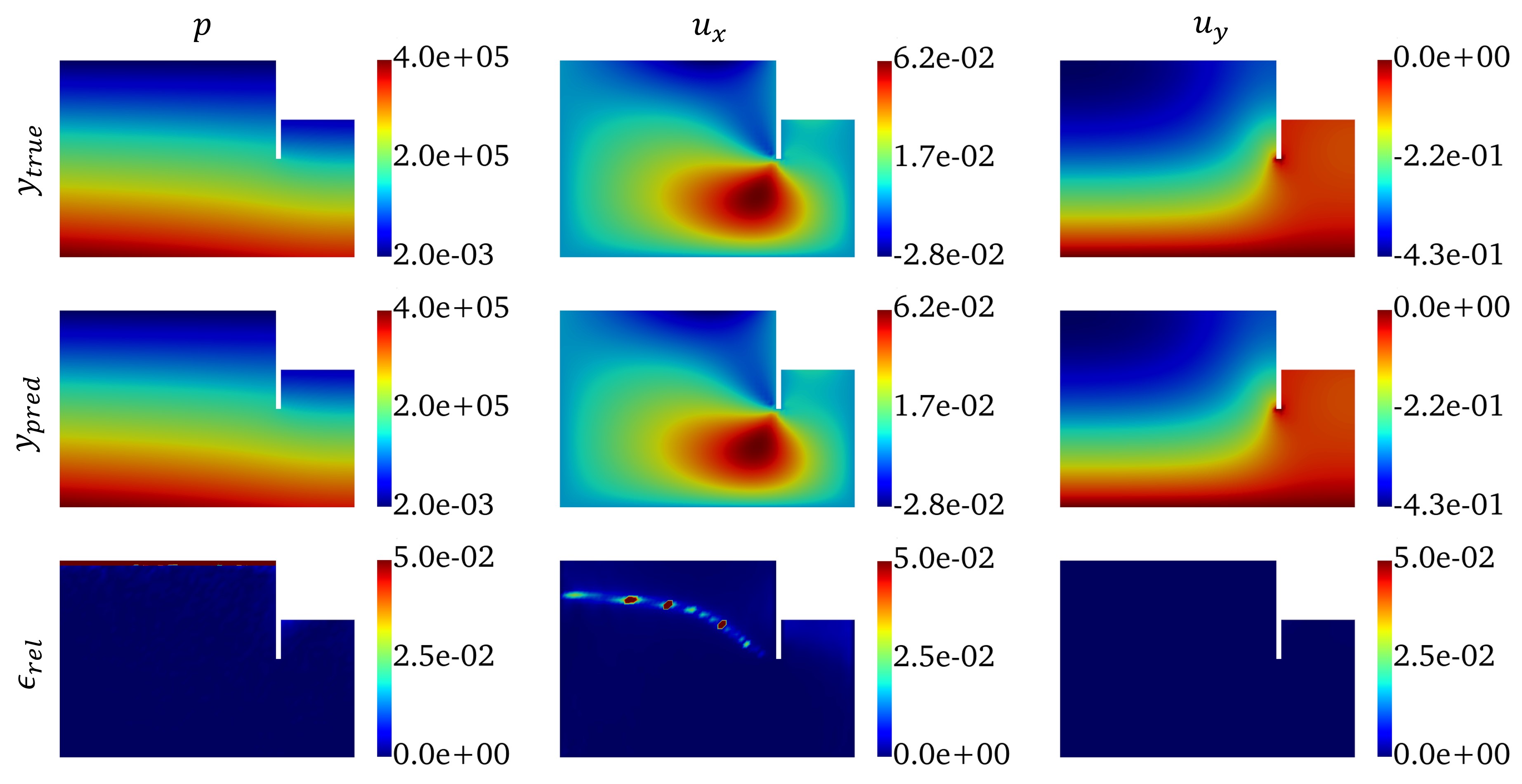}
\caption{Solution obtained using the fully coupled FEM solver ($y_{true}$) and I-FENN ($y_{pred}$) for the 90th error-percentile load case at the 60th time step for the 2D excavation example.}
\label{fig:excav_90th}
\end{figure}

I-FENN results for the 10th and 90th error-percentile load cases of the 2D excavation example are presented in Fig.~\ref{fig:excav_10th} and Fig.~\ref{fig:excav_90th}, respectively. A consistent accuracy is detected compared to the median load case presented in Fig.~\ref{fig:excav_median}, highlighting the I-FENN's reliability for different loading conditions.

\FloatBarrier
\bibliographystyle{elsarticle-num}
\bibliography{references/references}

\end{document}